\documentclass[reprint, superscriptaddress, amsmath, amssymb, aps]{revtex4-2}

\usepackage{graphicx} 
\usepackage{subcaption} 
\usepackage{ragged2e} 

\usepackage[usenames,dvipsnames]{xcolor} 
\usepackage[most]{tcolorbox} 

\usepackage{tabularx} 
\usepackage{ragged2e} 

\usepackage{tikz}

\makeatletter
\renewcommand{\@makecaption}[2]{%
  \vskip\abovecaptionskip
  \sbox\@tempboxa{#1: #2}%
  \ifdim \wd\@tempboxa > \hsize
    \justifying #1. #2\par
  \else
    \hbox to\hsize{\hfil\box\@tempboxa\hfil}%
  \fi
  \vskip\belowcaptionskip}
\makeatother

\usepackage[normalem]{ulem}

\usepackage{hyperref}

\hypersetup{
    colorlinks=true,
    linkcolor=blue,
    filecolor=red,      
    urlcolor=cyan,
    citecolor=red,
    pdftitle={Overleaf Example},
    pdfpagemode=FullScreen,
    }

\newcommand{\rev}[1]{#1}

\begin{document}

\preprint{APS/123-QED}

\title{The XY model with vision cone: non-reciprocal vs.~reciprocal interactions}

\author{Gabriele Bandini}
 \email{gbandini@sissa.it}
 \affiliation{SISSA --- International School for Advanced Studies and INFN, via Bonomea 265, 34136 Trieste, Italy }
 \author{Davide Venturelli}
\affiliation{Laboratoire de Physique Th\'eorique de la Mati\`ere Condens\'ee, CNRS/Sorbonne Universit\'e, 4 Place Jussieu, 75005 Paris, France}
\affiliation{Laboratoire Jean Perrin, CNRS/Sorbonne Universit\'e, 4 Place Jussieu, 75005 Paris, France}
\author{Sarah A.~M.~Loos}
 \affiliation{DAMTP, University of Cambridge, Wilberforce Rd, Cambridge CB3 0WA, United Kingdom} 
 \author{Asja Jelic}
 \affiliation{ICTP --- The Abdus Salam International Centre for Theoretical Physics, Strada Costiera 11, 34151 Trieste, Italy}
 \author{Andrea Gambassi}
 \affiliation{SISSA --- International School for Advanced Studies and INFN, via Bonomea 265, 34136 Trieste, Italy }

\date{\today}

\begin{abstract}
We study the behavior of the classical XY model on a two-dimensional square lattice, with interactions occurring within a vision cone of each spin. 
Via Monte Carlo simulations, we explore one non-reciprocal and two reciprocal implementations of these interactions.  
The corresponding energy involves couplings that depend non-trivially on the system's configuration, leading to both long-range and quasi-long-range ordered phases at low temperatures. 
Our results demonstrate that non-reciprocity is not essential for achieving long-range order at low temperatures. 
Using symmetry arguments, we provide a theoretical framework to explain these findings, and additionally we uncover an unexpected order-by-disorder transition. 
\end{abstract}

\maketitle
\tableofcontents
\section{Introduction}\label{sec:intro}

Among the lattice models 
studied in statistical physics, the 
XY model in spatial dimension $d=2$ occupies a prominent 
position, establishing itself as a paradigm for studying topological 
phase transitions. Unlike the Ising model, which in spatial dimensions $d\geq 2$ undergoes a second-order phase transition from a phase with long-range order (LRO) to a disordered (DO) phase~\cite{Onsager1944}, the two-dimensional 
XY model, with its continuous $O(2)$ symmetry, is constrained by the Mermin-Wagner theorem~\cite{Mermin1966}. Consequently, it does not exhibit true long-range order, but instead undergoes a topological phase transition from a quasi-long-range ordered (QLRO) phase to a DO phase~\cite{Kosterlitz_1973,kosterlitz1974}.

Vision cones (VC) have recently been introduced in the standard two-dimensional XY model~\cite{Loos2023} to explore the emergence of LRO in the context of 
two-dimensional flocks. Flocks consist of self-propelled, off-lattice agents which tend to align their 
velocities
with 
those of their neighbors. The pioneering studies of Vicsek~\cite{Vicsek1995} and Toner and Tu~\cite{Toner1995} demonstrated the emergence of LRO despite the continuous rotational symmetry of the velocity vectors. However, this fact does not contradict the Mermin-Wagner theorem 
because these systems are not in thermal 
equilibrium, a crucial assumption of the theorem.

A more detailed 
modeling of flock behavior involves incorporating non-reciprocal (NR) interactions, 
defined as those that violate Newton's third law (i.e.~the action-reaction principle). NR interactions lead to intriguing phenomena in active matter systems~\cite{Uchida2010, Shankar2022, Saha_2019, Nagy2010,cavagna2017nonsymmetric, Yllanes_2017,tan2022odd, Dadhichi2020,Besse_2022}, neural  networks~\cite{Montbri2018, Sompolinsky1986}, and metamaterials~\cite{Scheibner2020, Brandenbourger2019}. In biologically-inspired models, vision is often 
a source of non-reciprocity, for instance:~when agent $A$ observes agent $B$, $A$'s behavior may be influenced by $B$ (e.g., $A$ might align its motion with that of $B$); however, $A$ will 
influence $B$ only if $B$ reciprocates by looking 
at $A$. This asymmetric influence renders the interaction between $A$ and $B$ inherently non-reciprocal. 

Although the effects of the VC have already been considered 
in diverse models of self-propelled particles~\cite{Li_2011, Nguyen2015,Barberis2016,Durve2018,Couzin2002,Hildenbrandt2010,Dadhichi2020,Besse_2022}, 
they 
were only recently introduced in a lattice model in Ref.~\cite{Loos2023}. 
There, each two-component spin of the XY model in two spatial dimensions (thought as laying on the lattice plane) is equipped with a VC of amplitude $\theta$ centered around its orientation, and it interacts only with the neighboring spins within its field of view. This creates asymmetric microscopic couplings and renders the overall dynamics non-reciprocal. 
We shall refer to this system as the non-reciprocal XY model (NRXY). 
The non-reciprocal nature of the Monte Carlo dynamics
drives the system
into a non-equilibrium state, exhibiting a LRO phase at low temperatures. 
This behavior can be thought of as ``flocking without moving''~\cite{Dadhichi2020}, representing an abstract scenario where we observe, from a co-moving frame, a perfectly organized flock on a lattice, with no spatial fluctuations.
However, the existence of \textit{ bona-fide} long-range order in similar models is still under debate~\cite{Solon2022polarflocks,Chatterjee2022polarflocks,Besse_2022,solon-journal-club,rouzaire2024nonreciprocal}.

In this work, we show that the VC interaction can be designed also in a reciprocal way,
and we demonstrate that the emergence of a LRO phase in this context does \textit{not} actually depend on the non-reciprocal nature of the interactions. 
For this purpose, we introduce two reciprocal variations of the model: 
the asymmetric reciprocal XY model (ARXY), and the symmetric reciprocal XY model (SRXY).
Explanations of these names (in particular a subtle distinction we make between asymmetry and non-reciprocity) can be found in Sec.~\ref{sec:models}.
We will show that both models, whose couplings depend non-trivially on the system configuration,  turn out to feature 
an equilibrium stationary state with a LRO phase at low temperatures.


Our findings suggest that the emergence of LRO is not due to the non-equilibrium nature of the dynamics of the system, as seen in Vicsek-type models. Instead, LRO arises from the interplay between the VC and the lattice geometry, which breaks the continuous internal symmetry (e.g., for a square lattice, the $O(2)$ symmetry is effectively reduced to $\mathbb{Z}_4$). 

We study in detail the phase diagrams of the three models, and show that those of the NRXY and ARXY models are qualitatively analogous. 
In the latter, we find that the transition from LRO to DO exhibits different features depending on 
the 
value of $\theta$. 
The SRXY model, instead, exhibits a QLRO phase for $\theta> 180^{\circ}$.
However, for $\theta \gtrsim 180^{\circ}$ we find that, upon increasing the temperature from low values, a transition occurs from QLRO to LRO, 
a phenomenon similar to an order-by-disorder transition~\cite{Melko2005,Honecker2011,Perkins2015}.

Finally, we present a theoretical framework based on the symmetries of the models, which accounts for their peculiarities. 
These include couplings that depend on the system configuration and an internal degree of freedom intricately coupled to the lattice structure. Within this framework, we can 
rationalize 
our findings.

This work is organized as follows. In Sec.~\ref{sec:models} we introduce the three variants of the XY model on a square lattice which will be the subject of our study. Section~\ref{sec:methods} is dedicated to methods: in particular, Sec.~\ref{sec:observables} introduces the observables of interest, while Sec.~\ref{sec:toymodels} summarizes their behavior in three benchmark models (the XY model, the 4-state clock model, and the 6-state clock model). An analysis of the NRXY model, complementing the one of Ref.~\cite{Loos2023}, is presented in Sec.~\ref{sec:results-nrxy}. Section~\ref{sec:arxy-results} is focused on 
the ARXY model: its phase diagram is presented in Sec.~\ref{sec:arxy-pd}, while in Sec.~\ref{sec:symmetry} we introduce a theoretical framework based on internal symmetry which 
allows us to elucidate
the behavior observed for two specific values $\theta$ of the VC
in Sec.~\ref{sec:arxy100280-interpretation}.
In Sec.~\ref{sec:results-srxy}
we characterize the SRXY model by discussing its phase diagrams in Sec.~\ref{sec:srxy-pd}, and the mechanism responsible for the presence of QLRO for $180^\circ< \theta < 360^\circ$ in Sec.~\ref{sec:srxy-redundant}. In Sec.~\ref{sec:obd} we use the 
arguments previously introduced 
to rationalize the order-by-disorder transition observed for $\theta \gtrsim 180^\circ$. Details on the implementation of the Monte Carlo simulations, and 
additional results concerning the NRXY and ARXY models, can be found in the Appendices.

\section{The XY model with vision cone: three variations}\label{sec:models}

In this section, we introduce three variations of the classical XY model with short-range couplings  and vision-cone interactions on the square lattice with periodic boundary conditions (PBCs).

To each lattice site $i$ we associate a classical two-component spin variable $\mathbf{s}_i = (\cos\phi_i,\sin\phi_i)$, i.e., a unit-length vector parameterized by the angle $\phi_i$ that it forms with a reference direction on the plane. The spin is thought to belong to the same plane as the lattice (a crucial difference from models usually studied on the lattice, where the internal degrees of freedom are not 
coupled to the spatial 
structure), and the corresponding VC has an opening angle $\theta$ around the spatial direction of the spin. For convenience, the angle $\phi_i$ is determined with respect to one of the principal axes of the lattice.

The update protocol of the spins in the Monte Carlo (MC) dynamics
is defined as follows: let $\Delta_i E$
denote the variation of some energy 
$E$ (which we do not need to specify at this stage) caused by 
the spin update $\phi_i \to \phi_i^{\prime}$ attempted at lattice site $i$. This proposed update is then accepted with probability
\begin{equation}\label{eq:glaub}
    w_G\left(\phi_{i} \rightarrow \phi_{i}^{\prime}\right) =\frac{1}{2}\left[1-\tanh \left(\frac{\beta\,
    \Delta_i E}{2}
    \right)\right]
\end{equation}
in the Glauber protocol~\cite{Marro_Dickman_1999}.
The details of the implementation of the MC dynamics are reported in App.~\ref{sec:simulations}, where we also discuss the strategy that we adopt in order to avoid that the MC simulations get trapped in long-lived metastable states at low temperatures.
While for equilibrium systems this dynamics satisfies detailed balance --- and thus leads to a distribution in which a configuration with energy $E$ has probability $p \propto e^{-\beta E}$, where $\beta = 1/T$ is the inverse temperature --- the situation is different for non-reciprocal models. In these 
systems, the dynamics leads to a non-equilibrium steady state, where $\beta^{-1}$ does not correspond to the actual 
temperature of the system, and detailed balance is violated. Accordingly, 
the inherent arbitrariness in defining the update protocol leads 
to 
distinct steady states, as discussed in App.~\ref{sec:protocols}.

\begin{figure}[h!!]
   \includegraphics[width = \linewidth]{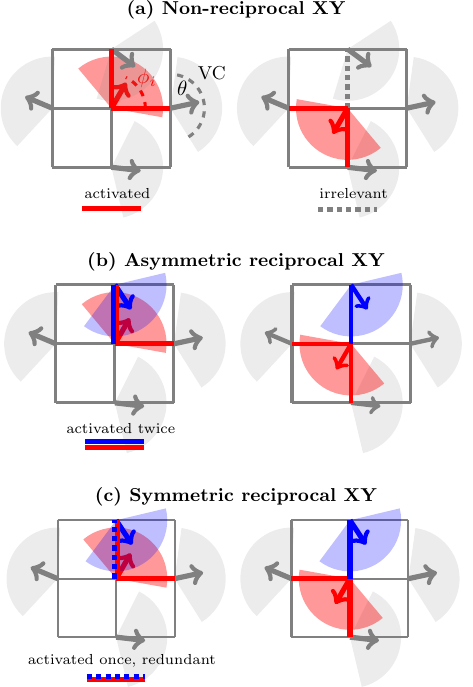}
    \caption{\small 
    Graphical illustration of the three models analyzed in this work.
    (a) The NRXY model features a \textit{selfish} energy (see Eq.~\eqref{eq:selfish}),
    in which the update of the (red) central spin accounts for the neighbors 
    that are within its VC (and that are connected to it by the red bonds, which we term ``activated''), while ignoring 
    those outside that cone, even if they ``see'' the central spin (and to which they are connected by the dashed bond on the right, termed ``irrelevant''). This choice results in a non-equilibrium dynamics. 
    (b) In the ARXY model, the update rule depends on the variation of the total energy 
    (see Eq.~\eqref{eq:arxy-energy}), hence it takes into account both the neighbors of the spin that are seen by it (red bonds), and those who see it (blue bonds). Each bond can thus be not activated (grey), activated once (as in the right panel), or twice (as in the left panel).  The resulting dynamics leads to an equilibrium state. 
    (c) In the SRXY model, the update also takes into account the neighbors of the spin that are seen by it (red bonds), and those who see it (blue bonds) --- see Eq.~\eqref{eq:hamLasso}. However, a bond cannot be activated twice, and thus the virtual second activation is \emph{redundant} (dashed blue and solid red bond). The resulting state is at equilibrium.} 
    \label{fig:update}
\end{figure}

The features of the 
actual update protocols for the three models, which we describe below, are summarized in Fig.~\ref{fig:update}.

\subsection{Non-reciprocal XY model --- NRXY}
\label{subsec:nrec}

The non-reciprocal XY model (NRXY), represented in Fig.~\ref{fig:update}(a), was introduced in 
Ref.~\cite{Loos2023}. In this model, a spin at a lattice site $i$ does not seek to minimize a global energy, 
but rather its own ``selfish'' local energy~\cite{VitellinonreciprocalIsing} $E_i^{\text{NR}}$, 
consisting of ferromagnetic interactions with the nearest neighbors within its vision cone. 
In particular, 
\begin{equation}\label{eq:selfish}
E_{i}^{\text{NR}}=-\sum_{j \in \mathcal N_i} J_{i j}\left(\phi_{i}\right) \cos \left(\phi_{i}-\phi_{j}\right),
\end{equation}
where $j \in \mathcal N_i$  indicates that the sum runs over all the nearest neighbors $j$ 
of spin $i$, and the (non-symmetric) couplings $J_{i j}\left(\phi_{i}\right)$ take into account the vision cone 
as 
\begin{equation}\label{eq:vision}
    J_{i j}\left(\phi_{i}\right)= \begin{cases}J, & \ \mbox{if} \ \min \left\{360^{\circ}-\Delta_{ij},\Delta_{ij}\right\} \leq \theta/2, \\ 0, & \ \mbox{otherwise}.
    \end{cases}
\end{equation}
Here $\Delta_{ij} = \left|\phi_{i}-\vartheta_{i j}\right|$ is the angle that the spin at site $i$ forms with the lattice vector connecting site $i$ with the nearest-neighbouring sites $j$, 
characterized by the angle $\vartheta_{i j} \in \{0^{\circ}, 90^{\circ}, 180^{\circ}, 270^{\circ}\}$. 
In practice, $J_{ij}(\phi_i)$ equals $J$ if the neighbor at site $j$ is within the vision cone of the spin at site $i$, while it vanishes otherwise. 

In what follows, the bonds connecting spin $i$ to spin $j$ for which $J_{ij}(\phi_i)$ does not vanish will be referred to as ``activated''. If, instead, $J_{ij}(\phi_i)$ vanishes but $J_{ji}(\phi_j)$ does not, the bond will be said to be ``irrelevant'' for the dynamics of the spin $i$. 
This occurs if the spin at site $i$ does not see the spin at site $j$, but the latter sees the former.  

In the NRXY model, the selfish energy $E_i^{\text{NR}}$ in Eq.~\eqref{eq:selfish} is then used to determine the probability to accept an attempted MC move, according to the update protocols in Eq.~\eqref{eq:glaub} with 
$\Delta_i E \mapsto \Delta_i E_i^{\text{NR}} $.
This implies that the decision to change the direction of spin $i$ depends solely on interactions with the spins that $i$ can see, without accounting for the energy cost associated with neighboring spins that see $i$ but are not seen by it. 
As a consequence, this model operates out of equilibrium. Its characteristics will be thoroughly examined in Sec.~\ref{sec:results-nrxy}, where we build upon the study conducted in Ref.~\cite{Loos2023}.

\subsection{Asymmetric reciprocal XY model --- ARXY}
\label{subsec:Irec}

For the asymmetric reciprocal XY model (ARXY), the update probabilities in Eq.~\eqref{eq:glaub} are calculated on the basis of
the 
difference 
$\Delta_i E \mapsto \Delta_i E^{\text{AR}} $
of the global 
energy 
obtained by summing all the non-reciprocal (``selfish'') energies in Eq.~\eqref{eq:selfish}, i.e.,
\begin{equation}\label{eq:arxy-energy}
\begin{split}
    E^{\text{AR}} &= \frac{1}{2}\sum_i E_i^{\text{NR}}\\
    & = - \frac{1}{2}\sum_{\langle ij\rangle} [J_{ij}(\phi_i) + J_{ji}(\phi_j)] \cos(\phi_i-\phi_j),
\end{split}
\end{equation}
where the sum runs over all pairs $\langle ij \rangle$ of nearest-neighbor sites $i$ and $j$ on the lattice. The factor $1/2$ ensures that both the NRXY and ARXY models reduce to the standard XY model with the same coupling when the vision cone is removed, i.e., for $\theta = 360^{\circ}$.

Note that, while the individual couplings $J_{ij}(\phi_i)$ and $J_{ji}(\phi_j)$ are asymmetric (i.e., $J_{ij}(\phi_i) \ne J_{ji}(\phi_j)$), the total coupling between $\phi_i$ and $\phi_j$ --- represented by the terms within the square brackets in Eq.~\eqref{eq:arxy-energy} --- is actually symmetric. 
Therefore, 
unlike
the NRXY model, the dynamics of the spin at site $i$ is influenced both by the spins it sees and by those that see it. This is depicted in Fig.~\ref{fig:update}(b), which illustrates how a bond $(ij)$ can be activated zero, one, or two times, 
 depending on $J_{ij}(\phi_i) + J_{ji}(\phi_j)$ being 0, $J$, or $2 J$.
We refer to this model as \emph{asymmetric} because the lattice couplings are asymmetric, even though the eventual interaction between the neighboring spins in Eq.~\eqref{eq:arxy-energy} (which enters the MC update) is symmetric.
In this way, we distinguish it from the SRXY model introduced in the next section, where the couplings are symmetric. 
We emphasize in particular that the asymmetry of $J_{ij}$ implies that 
$ \Delta_i E^\text{AR} \ne \Delta_i E_i^{\text{NR}}$, as pointed out in Ref.~\cite{Seara_2023}: this difference underscores the inherently non-equilibrium nature of NR systems, whose dynamics does not follow from a principle of 
global energy minimization.
On the contrary, the ARXY is a proper statistical equilibrium model, in which a \emph{bona-fide} temperature $\beta^{-1}$ is defined. 
Its characteristics are examined in Sec.~\ref{sec:arxy-results}.

\subsection{Symmetric reciprocal XY model --- SRXY}
\label{subsec:IIrec}

 Another reciprocal version of the XY model with VC is the symmetric reciprocal XY model (SRXY), whose couplings are devised to be 
 symmetric while still retaining a VC. 
This is achieved by defining the total energy 
as
\begin{equation}\label{eq:hamLasso}
    E^{\text{SR}}=- \sum_{\langle i j\rangle }  \Tilde{J}_{i j}\left(\phi_{i}, \phi_{j}\right) \cos \left(\phi_{i}-\phi_{j}\right),
\end{equation}
with 
\begin{align}\label{eq:coupLasso}
    \frac{\Tilde{J}_{i j}\left(\phi_{i}, \phi_{j}\right)}{J}=& \, 1+\operatorname{sgn}\big[
    \Theta\left(\phi_{i}-\vartheta_{i j}+\theta/2\right) \nonumber \\
    &-\Theta\left(\phi_{i}-\vartheta_{i j}-\theta/2\right) +\Theta\left(\phi_{j}-\vartheta_{j i}+\theta/2\right) \nonumber \\
    &-\Theta\left(\phi_{j}-\vartheta_{j i}-\theta/2\right)\big],
\end{align}
where $\Theta$ is the Heaviside function, while $\vartheta_{ij}$ are defined after Eq.~\eqref{eq:vision}. 
These couplings are symmetric, $\Tilde{J}_{i j}\left(\phi_{i}, \phi_{j}\right) = \Tilde{J}_{j i}\left(\phi_{j}, \phi_{i}\right)$.
In practice, as shown in Fig.~\ref{fig:update}(c), this means that if \emph{at least} one of the two nearest-neighbouring spins at sites $i$ and $j$ is looking at the other (i.e., one has the other within its vision cone), then the corresponding bond is activated in the sense that $\Tilde{J}_{i j}$ takes the finite value $J$. If the spins are both looking at each other, the bond is still activated only once: this is the main difference with respect to the ARXY model. 
Accordingly, one of the two activations can be deemed \emph{redundant}. If neither of the spins is looking at the other, then the bond is not activated and $\Tilde{J}_{i j}=0$. 
The resulting dynamics of this model is an equilibrium one. However, the intricate dependence of its couplings on the configuration of the system results in a behavior that is significantly distinct from those of the previously discussed NRXY and ARXY models, as we shall show in Sec.~\ref{sec:results-srxy}.

\subsection{Energetically unfavorable range --- EUR}
\label{sec:EUR}

\begin{figure*}
    \centering
    \includegraphics[width = 0.55\linewidth]{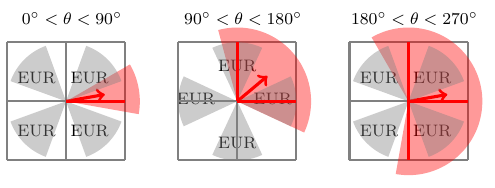}
    \put(-290,92.5){(a)}
    \put(-193,92.5){(b)}
    \put(-96,92.5){(c)}
    \hspace{30pt}
    \includegraphics[width = 0.29\linewidth]{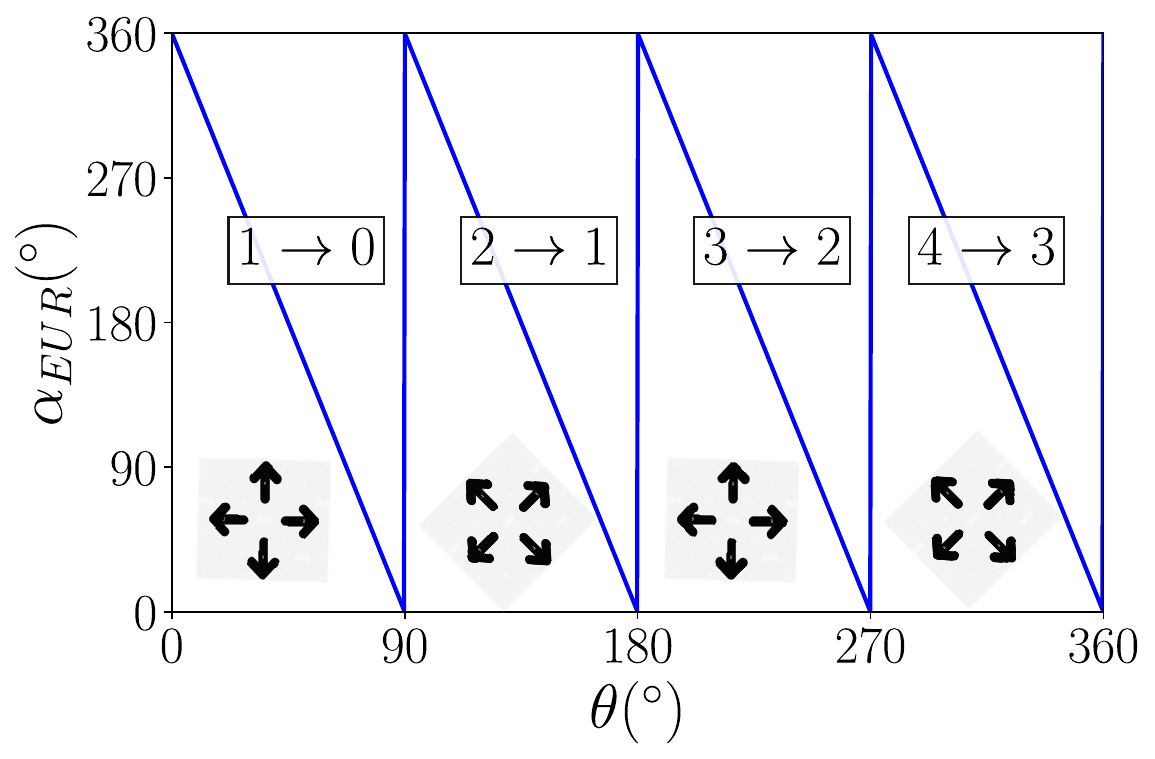}
    \put(-160,92.5){(d)}
\captionsetup{justification=RaggedRight}
    \caption{\small
    Illustration of the EUR. 
    (a--c) When the direction of a spin enters the EUR (gray zone), the number of visible nearest neighbors decreases by one compared to its maximal possible value determined by the angular opening $\theta$ of the VC (red fan). Correspondingly, the number of activated links (red) changes 
    \rev{(a) from 1 to 0 for $\theta \in [0^{\circ}, 90^{\circ}]$, (b) from 2 to 1 when $\theta \in [90^{\circ}, 180^{\circ}]$, (c) from 3 to 2 when $\theta \in [180^{\circ}, 270^{\circ}]$, and from 4 to 3 when $\theta \in [270^{\circ}, 360^{\circ}]$ (not shown)}. 
    Considering as a reference a state in which all spins are aligned in the same direction, this results in a discontinuous increase in energy, making the configuration energetically unfavorable, 
    whence
    the name or the region.  
    The angular position of the EUR depends on the value of $\theta$. (d) Size $\alpha_{\text{EUR}}$ of the EUR as a function of $\theta$, exhibiting a periodicity of $90^{\circ}$. For each period, the black arrows indicate the spatial directions that are \emph{favored} by a collective alignment of the spins on the lattice, resulting in the lowest possible energy. Upon varying $\theta$, the corresponding loss of a nearest neighbor, from $n$ to  $n-1$, with $n=1$, 2, 3, and 4 (indicated by $n\to n-1$ in the figure) has different effects depending on $n$ (even for the same size $\alpha_{\text{EUR}}$ of the EUR), which are discussed in Sec.~\ref{sec:arxy100280-interpretation}.
}
\label{fig:eur}
\end{figure*}

Before proceeding to the analysis of the various models introduced above, here we recall the concept
of energetically unfavorable range (EUR), introduced in Ref.~\cite{Loos2023}, that will prove crucial for 
understanding our results.  
The 
coupling $J_{ij}(\phi_i)$ with VC given in Eq.~\eqref{eq:vision}, shared by both the NRXY and ARXY models, implies that the number of nearest neighbors that are visible to the spin $i$ varies between the two consecutive integers $\lfloor \theta/{90^\circ}\rfloor$ and $\lceil \theta/{90^\circ}\rceil$ upon varying the spin orientation $\phi_i$\rev{, where $\lfloor x \rfloor = \text{max}(y\in \mathbb{Z},y\leq x)$ and $\lceil x \rceil=\text{min}(y\in \mathbb{Z},y > x)$ are the floor and ceiling functions, respectively. 
These values follow from the choice of a
square lattice: if the system were defined, e.g., on a triangular (hexagonal) lattice, then the number of visible spins would vary between $\lfloor \theta/{120^\circ}\rfloor$ and $\lceil \theta/{120^\circ}\rceil$ ($\lfloor \theta/{60^\circ}\rfloor$ and $\lceil \theta/60^\circ\rceil$)}. The EUR corresponds to the set of values of $\phi_i$ for which the number of visible neighbors takes its minimal value. 
\rev{This is illustrated in Fig.~\ref{fig:eur}(a)--(c)
for three different values of the amplitude $\theta$ of the VC (red fan).} 
The angular extension $\alpha_{\text{EUR}}$ of the EUR varies as a function of $\theta$ with a periodicity of $90^\circ$, as shown in Fig.~\ref{fig:eur}(d); \rev{its functional form reads
\begin{equation}\label{eq:alphaEUR}
    \alpha_{\text{EUR}} = 360^{\circ} - 4 (\theta - 90^{\circ}\lfloor\theta / 90^{\circ}\rfloor).
\end{equation}}
Moreover, both the 
\rev{extension} and the centering of the EUR change discontinuously at multiples of $90^\circ$
\rev{ (on the square lattice). 
To see this, consider, for example, a vision cone as in Fig.~\ref{fig:eur}(a), with $\theta \in [0^{\circ}, 90^{\circ}]$. Here,
the central red spin can see one neighbor only if its arrow points approximately
along one of the four lattice directions,
i.e., if
$|\phi_i - n \cdot 90^{\circ}|  <  (360^{\circ} - \alpha_{\text{EUR}})/8 $ for some value of 
$n\in \{0,1,2,3\}$. 
%
%
Conversely, the spin sees no neighbors if it points away from the spatial directions in which the neighbors are located, i.e., if its orientation differs approximately by $45^\circ$ from those of the lattice directions, with
$|\phi_i - n \cdot 90^{\circ} - 45^{\circ}| < \alpha_{\text{EUR}}/8$.
Note that $\alpha_\mathrm{EUR}$ quantifies the angular range within which the number of visible neighbors vanishes. 
When $\theta$ approaches $90^{\circ}$ from below, the spin sees almost always one neighbor regardless of the spatial direction in which it points, so that $\alpha_{\text{EUR}}\to 0^{\circ}$. Once $\theta$ exceeds $90^{\circ}$, as in Fig.~\ref{fig:eur}(b), the spin can see two neighbors only if it points approximately $45^{\circ}$ away from the lattice directions. 
This situation is actually opposite to that in the previous range. Consequently, the \textit{centering} of the EUR changes discontinuously whenever $\theta$ crosses $90^\circ$ and its integer multiples. 
In addition, $\alpha_{\text{EUR}}$ assumes its maximum value $360^{\circ}$ immediately after $\theta=90^\circ$, and then it decreases as $\theta$ increases --- until $\theta$ reaches the value $\theta=180^\circ$, where another analogous discontinuity occurs, as summarized in Fig.~\ref{fig:eur}(d).}

This implies that a perfectly aligned configuration of the spins maximizes the number of activated bonds and minimizes the energy when 
the system
is oriented 
either along the lattice directions or along the lattice directions plus $45^\circ$, depending on the value of $\theta$.
These preferential orientations are indicated by the arrows close to the horizontal axis in Fig.~\ref{fig:eur}(d). 
We anticipate here that the existence of this EUR actually shapes the phase diagrams of the NRXY and ARXY models, shown in, c.f., Figs.~\ref{fig:nrec-m} and \ref{fig:arxy-pd}, respectively. 
Note that the impact on the system of losing sight of one of the $n$ nearest neighbors by entering the $n \to n-1$ EUR depends not only on the size of this region, but also on how many other neighbors remain visible, i.e., on the value of $\theta$. 
This results in significantly different outcomes, which will be discussed in Sec.~\ref{sec:arxy100280-interpretation}.

Finally, we point out that the notion of EUR does not carry over straightforwardly to the SRXY model, because the corresponding coupling $\Tilde{J}_{i j}(\phi_i,\phi_j)$ (see Eq.~\eqref{eq:coupLasso}) depends on the two angles $\phi_{i,j}$, and thus the activation of the four bonds around each spin depends on the value of its angle and of those of the 4 neighboring spins. 
In this case, the number of activated bonds equals an integer between $\lfloor \theta/{90^\circ}\rfloor$ and 4. 
This leads to significant differences in the phase diagram, as will be shown in 
Sec.~\ref{sec:srxy-pd}.

\section{Methods}\label{sec:methods}

The aim of this study is to characterize the phase diagram of the models introduced in the previous Section, focusing on identifying regions that exhibit long-range order (LRO), quasi-long-range order (QLRO), and disorder (DO), as well as understanding the transitions between these phases. This is achieved via MC simulations of the dynamics, which allow one to
sample and analyze the observables we are interested in. 
Before conducting such analysis,
we recall
the results of MC simulations of systems with well-known behavior, specifically the $q$-state clock model with $q = 4$, 6 and the XY model, which exhibit the three aforementioned phases, and transitions of second and infinite order separating them.

Details on the numerical implementation of the MC dynamics are provided in App.~\ref{sec:simulations}.

\subsection{Observables}\label{sec:observables}

The three phases mentioned above, i.e., LRO, QLRO, and DO, can be
identified on the basis of the behavior of two key observables: the modulus of the magnetization and the correlation length. The (average) magnetization of the model is given by the vector
\begin{equation}\label{eq:vectorOP}
    \mathbf{m} = \frac{1}{N}\sum_{i=1}^N \mathbf{s}_i = \left(\frac{1}{N}\sum_{i=1}^N \cos{\phi_i}, \frac{1}{N}\sum_{i=1}^N \sin{\phi_i}\right),
\end{equation}
where the sum runs over all lattice sites of the model, 
which are $N=L^2$ in total on a square lattice with linear size $L$.
%
Its modulus $m = |\mathbf{m}|$ is then given by
\begin{equation}
    m=\frac{1}{N}\sqrt{\bigg(\sum_{i=1}^N \cos \phi_{i}\bigg)^{2}+\bigg(\sum_{i=1}^N\sin \phi_{i}\bigg)^{2}}.
\label{eq:def-mm}
\end{equation}
From the correlation function
\begin{equation}\label{eq:corr}
    G(r = |i-j|) = \langle \mathbf{s}_i\cdot\mathbf{s}_{j}\rangle = 
     \langle \cos{(\phi_i - \phi_{j})} \rangle, 
\end{equation}
we define the exponential correlation length $\xi$ as
\begin{equation}\label{eq:expcorrlength}
    \xi = \lim_{r \to \infty} \frac{r}{-\log{G(r)}}.
\end{equation}
In an infinitely extended system, LRO phases are characterized by a non-zero magnetization $m$ and an exponentially decaying correlation function, corresponding to a finite correlation length $\xi$. DO phases have zero magnetization $m=0$ and also exhibit finite correlation lengths $\xi$. In contrast, QLRO phases, as well as critical points (CPs), feature a power-law decaying correlation function $G(r)$, implying an infinite correlation length ($\xi\sim L$ for finite system sizes), with a vanishing 
magnetization $m$.
In the case of spatial dimension $d=2$
on which we focus here,
the above can be summarized as
\begin{align}
    G(r)&\sim
    \begin{cases}\label{eq:correlation-scaling}
    e^{-r/\xi}, & \text{for LRO, DO,} \\
     1/r^{2-\eta}, & \text{for QLRO,}
    \end{cases} \\[2mm]
    m &= 
    \begin{cases}
    \text{const.,} & \text{for LRO,} \\
    0, &  \text{for QLRO, DO.}
    \end{cases}
\end{align}
Notably, in the case of the XY model at low temperatures, the magnetization is subject to 
strong finite-size corrections due to slow spin-wave scaling~\cite{Tobochnik1979, Archambault1997}, described by
\begin{equation}\label{eq:SWscaling}
    m \sim N^{-T / (8 \pi J)}.
\end{equation}
This scaling makes it difficult to remove finite-size effects in simulations.
 Furthermore, determining the exponential correlation length $\xi$ from Eq.~\eqref{eq:expcorrlength} is practically 
 challenging
 for finite geometries.
 Instead, we will use the second-moment correlation length, defined as
\begin{equation}\label{eq:2ndcorrlength}
    \xi_2^2 =  \frac{\int d^dr\, r^2 \,G(r)}{\int\, d^dr\, G(r)} = -\frac{1}{\hat{G}(\mathbf{k})} \frac{\partial^2\hat{G}(\mathbf{k})}{\partial k^2}\bigg|_{k = 0},
\end{equation}
where $\hat{G}(\mathbf{k})$ is the Fourier transform of the correlation function:
\begin{equation}\label{eq:gkmin}
\hat{G}(\mathbf{k})=\left\langle\sum_{\mu=x, y}\left|\frac{1}{N} \sum_{i=1}^{N} s_{i \mu} \, \mathrm e^{i\mathbf{k}\cdot \mathbf{r}}\right|^{2}\right\rangle,
\end{equation}
with $s_{ix}= \cos{\phi_i}$ and $s_{iy}= \sin{\phi_i}$.
On the lattice, $\xi_2^2$ in Eq.~\eqref{eq:2ndcorrlength} is expressed as~\cite{cooper1982}
\begin{equation}\label{eq:corrLength}
\xi^{(2)}_L=\frac{1}{2 \sin \left(k_{m} / 2\right)} \sqrt{\frac{ \hat{G}(0)}{ \hat{G}(\mathbf{k}_m)}-1},
\end{equation}
where $k_m = 2\pi / L$, and $\mathbf{k}_m = (k_m, 0)$.
Since in the LRO phase $\hat{G}(0)\sim L^{d}$ and $\hat{G}(\mathbf{k}_m)\sim L^{0}$, the behavior of $\xi_L^{(2)}/L$ in $d=2$ is given by~\cite{amit2005} 
\begin{equation}\label{eq:corrlengthScaling}
    \xi^{(2)}_L/L \sim \left\{ \begin{array}{ll}
        L, & \text{for LRO,} \\
        \text{const.,} &  \text{for QLRO or CPs,} \\
        1/L, & \text{for DO.} 
    \end{array}
    \right.
\end{equation}
Accordingly, inspecting the behavior of $\xi_L^{(2)}/L$ upon increasing $L$ turns out to be a reliable approach for identifying the three phases.

Another useful observable for characterizing phase transitions is the variance of the magnetization,
defined as
\begin{equation}
    \chi_m = \frac{N}{T} \left(\langle m^2 \rangle -  \langle m \rangle^2\right),
    \label{eq:def-chi}
\end{equation}
which is equal to the magnetic susceptibility under equilibrium conditions.
Similarly, the variance of the internal energy
\begin{equation}
    C_v = \frac{1}{N T^2} \left(\langle E^2 \rangle - \langle E \rangle^2\right)
    \label{eq:def-C}
\end{equation}
is equal to the specific heat at equilibrium.

In order to distinguish between the three phases, it 
often proves
useful to consider also the vectorial magnetization in Eq.~\eqref{eq:vectorOP}. 
\rev{This 
quantity is sampled 
along a MC run
and visualized in a scatter plot, 
as an informative approximation of its actual distribution. In fact, the analysis of this distribution provides insight into the phase behavior of the system, which will be discussed in Sec.~\ref{sec:toymodels}.} 

Finally, to quantify the irreversibility of the dynamics in the cases where the system is out of equilibrium, we consider the mean entropy production rate (EPR) per spin associated with a certain realization of the process $\left\{\phi_i(t)\right\}_{i,t}$, defined as~\cite{Martynec2020,noa2019, Tania2012,Schnakenberg1976}
\begin{equation}\label{eq:epr}
\mathrm{EPR}=\left\langle\ln \frac{w\left(\phi_{i}(t) \rightarrow \phi_{i}(t+d t)\right)}{w\left(\phi_{i}(t+d t) \rightarrow \phi_{i}(t)\right)}\right\rangle / d t.
\end{equation}
Here
$w\left(\phi_{i}(t) \rightarrow \phi_{i}(t+d t)\right)$ is the transition rate for updating the angle 
$\phi_i$ 
from the value $\phi_{i}(t)$ at
time $t$ to the value $\phi_{i}(t+d t)$ at time $t+dt$ (where $dt = 1$ in the MC simulation), as described by the Glauber algorithm of Eq.~\eqref{eq:glaub}. The EPR is computed by averaging the logarithm of the ratio of forward to reverse transition rates over the entire MC history. This measure provides valuable information about the deviation of the dynamics of the system from equilibrium.

\subsection{Phase diagrams of reference models}
\label{sec:toymodels}

Here we recall the behavior of the quantities defined in Sec.~\ref{sec:observables} for three paradigmatic models with well-known properties that we will use for comparison: the 4- and 6-state clock models and the XY model, all in spatial dimension $d=2$. 
These models are all characterized by the same formal expression for the energy $E$, i.e., 
\begin{equation}
    E = -J \sum_{\langle ij \rangle}  \cos{(\phi_i - \phi_j)},
\end{equation}
where the angular variable 
$\phi_i$ takes different values depending on the model. For the XY model, 
$\phi_i$ is continuous between 0 and $2\pi$, while for the clock models, 
$\phi_i$ takes discrete values $\phi_i = 2\pi k/n$ for $k = 0, 1, \ldots, n-1$, with $n = 4$ or $n = 6$ for the 4-state and 6-state clock models, respectively.

\begin{figure*}
       \centering
    \includegraphics[width=0.96\linewidth]{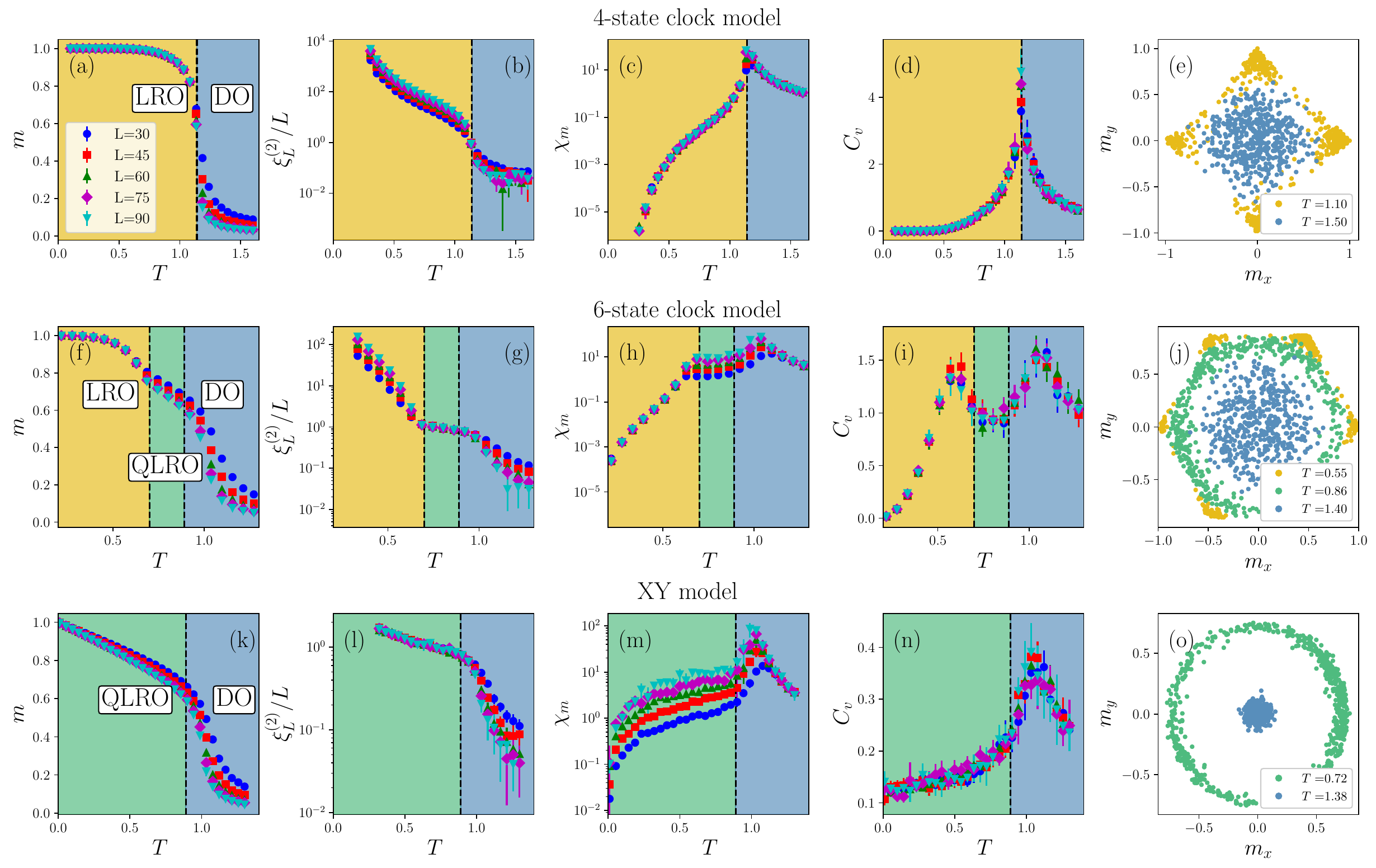}
    \captionsetup{justification=RaggedRight}
    \caption{\small Temperature dependence of the observables defined in Sec.~\ref{sec:observables} for three benchmark models, namely: (a)--(e) the 4-state clock model, (f)--(j) the 6-state clock model, and (k)--(o) the XY model. The quantities $m$, $\xi_L^{(2)}/L$, $\chi_m$, and $C_v$ are shown for various system sizes $L$ (indicated by the legend in panel (a)), while scatter plots of $\mathbf{m}$ are presented at selected temperatures for a fixed size $L = 90$. Note that the plots of $\xi_L^{(2)}/L$ (exhibiting the distinct behaviors outlined in Eq.~\eqref{eq:corrlengthScaling}) and $\mathbf{m}$ are particularly effective for distinguishing the three phases: LRO (yellow), QLRO (green), and DO (blue). See Sec.~\ref{sec:toymodels} for a detailed description of the observed behaviors. 
    }
    \label{fig:benchmarks}
\end{figure*}

\textit{The 4-state clock model. --- }
The behavior of the 4-state clock model is summarized in Fig.~\ref{fig:benchmarks}(a)--(e), which shows the observables $m$, $\xi_L^{(2)}/L$, $\chi_m$, and $C_v$ for different system sizes 
$L$, and as a function of temperature, as well as $\mathbf{m} = (m_x,m_y)$ for two temperatures within the two phases. (Hereafter, $J=1$ is assumed in all models.) This model has an internal $\mathbb{Z}_4$ symmetry and it undergoes a second-order phase transition from a LRO phase to a DO phase, belonging to the Ising universality class~\cite{li2020}. 
The magnetization $m$ in panel (a) thus goes from a finite value at low temperatures to zero as the temperature is increased beyond the transition at $T_c =  1 / \log{(1 + \sqrt{2})}\simeq 1.13 $~\cite{baxter2007exactly},
indicated by the dashed vertical line in all panels. 
The curves of the second-moment correlation length $\xi_L^{(2)}$ normalized by $L$, shown in panel (b), are size-dependent in the LRO and DO phases (scaling, respectively, as $L$ and $1/L$ upon increasing $L$ at fixed temperature, see Eq.~\eqref{eq:corrlengthScaling}), and cross at the transition point.

The susceptibility in panel (c) is size-independent in both the LRO and DO phases but exhibits a size-dependent peak at the transition point. The same holds for the specific heat in panel (d).

Finally, the scatter plot of the vectorial magnetization $\mathbf{m}$ in panel (e) shows that, within the LRO phase, most of the spins align along one of the four allowed directions, with configuration changes occurring only due to finite-size effects. In contrast, in the DO phase, the magnetization no longer favors any of the four directions, forming an isotropic cloud around the origin.

\textit{The 6-state clock model. --- }
Next,
the behavior of the 6-state clock model is summarized in Fig.~\ref{fig:benchmarks}(f)--(j). With its internal $\mathbb{Z}_6$ symmetry, this model exhibits two infinite-order transitions: one from the LRO to the QLRO phase, and the other from the QLRO to the DO phase~\cite{Surungan_2019}. 
The corresponding transition temperatures, $T_c^{(1)}\simeq 0.7$ and $T_c^{(2)}\simeq 0.89$, indicated by the vertical dashed lines, are those numerically estimated in Ref.~\cite{tuan2022}.
The magnetization $m$ shown in panel (f) is finite in the LRO phase, remains finite in the QLRO phase due to finite-size effects --- similarly to what happens in the XY model (see Eq.~\eqref{eq:SWscaling}) --- while it vanishes in the DO phase.
The curves of the second-moment correlation length $\xi_L^{(2)}$, normalized by $L$ and shown in panel (g),
scale as $L$ in the LRO phase, collapse onto the same value within the finite range of temperatures  corresponding to the QLRO phase, and then separate again, scaling as $1/L$, in the DO phase. 
\rev{These different scalings are shown, for selected temperatures, in Fig.~\ref{fig:scalingCorrelationLength}(a) in App.~\ref{sec:corr-length-scaling}.} 
The magnetic susceptibility in panel (h) does not scale with the system size in the LRO and DO phases, but it features a size dependence in the QLRO phase. The position of the scale-dependent peak at the transition from QLRO to DO tends logarithmically to the infinite-size transition temperature $T_{\rm BKT}\simeq 0.89$~\cite{tuan2022}. 

The specific heat $C_v$ in panel (i) shows size-independent peaks that do not correspond directly to the transition temperature, as it is also observed in other infinite-order transitions such as the one of the XY model~\cite{Drouin-Touchette2022}.
The distribution of the vectorial magnetization $\mathbf{m}$ in panel (j) reveals that, in the LRO phase, $\mathbf{m}$ tends to align  along one of the six available directions, with fluctuations due to finite-size effects. In the QLRO phase, the system no longer prefers any particular direction, and the magnetization remains finite due to finite-size effects, resulting in the observed ring-shaped distribution. In the DO phase, the distribution forms a cloud around the origin.

\textit{The XY model. --- }
Finally, the behavior of the XY model is summarized in Fig.~\ref{fig:benchmarks}(k)--(o). With its internal $O(2)$ symmetry, this model exhibits a single infinite-order transition, from the QLRO phase to the DO phase~\cite{Kosterlitz_1973,kosterlitz1974}. The transition temperature, indicated with a dashed line, is numerically estimated in Ref.~\cite{Hasenbusch_2005}.
The magnetization $m$ in panel (k)
remains finite in the QLRO phase due to significant finite-size effects, as described by Eq.~\eqref{eq:SWscaling}, and then it decreases to zero in the DO phase.
Here $\xi_L^{(2)}$,
normalized by $L$ and shown in panel (l), collapses onto the same value in the QLRO phase and then separates as $1/L$ in the DO phase. The magnetic susceptibility in panel (m) shows a size dependence in the QLRO phase while it is size-independent in the DO phase. 
The specific heat in panel (n) presents a size-independent peak that does not occur at the transition temperature. The distribution of the vector magnetization in panel (o) forms a ring in the QLRO phase, with a finite radius due to the significant finite-size effects. Finally, in the DO phase, the distribution becomes an isotropic cloud centered around the origin.

In the next three sections, the comparison with the behaviors described above will be used 
to analyze the phase diagram of the lattice models with VC introduced in Sec.~\ref{sec:models}.

\begin{figure*}
    \centering
        \includegraphics[width = 0.44\linewidth]{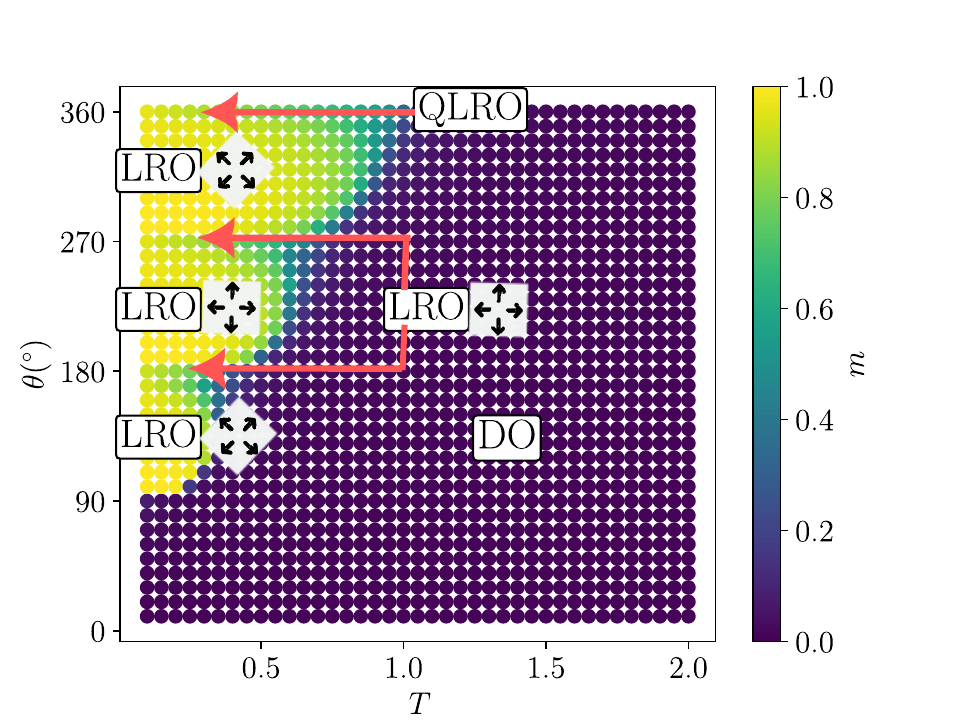}
        \put(-234,141){(a)}
        \!\!\!\!\!\!\!\!\!
        \includegraphics[width = 0.44\linewidth]{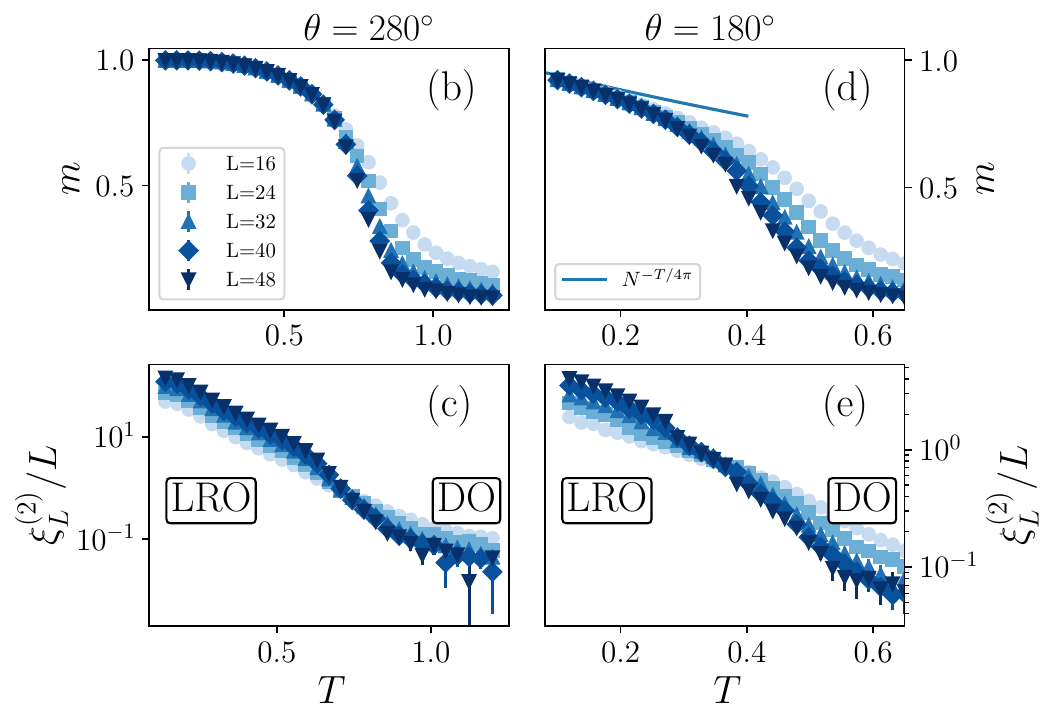}
         \captionsetup{justification=RaggedRight} 
    \caption{\small Results of the MC simulations for the NRXY model in the non-equilibrium steady state. 
(a) The magnetization $m$ as a function of temperature $T$ and VC angle $\theta$ reveals a phase diagram characterized by three distinct lobes. At low temperatures and for $\theta > 90^\circ$, a LRO phase (in yellow) is observed, with the spins showing a collective alignment along one of the spatial directions indicated by the black arrows. At higher temperatures, the system undergoes a transition into a DO phase (dark blue). 
(b)--(c) Magnetization $m$ and ratio $\xi_L^{(2)}/L$ as functions of $T$, for a vision cone angle $\theta = 280^\circ$, showing a clear transition between LRO and DO phases. 
(d)--(e) Magnetization $m$ and $\xi_L^{(2)}/L$ as functions of $T$, for $\theta = 180^\circ$. In panel (d), the scaling of $m$ described by Eq.~\eqref{eq:SWscaling} for the XY model is shown for
$L = 48$ (solid blue line),
exhibiting consistency with the numerical data at low temperature (symbols). However, the behavior of the data for $\xi_L^{(2)}/L$ in panel (e) --- qualitatively similar to panel (c) --- indicates that the model does not actually develop QLRO at low temperatures but that it undergoes a transition between a LRO and a DO phase, as in the previous case.}
\label{fig:nrec-m}
\end{figure*}

\section{The non-reciprocal XY model}\label{sec:results-nrxy}

We begin by analyzing the NRXY model of Sec.~\ref{subsec:nrec},
previously studied in Ref.~\cite{Loos2023}. Our goal is to further investigate certain aspects of the model, highlighting that its fundamental properties are equivalent to those of its reciprocal counterpart, that is, the asymmetric reciprocal model ARXY, whose analysis will be the subject of Sec.~\ref{sec:arxy-results}. 
In particular, we will show that the existence of a phase with LRO is due to the interaction being characterized by a vision cone, rather than its non-reciprocal nature.

The effects of the chosen update protocol on the stationary state are discussed in App.~\ref{sec:protocols}.
In general, the MC dynamics drives the system into a non-equilibrium stationary state with a non-vanishing EPR;
we refer to App.~\ref{sec:epr} 
for a detailed analysis of the EPR for different points in the phase diagram of the model.

In Fig.~\ref{fig:nrec-m}(a), we report
the magnetization $m$ in the steady state as a function of the temperature $T$ and the VC angle $\theta$, for a system of size $L=100$. 
In particular, this plot reveals
that the phase diagram of the model features a structure with three lobes, characterized by an enhanced $m$, which start and end at angles that are multiples of $90^{\circ}$. This is a consequence of the periodicity of the
occurrence of EUR, highlighted in Fig.~\ref{fig:eur}. Within these lobes, a LRO phase is present, with the spins preferentially and collectively aligned along one of the four directions indicated by the black arrows in Fig.~\ref{fig:nrec-m}(a) for each lobe. 
An account of the origin of these lobes and why there is no LRO for $\theta < 90^{\circ}$ is given in the 
supplemental material of Ref.~\cite{Loos2023}. 
For $\theta = 360^{\circ}$, the standard XY model is recovered and we find a QLRO phase at low temperature.

In Figs.~\ref{fig:nrec-m}(b)--(e), we report $m$ and 
$\xi_L^{(2)}/L$ 
as functions of the temperature $T$ for various system sizes, and for two representative values of 
$\theta=280^\circ$ and $180^\circ$. 
As already suggested in Ref.~\cite{Loos2023}, for all values of the VC angle that have a finite EUR, such as 
$\theta = 280^\circ$, we expect a LRO phase. Indeed, the magnetization $m$ in panel (b) approaches 1 for $T\to 0$, remaining flat at low temperatures, and goes to zero above the transition temperature. Similarly, the curves of $\xi_L^{(2)}/L$ for the various values of $L$, reported in panel (c), intersect at a single point, marking the transition from LRO to DO. In these two phases, the curves of $\xi_L^{(2)}/L$ scale as $L$ and $1/L$, respectively, consistently with Eq.~\eqref{eq:2ndcorrlength}.

The case with $\theta = 180^\circ$
(similar to $\theta = 270^\circ$) --- suggested to exhibit QLRO at low temperatures in Ref.~\cite{Loos2023} due to disappearance of EUR --- is more nuanced. In panel (d) of Fig.~\ref{fig:nrec-m}, we observe that at low temperatures the dependence of the magnetization on $T$ is not flat, but instead follows  the scaling behavior of the XY model described by Eq.~\eqref{eq:SWscaling} (with the coupling $J$ 
effectively halved because, for $\theta = 180^\circ$, only half of the bonds are activated). However, in panel (e), the curves of the temperature dependence of $\xi_L^{(2)}/L$ for various $L$ do not collapse on a single curve at low temperatures, as expected in the presence of QLRO, but instead they intersect at a single point. 
\rev{For temperatures below and above the temperature of this intersection point, it actually turns out that $\xi_L^{(2)}/L$ scales as $L$ and $1/L$, respectively, as shown in Fig.~\ref{fig:scalingCorrelationLength}(b) of App.~\ref{sec:corr-length-scaling}}.
This behavior, combined with the observation that the system prefers alignment along the lattice directions at low temperatures
(indicated by the black arrows in Fig.~\ref{fig:nrec-m}(a)), suggests that, for these values of $\theta$, the phase has LRO at low temperatures despite a vanishing EUR. This phenomenon, also observed for the ARXY model, will be explained in Sec.~\ref{sec:arxy180-interpretation}.

\section{The asymmetric reciprocal XY model}\label{sec:arxy-results}

In this section, we analyze the ARXY model introduced in Sec.~\ref{subsec:Irec}, and we put forward 
an argument based on internal symmetries that allows us 
to explain the observed numerical results. 

\subsection{Phase diagrams}\label{sec:arxy-pd}

\begin{figure*}
    \centering
        \includegraphics[width = 0.33\linewidth]{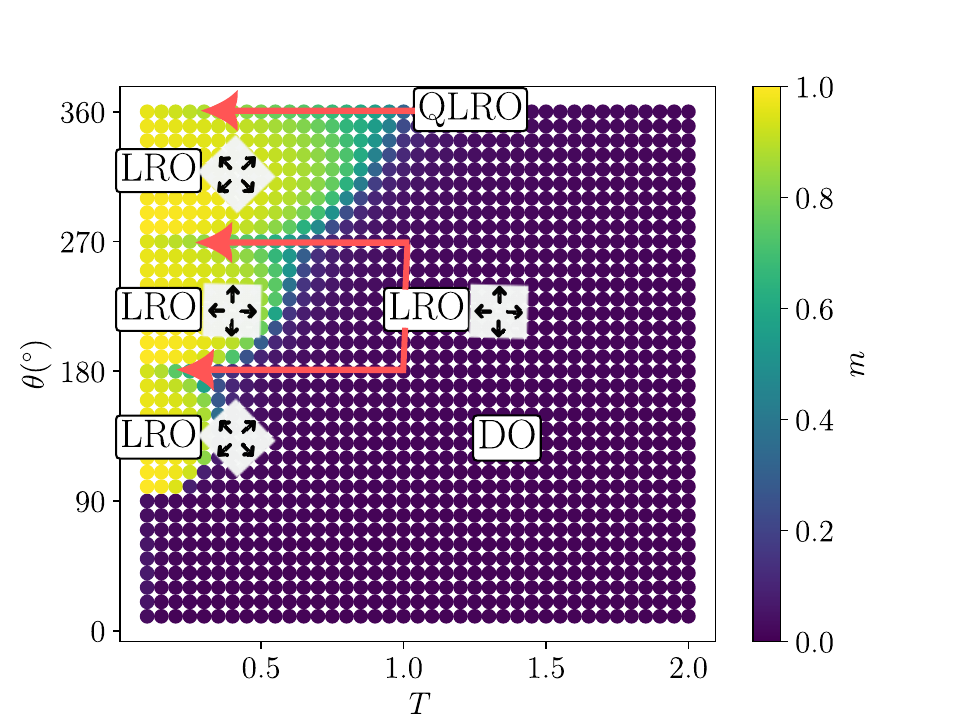}
        \put(-174,105){(a)}
        \!\!
        \includegraphics[width = 0.33\linewidth]{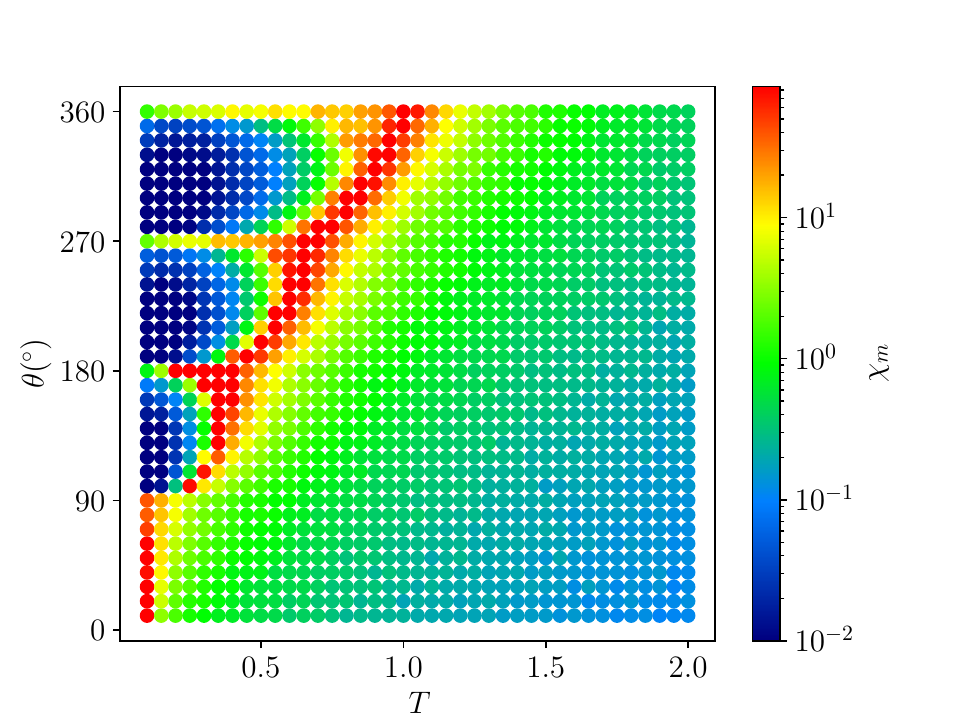}
        \put(-174,105){(b)}
        \!\!
        \includegraphics[width = 0.33\linewidth]{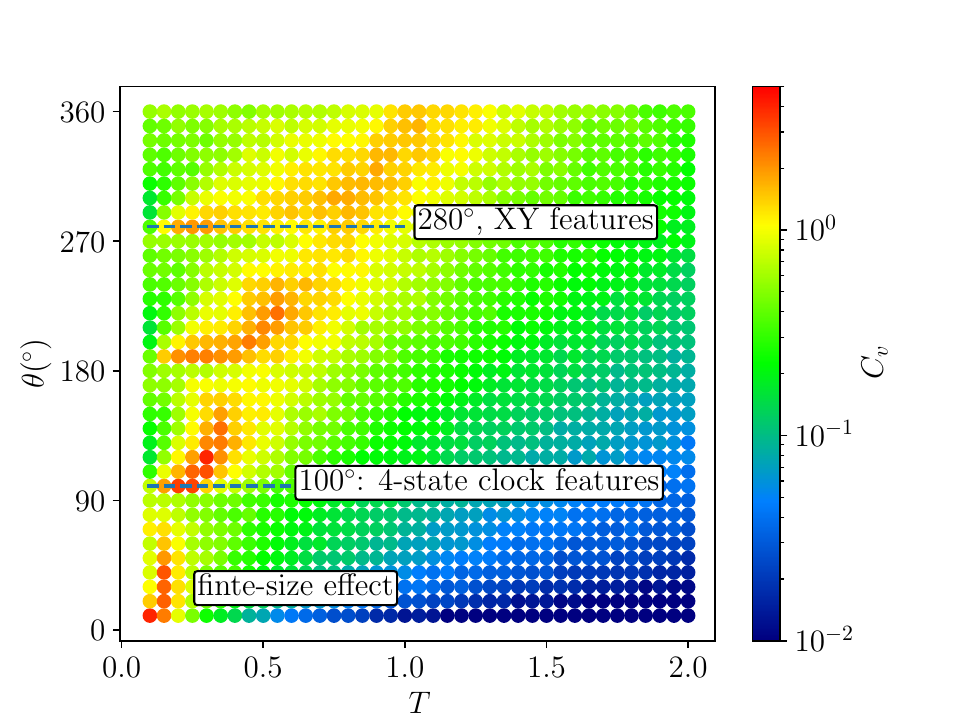}
        \put(-174,105){(c)}
         \captionsetup{justification=RaggedRight} 
    \caption{\small 
Results of the MC simulations for the equilibrium state of the ARXY model
obtained for a system size $L=100$. 
(a) The behavior of the magnetization $m$ is characterized by the presence of three lobes, qualitatively indistinguishable from those of the NRXY model (see Fig.~\ref{fig:nrec-m}(a)), with the LRO phase present within the lobes and the DO phase outside them. (b) The low  values of the magnetic susceptibility $\chi_m$ within the three lobes confirms the presence of the LRO phase
(see the discussion in, c.f., Sec.~\ref{sec:arxy-results}).
However, from this diagram it is not 
possible to conclude 
whether the system exhibits LRO or QLRO at low temperatures for $\theta = 180^{\circ}$ and $270^{\circ}$. 
The susceptibility $\chi_m$ as a function of temperature features a peak (in red) at the transition that separates the LRO phase from the DO phase for $\theta > 90^\circ$, while it increases monotonically upon decreasing the temperature towards $T=0$ for $\theta < 90^\circ$. 
(c) The specific heat $C_v$ shows peaks in correspondence of the transitions from LRO to DO when $\theta > 90^{\circ}$, and at low temperatures for $\theta < 90^{\circ}$.
The origin of these low-temperature peaks is addressed in App.~\ref{sec:M2-60}.
} \label{fig:arxy-pd}
\end{figure*}

The behavior of the magnetization $m$ of this model as a function of temperature $T$ and the VC angle $\theta$ is reported  in Fig.~\ref{fig:arxy-pd}(a). In particular, we note that it is qualitatively very similar to that of the NRXY, presented in Fig.~\ref{fig:nrec-m}(a).
This is rather remarkable, since in the ARXY discussed here
the MC averages of the magnetization (and of the other observables defined in Sec.~\ref{sec:observables}) 
provide accurate estimates of the averages sampled from the Boltzmann distribution, which characterizes equilibrium stationary states, whereas the NRXY is out of equilibrium. 
However, at a closer inspection, some quantitative differences emerge in the temperature dependence of the magnetization, as we show
in Fig.~\ref{fig:EPR-protocols}(b) of App.~\ref{sec:protocols}.

The susceptibility $\chi_m$ as a function of $T$ and $\theta$, obtained for a system of size $L=100$, is shown in Fig.~\ref{fig:arxy-pd}(b). We observe that $\chi_m$ is small in the LRO phase, while it increases upon increasing the temperature $T$ for a fixed value of $\theta$, reaching its maximum along the boundaries of the three lobes identified in panel (a), which separate the LRO phase from the DO phase. 
The higher values of the susceptibility for $\theta = 180^{\circ}$ and $270^{\circ}$, compared to those observed at similar temperatures for angles immediately above and below, raise the question of whether QLRO is present here, as it is for $\theta = 360^{\circ}$. However, as demonstrated for the NRXY model, this is not the case, as will be further elaborated in Sec.~\ref{sec:arxy180-interpretation}.

The behavior of the specific heat $C_v$ for this model is shown in Fig.~\ref{fig:arxy-pd}(c). Also in this case, we observe that, for a fixed value of $\theta$, $C_v$ as a function of $T$ reaches its 
maxima along the boundaries of the three lobes separating the LRO phase from the DO phase. The values attained at these maxima actually depend on $\theta$: in Sec.~\ref{sec:arxy100280-interpretation} we address the origin of this dependence, 
focusing in particular on $\theta = 100^{\circ}$ and $\theta = 280^{\circ}$. 
The peculiar behavior for $\theta < 90^{\circ}$, where the system does not magnetize for the same reasons presented in the supplemental material of Ref.~\cite{Loos2023}, is further investigated for $\theta = 60^{\circ}$ in App.~\ref{sec:M2-60}.

Below, we introduce the theoretical argument on the basis of which we are able to 
explain the numerical results presented above.

\begin{figure*}
    \centering
        \includegraphics[width = 0.33\linewidth]{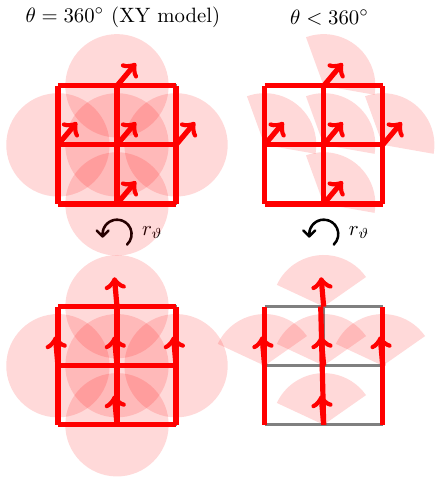}
        \put(-180,150){(a)}
        \hspace{30pt}
        \includegraphics[width = 0.33\linewidth]{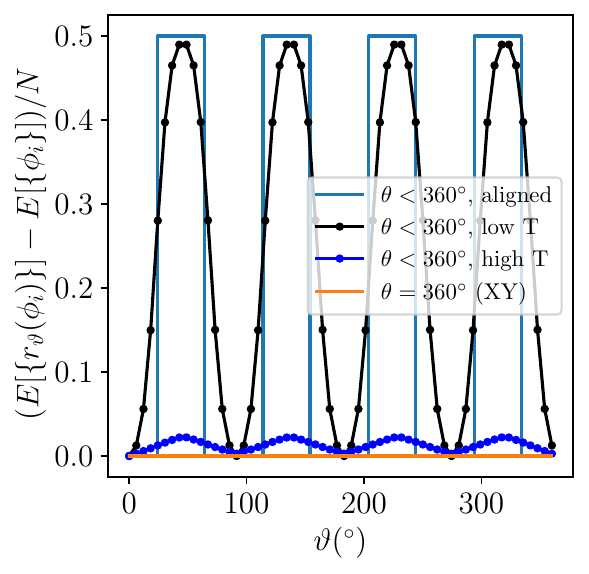}
        \put(-180,150){(b)}
        \captionsetup{justification=RaggedRight}
         \caption{\small 
    Illustration of the effects of a global rotation $r_{\vartheta}$ on the energy of a configuration, used as a tool to probe the symmetry of a system. (a) In the case of the XY model (i.e., of the ARXY with $\theta = 360^{\circ}$), represented on the left, $r_{\vartheta}$ does not change the energy of the system, independently of the initial configuration. In fact, all spins continue to interact with their nearest neighbors, and the differences between angles (i.e.,~$\phi_i-\phi_j$) remain constant. For the ARXY with $\theta  < 360^\circ$, represented on the right, the angle differences between neighboring spins are again not altered by $r_{\vartheta}$, but the structure of the activated bonds (highlighted in red) changes, resulting in a change of the energy and in the absence of an $O(2)$ symmetry in the system. (b) Dependence on the rotation angle $\vartheta$ of the energy difference per spin $(E[\{r_{\vartheta}(\phi_i)\}] - E[\{\phi_i\}])/N$ (displayed data corresponds to $\theta=315^\circ$). 
    For the XY model, the energy remains constant upon rotation, as expected. Conversely, for ARXY with $\theta < 360^\circ$, the change in energy depends on the starting configuration. For a perfectly aligned configuration, the rotation results into a square wave with a periodicity of $90^\circ$. For a low-temperature thermal configuration (not perfectly aligned), 
    the curve is smoothed but retains the $90^\circ$ periodicity. At 
    higher temperatures,
    corresponding to the DO phase, the curve is again almost independent of $\vartheta$, similar to the XY model.
}  \label{fig:rotation}
\end{figure*}

\subsection{Internal symmetry}\label{sec:symmetry}

In this Section, we provide an interpretation of the results of the MC simulations presented above, which requires adapting the notion 
of internal symmetry to the types of models we are studying.

In the models typically examined in the context of statistical physics, the notion of internal symmetry plays a crucial role in characterizing phase transitions. In particular, the XY model has an $O(2)$ symmetry, which means that by applying a global rotation $r_{\vartheta}(\phi_i) = \phi_i + \vartheta \ (\text{mod}\ 2\pi)$ of an angle $\vartheta$ to all spins (i.e., the internal degrees of freedom) of the system, the total energy does not change:
\begin{equation}
    E_{XY}[\{\phi_i\}] = E_{XY}[\{r_{\vartheta}(\phi_i)\}].
\end{equation}
Crucially,
in the models usually studied in statistical physics (including the XY model that is relevant here), the internal degrees of freedom are not 
coupled to the spatial structure of 
the system, characterized by the dimensionality and the lattice type. 
By contrast, in the models we are currently interested in, the vision cone acts as a bridge between the internal degrees of freedom and the lattice structure, rendering the actual couplings~$J_{ij}$
configuration dependent. In particular, when the vision cone is $\theta < 360^{\circ}$, applying a rotation $r_{\vartheta}$  has also the effect of changing the structure of the activated bonds, as shown in Fig.~\ref{fig:rotation}(a), where the active bonds are highlighted in red. 

Plotting the energy difference per spin, 
$(E[\{r_{\vartheta}(\phi_i)\}] - E[\{\phi_i\}])/N$, 
obtained upon rotating the system, provides information about the symmetry emerging from this interplay between the vision cone and the lattice structure. 
In fact, for the XY model (with vision cone $\theta = 360^\circ$), the energy difference due to such a rotation is zero, as shown in Fig.~\ref{fig:rotation}(b), as expected due to the $O(2)$ symmetry of the model. Conversely, in the presence of a vision cone $\theta < 360^{\circ}$ (here for the ARXY model), 
the energy difference obtained by rotating a perfectly aligned initial spin configuration takes the form of a square wave, with peaks of value $1/2$
(see Eq.~\eqref{eq:arxy-energy}).  This corresponds to the energy gained when each spin stops interacting with a neighbor, and is due to the presence of the EUR discussed in Sec.~\ref{sec:EUR}.
In the absence of symmetry, the difference between $E[\{r_{\vartheta}(\phi_i)\}]$ and $E[\{\phi_i\}]$ actually depends on the configuration $\{\phi_i\}$. For this reason, in Fig.~\ref{fig:rotation}(b)
we apply the rotation $r_{\vartheta}$ also to the case of a thermal configuration of the system at a certain temperature $T$. In practice, at each step of the MC evolution, we rotate the configuration $\{\phi_i\}$ of the system by an angle $\vartheta$, calculate the energy difference per spin $(E[\{r_{\vartheta}(\phi_i)\}] - E[\{\phi_i\}])/N$, and then we average over many MC realizations. The resulting curve, obtained at low temperatures in the LRO phase, is smoother compared to the one corresponding to an initial configuration with perfectly aligned spins, but it still exhibits a periodicity of $90^\circ$. We also note that, as the temperature $T$ increases, the system is progressively less influenced by the presence of the EUR: the energy-difference curve becomes increasingly flatter, corresponding to the DO phase. This procedure will be used to rationalize the order-by-disorder transition observed for the SRXY model in, c.f., Sec.~\ref{sec:obd}. 

In the following, we apply these notions for elucidating different features 
of the transition from LRO to DO for
two selected values of the VC angle.
%


\subsection{Two different manifestations of $\mathbb{Z}_4$ symmetry: $\theta = 100^{\circ}$ and $\theta =280^{\circ}$}
\label{sec:arxy100280-interpretation}

\begin{figure*}
    \centering
        \includegraphics[width=0.99\linewidth]{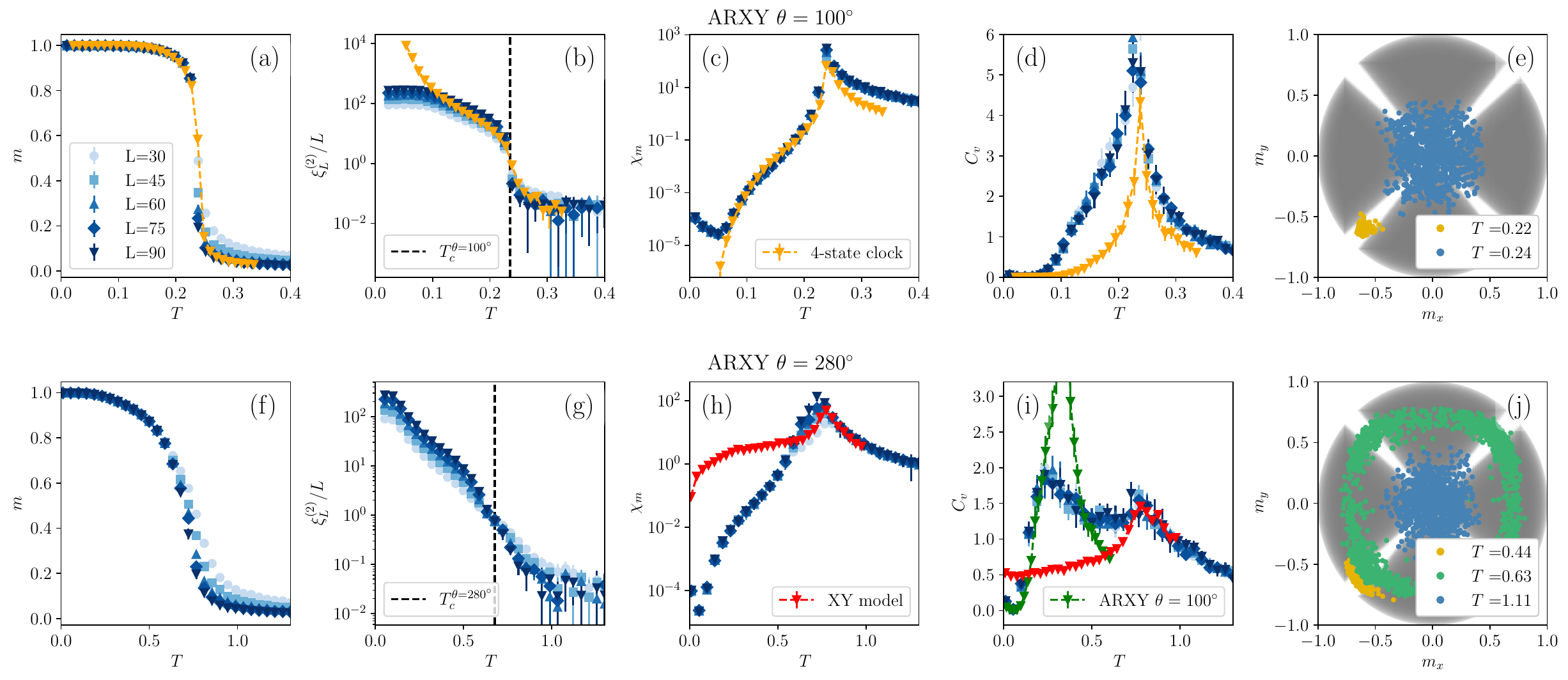}
         \captionsetup{justification=RaggedRight} 
    \caption{\small
     Behavior
     of the quantities introduced in Sec.~\ref{sec:observables} for the
     ARXY model
     with VC angles (a)--(e) $\theta = 100^{\circ}$, and (f)--(j) $\theta = 280^{\circ}$. Despite both VC angles sharing the same EUR size, the plots show marked differences.
     Notably, for $\theta = 100^{\circ}$ (upper row of plots), the behavior is 
     similar to that of the 4-state clock model (gold diamond symbols), as evidenced in panels (a)--(c) by the similar profile of $m$, $\xi_L^{(2)}$ and $\chi_m$ (upon rescaling of temperature), and by the comparable peak height of the specific heat $C_v$ in panel (d). For $\theta = 280^{\circ}$ (lower row of plots), we observe a hybrid behavior, as seen particularly in panel (i), where the specific heat initially grows similarly to the case $\theta = 100^{\circ}$ (green triangles), then decreases, and later exhibits a second peak with the same profile as the XY model (red triangles in panels (h) and (i)). 
} \label{fig:M2-theta100280}
\label{fig:100vs180}
\end{figure*}

\begin{figure}
    \centering
        \includegraphics[width = 0.17\textwidth]{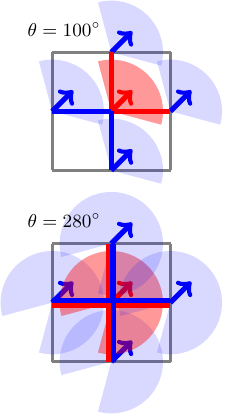}
        \put(-90,130){(a)}
        \hspace{5pt}
        \includegraphics[width = 0.28\textwidth]{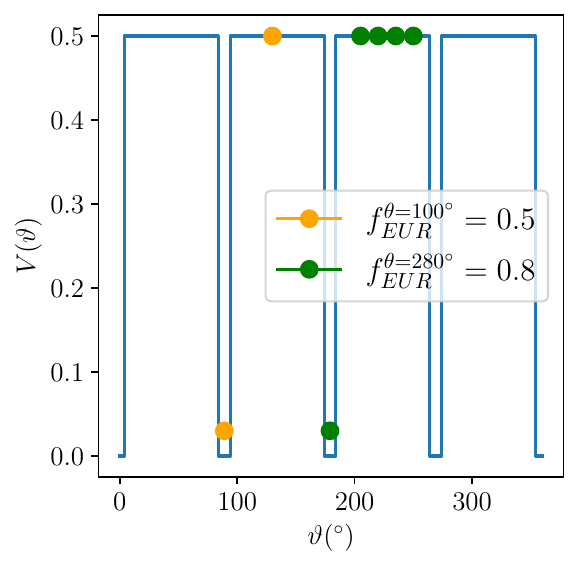}
        \put(-150,130){(b)}
         \captionsetup{justification=RaggedRight} 
    \caption{\small 
    Illustration of the difference in the behavior of the
    ARXY model with VC angles $\theta = 100^{\circ}$ and $\theta = 280^{\circ}$ at the transition from the LRO to the DO phase. (a) Perfectly aligned configuration $\{\bar{\phi}_i = 45^{\circ}\}$ --- as an approximation for low-temperature configurations --- for both 
    VC angles.
    Even though the EUR has the same size, for $\theta = 280^\circ$ the activated bonds (in red) are 
    twice as many 
    as
    for 
    $\theta = 100^{\circ}$, which doubles the energy scale of the model. 
    (b) Profile of the energy variation per spin, obtained by applying a rotation $r_{\vartheta}$ to the aligned configurations shown in (a). The maxima correspond to the energy gain of 1/2 when a spin enters the EUR and loses one nearest neighbor. 
    Within our approximation, we consider this profile as a potential $V(\vartheta)$ (see Eq.~\eqref{eq:v-theta}), to be used at finite temperatures to compute the fraction of spins $f_{\text{EUR}}$ that are within the EUR (see Eq.~\eqref{eq:fraction-eur}). 
    At the transition temperature for $\theta = 100^\circ$, only half of the spins are within the EUR, and the $\mathbb{Z}_4$ symmetry is more pronounced than for $\theta = 280^\circ$, where 4 out of 5 spins are in the EUR, and the potential effectively seen by the system is flat, resembling that of the XY model. 
    }
    \label{fig:zeroT-mf}
\end{figure}

\textit{Numerical results. --- } 
Figure~\ref{fig:100vs180} reports the results of the MC simulations of the ARXY for the two values $\theta= 100^{\circ}$ and $\theta = 280^{\circ}$ of the VC angle. In both cases, the size $\alpha_\text{EUR}$ of the EUR is the same, i.e., $\alpha_{\text{EUR}} = 320^\circ$. However, when a spin enters the EUR, the number of activated bonds changes from $2$ to $1$ for $\theta = 100^{\circ}$, and from $4$ to $3$ for $\theta = 280^{\circ}$. 
This 
entails a
significantly different behavior of the observables for $\theta= 100^{\circ}$ and $\theta = 280^{\circ}$, shown in the upper and lower rows of Fig.~\ref{fig:100vs180}, respectively. 

For $\theta = 100^{\circ}$, the behavior observed in Figs.~\ref{fig:100vs180}(a)--(e) qualitatively resembles that of the 4-state clock model shown in Figs.~\ref{fig:benchmarks}(a)--(e), with an appropriate rescaling of the temperature. In particular, 
the magnetization $m$ in panel 
(a), 
the curves of $\xi_L^{(2)}/L$ in panel (b) (which, for different system sizes $L$, do not collapse below the transition temperature, but instead intersect at the onset of the DO phase), and the magnetic susceptibility $\chi_m$ in panel (c) (that is size-independent below the transition temperature), indicate the presence of LRO at low temperature. 
Numerical data for the 4-state clock model with $L=60$ are included for comparison, and they are represented by gold symbols and a dashed line in panels (a)--(d) of Fig.~\ref{fig:100vs180}. 
By locating the crossing of the
curves
$\xi_L^{(2)}/L$~vs.~$T$ with different $L$, 
we estimate the transition temperature $T_c^{\theta = 100^{\circ}} \simeq 0.23$,
which is indicated by a vertical dashed line in panel (b). 
This temperature is approximately four times smaller than that of the standard XY model. 
This fact can be rationalized by considering that,
when the system is disordered, approximately one bond out of four is activated, and thus the coupling energy rescales accordingly.

The temperature dependence of the specific heat $C_v$ in Fig.~\ref{fig:100vs180}(d) is also qualitatively very similar to that of the 4-state clock model, and it features a peak of comparable magnitude. 
The vectorial magnetization $(m_x,m_y)$ in Fig.~\ref{fig:100vs180}(e) illustrates that, at low temperatures, the spins collectively align along one of the four preferred directions (complementary to the EUR), while at higher temperatures, in the DO phase,  they form a rotationally symmetric cloud centered around $(0,0)$.

For $\theta = 280^{\circ}$ in Fig.~\ref{fig:100vs180}, the size dependence of the magnetization $m$ in panel (f),
of $\xi_L^{(2)}/L$ in panel (g), 
and of $\chi_m$ in panel (h)
are similar to those of the corresponding variables for the VC angle $\theta = 100^{\circ}$, shown in the panels (a)--(c) above. This indicates the presence of a phase with LRO at low temperatures, and a DO phase at higher temperatures. 

However, the specific heat $C_v$ in panel (i) of Fig.~\ref{fig:100vs180} shows a rather peculiar dependence on temperature $T$: upon increasing $T$, there is a first rise in $C_v$ that matches the increase of $C_v$ for the case $\theta = 100^{\circ}$ of panel (d) (displayed again for convenience, with a temperature rescaling, with green symbols in panel (i)). 
%
%
Upon further increasing $T$, a second peak gradually appears, with the same profile as that of the XY model (red symbols) reported in Fig.~\ref{fig:benchmarks}(n). 
%
This suggests that, contrary to the case $\theta = 100^{\circ}$, for $\theta = 280^{\circ}$ the system displays also a behavior which is somehow reminiscent of the XY model. From the crossing of the $\xi_L^{(2)}/L$ curves, we estimate the transition temperature $T_c^{\theta = 280^{\circ}} \simeq 0.68$,
which is indicated by a vertical dashed line in panel (g). This temperature is approximately 3/4 of that in the XY model. We rationalize this observation by noting that in an almost disordered phase, for $\theta = 280^{\circ}$, an average of three out of four bonds are activated, leading to a corresponding rescaling of the effective coupling. 
The scatter plot of the vectorial magnetization $\mathbf{m} = (m_x,m_y)$ reported in Fig.~\ref{fig:100vs180}(j) also suggests that an intermediate case between 
disorder and
the alignment in one of the four minima 
--- such as those observed in panel (e) --- appears. In this case, the magnetization $m$ does not vanish, but $\mathbf{m}$ lacks a preferred direction and the system becomes isotropic, similarly to what happens for the XY model in the phase with QLRO (see Fig.~\ref{fig:benchmarks}(o)). 
The emergence of this QLRO-like behavior in a phase which actually admits a LRO can be explained as discussed below.

\textit{The role of internal symmetry. --- } 
To motivate the observations listed above, let us consider
a zero-temperature configuration in which all spins are aligned, as illustrated in Fig.~\ref{fig:zeroT-mf}(a) for both values of the VC angle, e.g., forming an angle $\bar{\phi}_i = 45^\circ$ at site $i$. In the case of $\theta = 100^\circ$, each spin participates in the activation of two bonds (thus the energy per spin is $E^{\rm AR}/N = -1$, according to the normalization of Eq.~\eqref{eq:arxy-energy}), while in the case of $\theta = 280^\circ$, each spin activates four bonds (and thus $E^{\rm AR}/N = -2$). This estimate of the energy magnitudes suggests that the transition temperature for the model (that is proportional to the inverse of the effective coupling strength) with $\theta = 100^\circ$ is expected to be approximately half of the transition temperature of the model with $\theta = 280^\circ$. Indeed, as discussed above, we estimated them to be $T_c^{\theta=100^{\circ}} \simeq 0.23$ and $T_c^{\theta=280^{\circ}} \simeq 0.68 $ from panels (b) and (g) of Fig.~\ref{fig:100vs180}, respectively.
By applying a rotation $r_{\vartheta}$ to the configuration $\{\bar{\phi}_i\}$, the energy variation per spin, i.e., 
\begin{equation}\label{eq:v-theta}
    V(\vartheta) = (E[\{r_{\vartheta} (\bar{\phi}_i)\}] - E[\{\bar{\phi}_i\}])/N,
\end{equation}
shown in Fig.~\ref{fig:zeroT-mf}(b) as a function of $\vartheta$, exhibits a periodicity of $90^\circ$. 
The upper plateaus of the profile correspond to the values of $\vartheta$ for which the rotated spins enter the EUR, losing one bond and thus increasing the energy per spin by 1/2. This $90^\circ$ periodicity implies that the interplay between the arrangement of the spins according to a square lattice and the vision cone actually reduces the symmetry of the model to $\mathbb{Z}_4$. This explains the features of the collective behavior of the model with $\theta = 100^\circ$, characterized by the fact that all observables reported in Fig.~\ref{fig:100vs180} are very similar to those of the 4-state clock model.

However, the same argument does not carry over to the case $\theta = 280^\circ$. In fact, while the EUR is not energetically favored, it is not strictly forbidden. Accordingly, at sufficiently high temperatures, a significant fraction of the spins
actually
occupies the EUR. 
In order to calculate the probability for this to occur at finite temperature, let us consider the profile shown in Fig.~\ref{fig:zeroT-mf}(b), that was obtained for an initial zero-temperature configuration.
In a first approximation, we assume 
that most of the spins remain 
roughly
aligned at finite temperature, with only a finite fraction $f_\text{EUR}$ of them entering the EUR. This effectively reduces the problem to that of a single particle at inverse temperature $\beta=1/T$ in the one-dimensional potential provided by Fig.~\ref{fig:zeroT-mf}(b), with $V(\vartheta)\simeq 0$ or $\simeq 1/2$ (see Eq.~\eqref{eq:v-theta}). 
In particular, $e^{-\beta V(\vartheta)}$ is approximately equal to 1 or $e^{-\beta / 2}$, and is proportional to the probability
for such particle to be, respectively, outside or inside the EUR.
\rev{In terms of the total extension $\alpha_\text{EUR} \in [0^\circ, 360^\circ)$ of the latter, 
the fraction $f_\text{EUR}$ of spins within the EUR is proportional to $\alpha_\text{EUR}\,\, e^{-\beta / 2}$, which has to be normalized by the constant $Z = \alpha_\text{EUR}\cdot e^{-\beta / 2} + (360^{\circ} - \alpha_{\text{EUR}}) \cdot 1$. 
Accordingly, one has
\begin{align} 
    f_\text{EUR}  =  \frac{\alpha_\text{EUR}\,\, e^{-\beta / 2}}{Z}  =\frac{\alpha_\text{EUR}\,\, e^{-\beta / 2}}{360^{\circ} - \alpha_\text{EUR} ( 1- e^{-\beta / 2})}. 
    \label{eq:fraction-eur}
\end{align} 
}
This relation allows us to determine the value of $f_\text{EUR}$ at the transition temperature for $\theta=100^\circ$ and $\theta=280^\circ$ on the basis of the knowledge of the transition temperatures, previously estimated from the crossing points in  Figs.~\ref{fig:100vs180}(b) and (g). Correspondingly, we obtain
$f_\text{EUR}^{\theta = 100^\circ} \simeq 0.5$ and $f_\text{EUR}^{\theta = 280^\circ} \simeq 0.8$. 
This implies that  the number of spins within the EUR for $\theta = 100^\circ$ equals that outside of it, meaning that the statistical weight of the four minima is actually relevant, and the behavior of the spins is influenced by their presence. On the contrary, for $\theta = 280^\circ$, the fraction of spins within the EUR is significantly larger, indicating that the statistical relevance of the four minima is reduced. As a result, the effective potential experienced by a significant fraction of the spins of the system, and resulting from that in Fig.~\ref{fig:zeroT-mf}(b), becomes essentially flat. 
Consequently, the system behaves like an XY model, for which a rotation $r_{\vartheta}$ leaves the energy unchanged. 
Note, however, that different quantities are influenced to a different extent by this approximate recovery of the $O(2)$ symmetry: the specific heat in Fig.~\ref{fig:100vs180}(i), in fact, displays the same behavior observed for the XY model at the transition, while the ratio $\xi_{L}^{(2)}/L$ reported in Fig.~\ref{fig:100vs180}(c) continues to feature the intersection typical of the transition from LRO to DO, indicating the presence of an underlying $\mathbb{Z}_4$ symmetry.

The above argument explains why the ARXY can behave very differently even for two values of $\theta$ for which the extension $\alpha_\text{EUR}$ of the EUR is the same: the change in the number of activated bonds per spin --- going from 2 to 1 for $\theta = 100^\circ$ and from 4 to 3 for $\theta = 280^\circ$ --- produces a significant difference. Specifically, it alters the energy scale of the systems relative to the energy lost upon deactivating a bond, which is always $1/2$.

Finally, for $\alpha_\text{EUR} = 0^\circ$ (which is equivalent to $\alpha_\text{EUR} = 360^\circ$, see Fig.~\ref{fig:eur}(d)), 
the argument presented above would suggest that the model should behave like a standard XY model with $O(2)$ symmetry. In the next section, we will explain why this is not really the case for  $\theta = 180^{\circ}$.

\subsection{ARXY for $\theta = 180^{\circ}$: LRO without EUR}\label{sec:arxy180-interpretation}


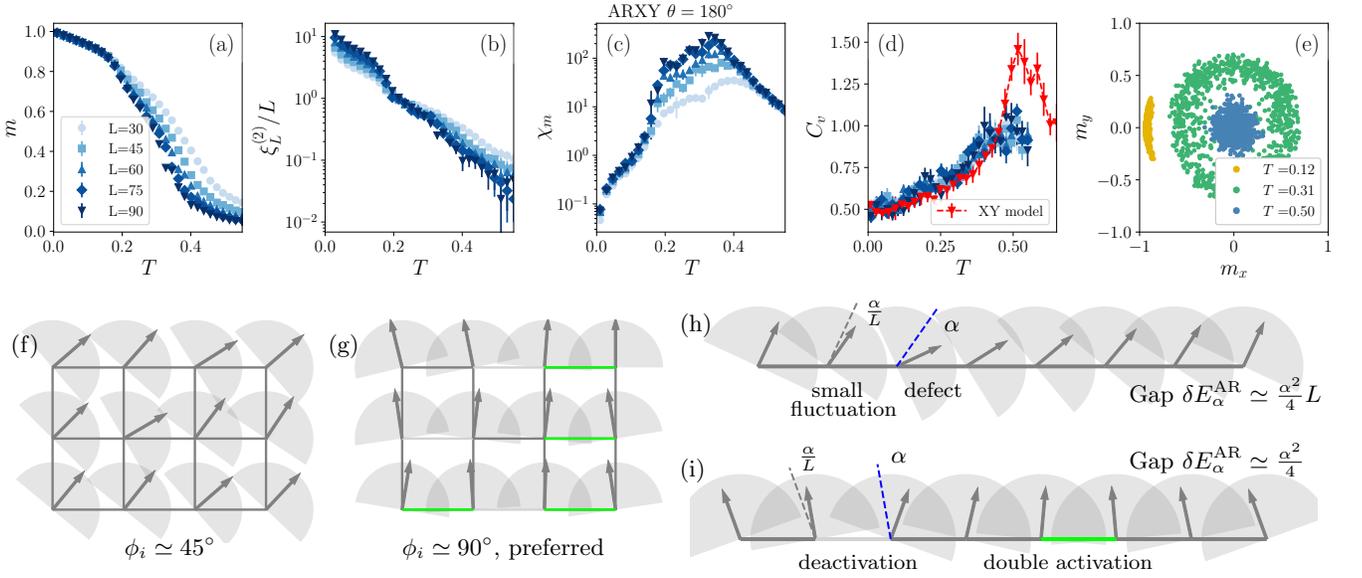
\begin{figure*}
    \centering
        \includegraphics[width=\linewidth]{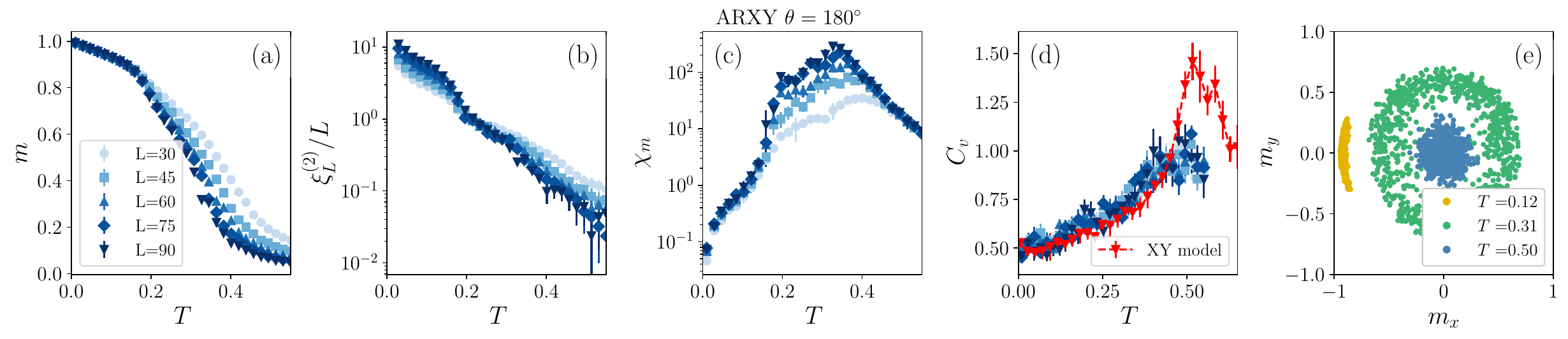}
        \raisebox{1cm}{\includegraphics[width = 0.5\linewidth]{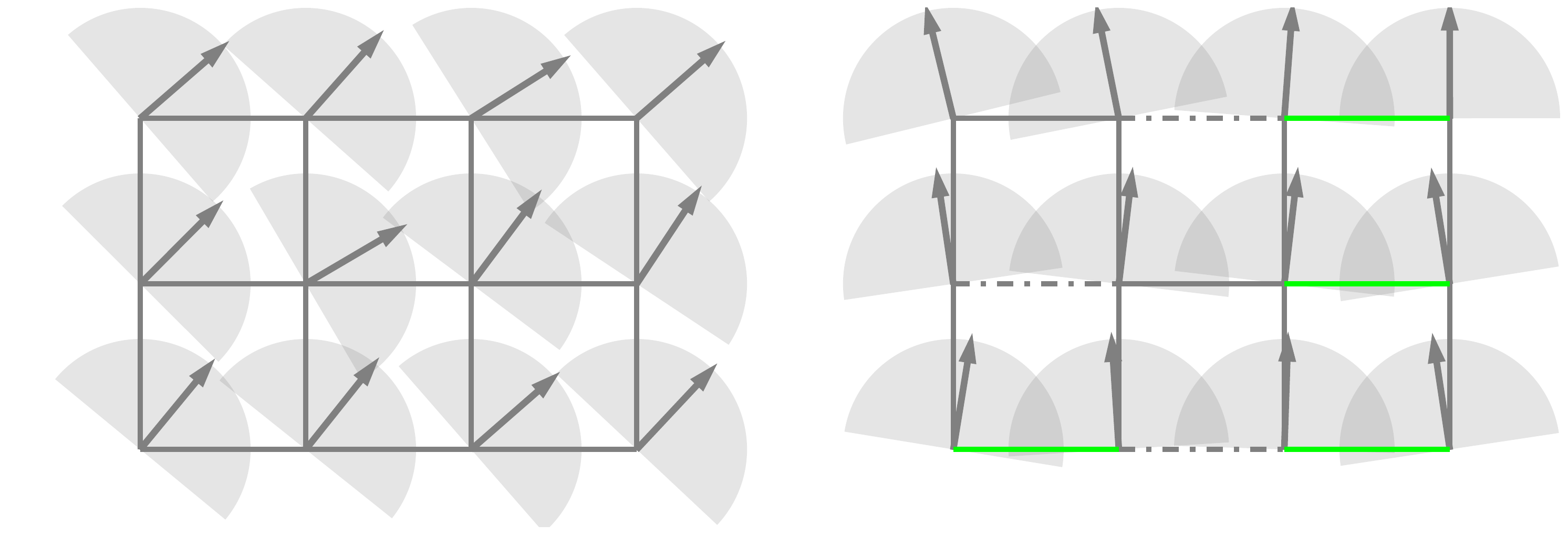}}
        \put(-248,102){(f)}
        \put(-128,102){(g)}
        \put(-205,25){$\phi_i\simeq 45^{\circ}$}
        \put(-100,25){$\phi_i\simeq 90^{\circ}$, preferred}
        \includegraphics[width = 0.5\linewidth]{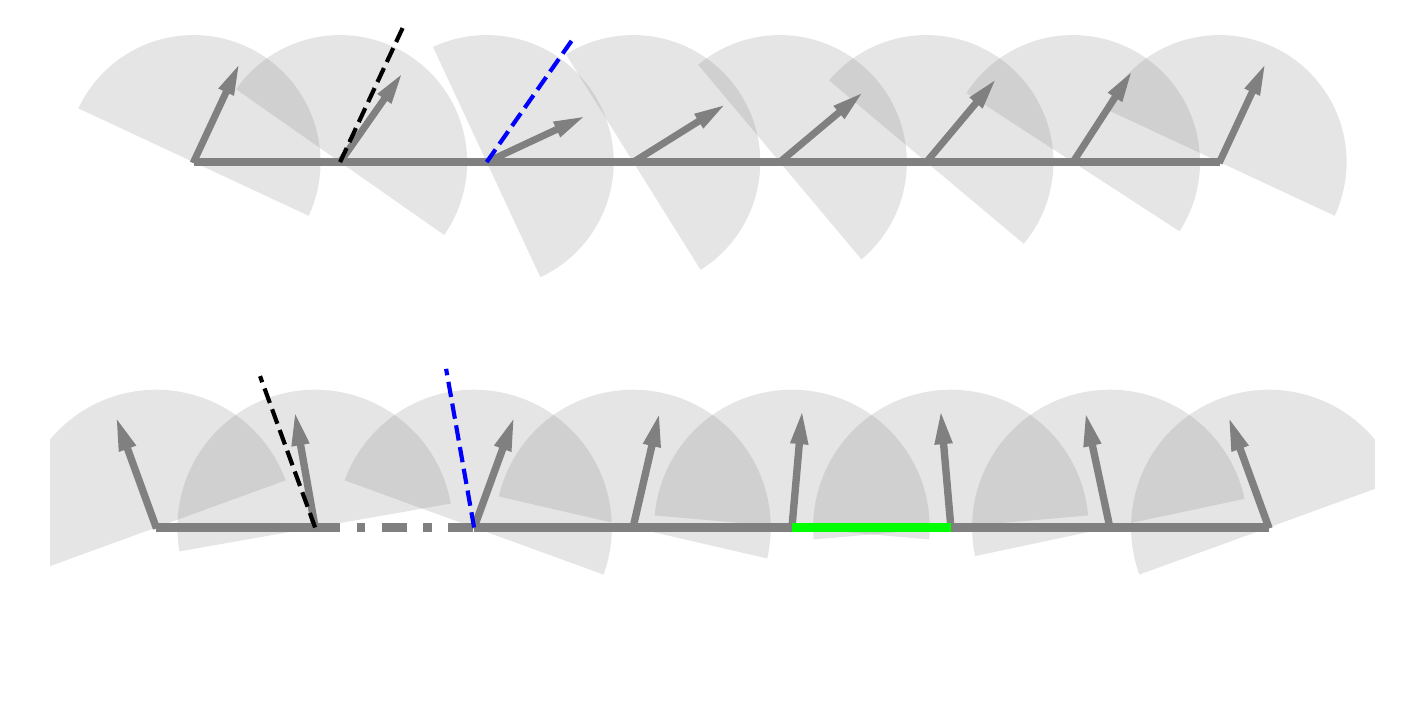} 
        \put(-250,110){(h)}
        \put(-180,115){$\frac{\alpha }{ L}$}
        \put(-150,110){$\alpha$}
        \put(-200,85){\footnotesize small }
        \put(-208,78){\footnotesize fluctuation}
        \put(-165,85){\footnotesize defect}
        \put(-80, 83){Gap $\delta E_\alpha^\text{AR} \simeq \frac{\alpha^2}{4}L$}        
        \put(-250,55){(i)}
        \put(-205,60){$\frac{\alpha }{ L}$}
        \put(-170,60){$\alpha$}
        \put(-205,20){\footnotesize deactivation}
        \put(-135,20){\footnotesize double activation}
        \put(-80, 58){Gap $\delta E_\alpha^\text{AR}\simeq \frac{\alpha^2}{4}$} 
        \vspace{-0.5cm}
        \captionsetup{justification=RaggedRight} 
    \caption{\small ARXY model for $\theta = 180^\circ$. (a)--(e) Result of the MC simulations for the quantities introduced in Sec.~\ref{sec:observables}. In panel (b), the behavior of the curves of $\xi_L^{(2)}/L$ upon varying $L$ indicates the presence of LRO at low temperatures. The vectorial magnetization in panel (e) demonstrates that, at low temperatures, the spins preferentially align along one of the four (equivalent) lattice directions.
(f)--(i) Illustration of the mechanism favoring configurations aligned along the lattice directions at low temperatures. 
(f) For a configuration aligned around $45^\circ$, all bonds remain activated \rev{(solid gray)} even in the presence of small fluctuations. 
(g) For a configuration aligned around $90^\circ$, small fluctuations can deactivate one bond \rev{(dash-dotted gray)} and doubly activate another one (in green). 
(h) For a gapless spin wave, the angular change between neighboring sites $i$ and $j$ scales as $\phi_i - \phi_j \sim 1/L$. A finite fluctuation $\alpha$, indicated as a defect in the figure, increases the energy by $\alpha^2 / 2$, and the corresponding configuration has an energy that scales subextensively. 
(i) If the bond at the defect is deactivated, the energy increase is finite ($1/2$), but is compensated by the double activation of another bond further along the chain. The corresponding configuration (identical to that in panel (h), but rotated by $45^\circ$) is gapless in the thermodynamic limit.
}  \label{fig:180-explanation}
\end{figure*}

As discussed at the end of the previous section, the argument presented there indicates that, in the absence of an EUR, the model should display an $O(2)$ symmetry. This was also suggested in Ref.~\cite{Loos2023} for the NRXY model. Accordingly, one would expect that, in this case, the transition occurs between a QLRO phase and a DO phase. However, this is not what we find, as observed in Fig.~\ref{fig:nrec-m}(e) for the NRXY model, and as will be shown for the ARXY model in the present section.

\textit{Numerical results. ---} The numerical results for the ARXY model at $\theta = 180^{\circ}$ are shown in Fig.~\ref{fig:180-explanation}(a)--(e). By looking at the temperature dependence of $\xi_L^{(2)}/L$ in panel (b), one notices that the curves corresponding to different sizes $L$ do not show the collapse expected for an extended critical phase (such as the one with QLRO) at low temperatures. On the contrary, the behavior is similar to that seen for the NRXY model in Fig.~\ref{fig:nrec-m}(e), indicating LRO.
In addition, from the scatter plot of the vectorial magnetization $\mathbf{m}$ in panel (e) it seems that at low temperatures the system preferentially aligns along one of the four (actually equivalent) lattice directions. Upon increasing the temperature, a behavior similar to that observed for the ARXY model at $\theta = 280^{\circ}$ emerges, where a flavor of QLRO appears, and no spatial orientation seems to be preferred. This is evident from the magnetic susceptibility in panel (c) as well, which scales with the system size within a finite range of temperatures before the transition.

\textit{Heuristic explanation. ---}
We aim to explain the fact that, at least at low temperatures, the various spatial orientations of the spins are not equivalent. This leads to a mechanism that reduces the symmetry from $O(2)$ to $\mathbb{Z}_4$, even in the absence of the EUR. 
To do so, we analyze the configurations with small fluctuations around a perfectly aligned state. In such a state, for $\theta = 180^\circ$, the number of activated bonds per spin is at most 2 (except for a perfect alignment with $\phi \in \{ 0^\circ, 90^\circ, 180^\circ, 270^\circ\}$, which is physically unrealistic). 
As shown in Fig.~\ref{fig:180-explanation}(f), if the system predominantly aligns around $45^\circ$ (or its odd multiples),
such small oscillations do not alter the configuration of the bonds activated once by the interaction (these bonds are depicted in dark gray in the figure). 
In contrast, if the spins of the system align along one of the four lattice directions, as depicted in Fig.~\ref{fig:180-explanation}(g), then even small fluctuations may lead to the double activation of one bond (green line) at the expense of the complete deactivation of another bond \rev{(dash-dotted gray)}. Note that the
PBCs
imposed on the lattice ensure that the number of
bonds that are activated twice is equal to that of the bonds that are deactivated, such that the energy of the configuration, up to fluctuations, is almost unaltered. 


As we are about to show, the two reference configurations discussed above are actually characterized by different spectra of gapless excitations, and thus they are not equivalent in determining the thermodynamics of the system, therefore breaking the rotational symmetry. 

Let us first consider the standard XY model, and its gapless spin-wave excitations: these are fluctuations where the angular change between neighboring sites $i$ and $j$ scales as $\phi_i - \phi_j \sim 1/L$,
which lead to a local increase of energy $\sim 1/L^2$. Summing this increase over the entire system amounts to multiplying it by $L^2$, which renders an intensive energy increase $\propto L^0$, corresponding to an excitation that is gapless in the thermodynamic limit. 
These kind of excitations are also present for the ARXY model, irrespective of the direction of global alignment of the spins (i.e., for both $45^\circ$ and $90^\circ$, and their multiples) --- provided that we consider, in these cases, 
fluctuations that do not alter the global alignment 
of the configurations, such that it still possible to identify configurations for which $\phi_i\simeq 45^{\circ}$ or $\phi_i\simeq 90^{\circ}$, as shown in panels (f) and (g) of Fig.~\ref{fig:180-explanation}, respectively.

Still considering the standard XY model, we now focus on the energy increase $\delta E_{\alpha}$ that results from the formation of a 
line of defects, where the angle of a single spin 
along a row changes by a finite value $\alpha$ (potentially small, but not scaling as $1/L$) compared to the direction of collective alignment of the remaining spins. Referring to
Fig.~\ref{fig:180-explanation}(h), assuming that one defect per lattice row is formed, the excitation energy is
\begin{equation}\label{eq:gapped-exitation}
    \delta E_{\alpha} = L \left[\frac{\alpha^2}{2L^2} (L-1) + \frac{\alpha^2}{2}\right] \simeq  \frac{\alpha^2}{2} L,
\end{equation}  
for large $L$. This results in a subextensive excitation.

Turning to the ARXY model, if the spins in the original configuration are aligned around $45^\circ$, and the fluctuations are not sufficiently large to alter the spatial arrangement of the activated lattice bonds, the excitation energy of the configuration in panel (h) is the same (up to a factor $1/2$ from Eq.~\eqref{eq:arxy-energy}), given by $\delta E^\text{AR}_{\alpha} \simeq (\alpha^2/4)L$.

On the other hand, if the configuration aligns around $90^\circ$, the same configuration actually becomes gapless. In panel (i), we show the same configuration 
as in panel (h) but rotated by $45^\circ$. At the defect point (along a row), one bond deactivates, gaining an energy of $1/2$ (independently of $\alpha$, thus the functional dependence on the defect angle disappears). This gain is 
compensated
by the double activation of another bond further along the row (because every deactivation must correspond to an activation of some other bond, due to PBCs). Consequently, the energy of this activation is, in this case,
\begin{equation}\label{eq:gapless-exitation}
    \delta E^\text{AR}_{\alpha} = L \left[\frac{\alpha^2}{4L^2} (L-2) + \frac{1}{2} - \frac{1}{2}\left(1 - \frac{\alpha^2}{2L^2}\right) \right] \simeq \frac{\alpha^2}{4}.
\end{equation}
This results in an intensive excitation $\propto L^0$.

In other words, in the second case, the spectrum of strongly gapless excitations (where the global energy required for excitation does not scale with system size) is broader. By contrast, in the first case, the same configurations have an energy 
that scales
subextensively. 
Our heuristic argument is then based on entropic reasoning: 
the system favors alignments 
around the spatial direction from which more microstates can be explored via fluctuations at
no energy cost.
We demonstrated that configurations near the lattice directions exhibit a higher number of strongly gapless states: consequently, the four lattice directions are preferred, breaking the $O(2)$ symmetry of the system and leaving a residual $\mathbb{Z}_4$ symmetry. Notably, this symmetry breaking occurs without the involvement of an EUR.

At present,
we are unable to make a more direct comparison with the behavior of the 4-state clock model or the XY model, as was done in Sec.~\ref{sec:arxy100280-interpretation}.

\section{The symmetric reciprocal XY model}\label{sec:results-srxy}

In this Section, we first present the results of MC simulations of the SRXY model introduced in Sec.~\ref{subsec:IIrec}, and then use symmetry arguments to rationalize the behavior observed for the various quantities.

\begin{figure*}
    \centering
        \includegraphics[width = 0.33\linewidth]{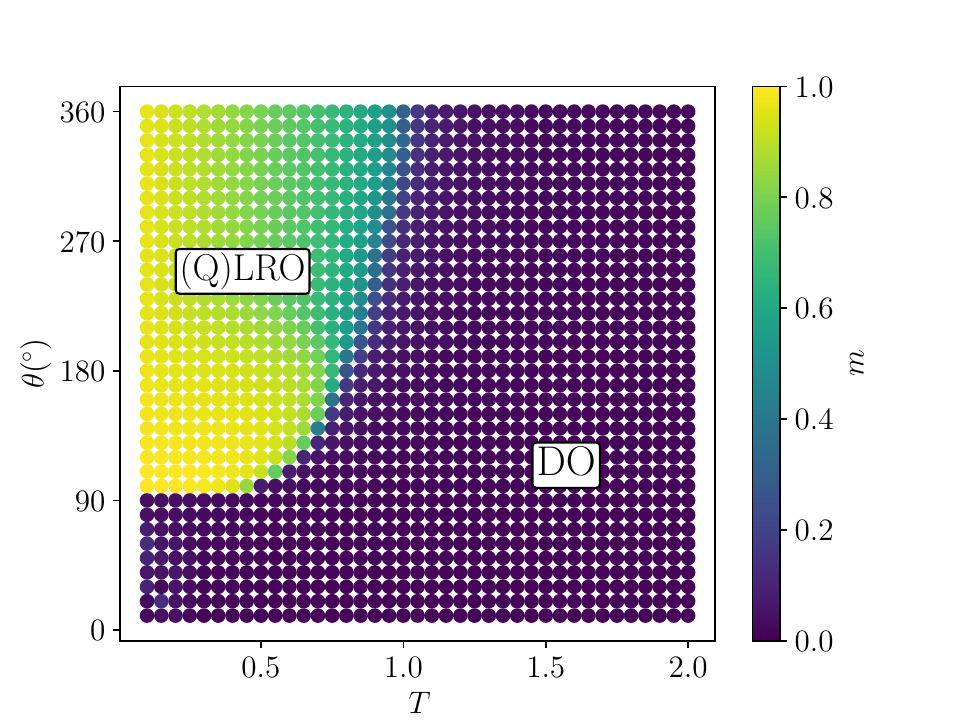}
        \put(-174,105){(a)}
        \!\!
        \includegraphics[width = 0.33\linewidth]{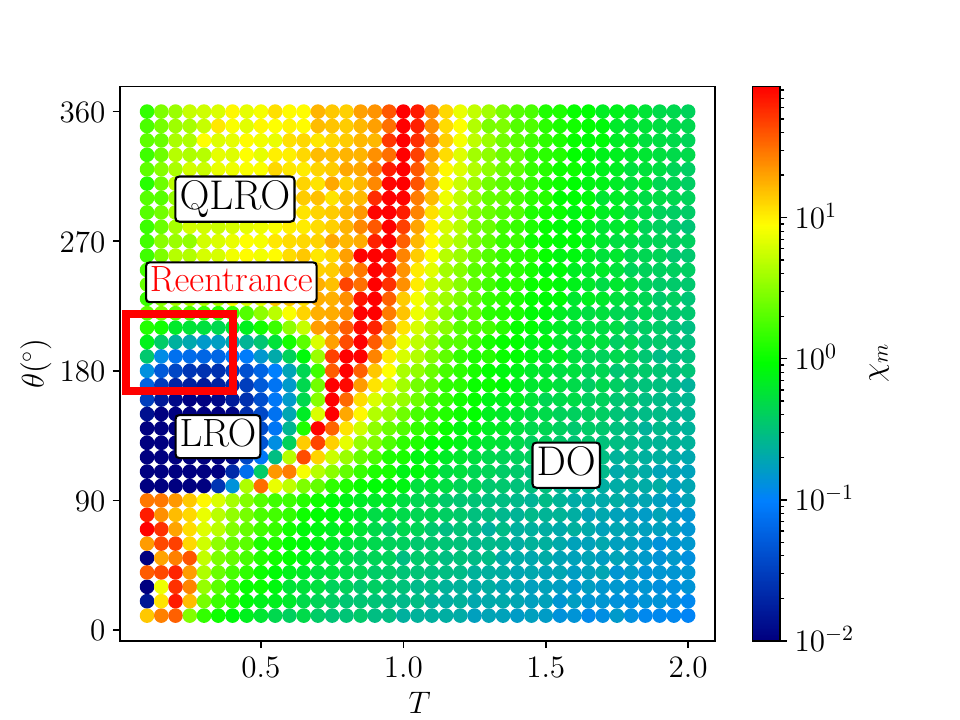}
        \put(-174,105){(b)}
        \!\!
        \includegraphics[width = 0.33\linewidth]{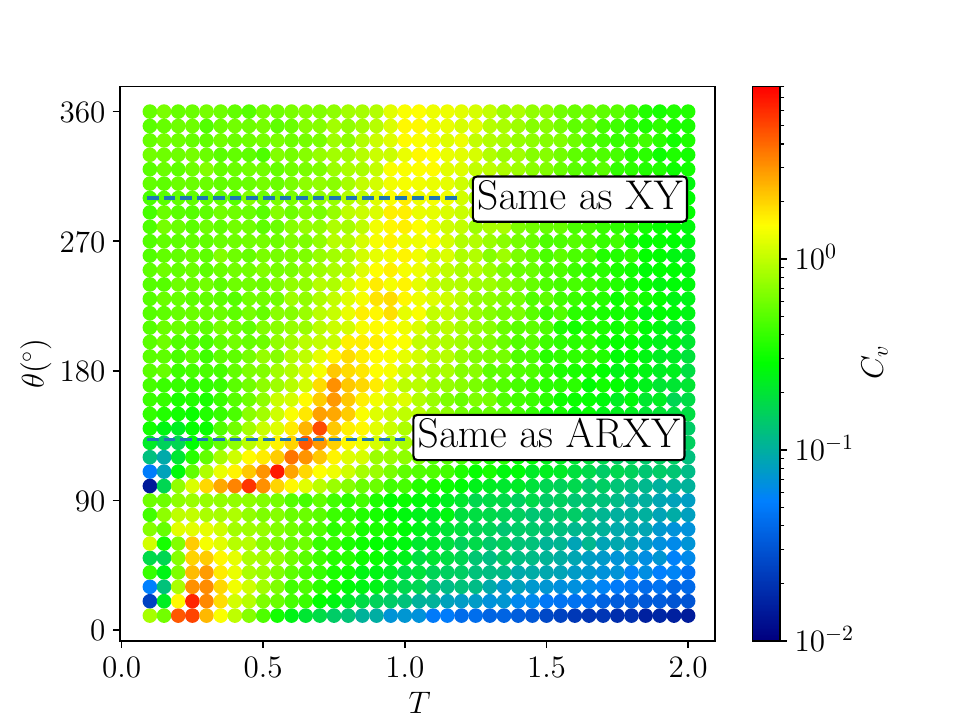}
        \put(-174,105){(c)} \captionsetup{justification=RaggedRight} 
    \caption{\small 
Results of the MC simulations for the equilibrium state of the SRXY model, obtained for a system of size $L=100$. (a) The magnetization $m$ as a function of the temperature $T$ and the VC angle $\theta$ does not exhibit the three lobes seen in the previous two models, see Figs.~\ref{fig:nrec-m}(a)~and~\ref{fig:arxy-pd}(a). The behavior of $m$, however, is not sufficient to distinguish
whether the low-temperature region with non-zero magnetization corresponds to a LRO or a QLRO phase. 
(b) The magnetic susceptibility $\chi_m$, below the transition to the DO phase, is characterized by a region within which it is relatively small, indicating the presence of a LRO, and another one within which it takes larger values, corresponding to the QLRO phase. 
Moreover, for $\theta \gtrsim 180^{\circ}$ we note the existence of a reentrance region (highlighted  by a red box), 
where the system transitions from QLRO to LRO as the temperature $T$ increases. 
(c) The specific heat $C_v$ shows peaks corresponding to the transitions from 
the low-temperature region
to DO, with higher peaks where the transition is from LRO to DO (as in the ARXY model), and lower peaks where the transition is from QLRO to DO (as in the XY model).
}
\label{fig:srxy-pd}
\end{figure*}

\subsection{Phase diagrams}\label{sec:srxy-pd}

Figure~\ref{fig:srxy-pd} shows the behavior of $m$, $\chi_m$ and $C_v$ as a function of the temperature $T$ and the VC angle $\theta$, obtained for a square system with $L=100$.
In particular,  panel (a) presents the magnetization $m$, which, contrary to the previous two models, does \emph{not} feature the three lobes which were visible in Figs.~\ref{fig:nrec-m}(a)~and~\ref{fig:arxy-pd}(a). 

To determine whether the low-temperature region with finite magnetization corresponds to a LRO or QLRO phase, we analyze the magnetic susceptibility $\chi_m$ shown in Fig.~\ref{fig:srxy-pd}(b). The plot reveals the presence of a lobe characterized by low values of the susceptibility, which suggests the possible presence of a LRO phase, and a region with higher susceptibility, associated with the QLRO phase, as we will discuss further below.
Moreover, the plot in Fig.~\ref{fig:srxy-pd}(c) shows that the specific heat $C_v$ takes larger values along the peaks for $90^{\circ} < \theta \lesssim 180^{\circ}$ (corresponding to the transition from LRO to DO), and smaller values for $\theta \gtrsim 180^{\circ}$ (corresponding to the transition from QLRO to DO). 
As we shall discuss in Sec.~\ref{sec:srxy-redundant}, these different behaviors observed for smaller or larger VC angles can be attributed to the presence of the redundant bonds introduced in Sec.~\ref{subsec:IIrec} for 
$\theta>180^{\circ}$.

Interestingly enough, the lobe characterized by LRO extends beyond the range $\theta \in [90^{\circ}, 180^{\circ}]$ and, in particular, it exceeds $180^{\circ}$, before showing a reentrant behavior at lower temperatures. The reentrance is highlighted by the red box in Fig.~\ref{fig:srxy-pd}(b). This suggests a rather unusual phenomenon: for values of $\theta$ slightly above $180^{\circ}$,
increasing the temperature $T$ drives the system from a QLRO phase into a LRO phase, resembling a type of order-by-disorder transition, which shall be discussed in Sec.~\ref{sec:obd}. 

\begin{figure*}
    \centering
        \includegraphics[width=\linewidth]{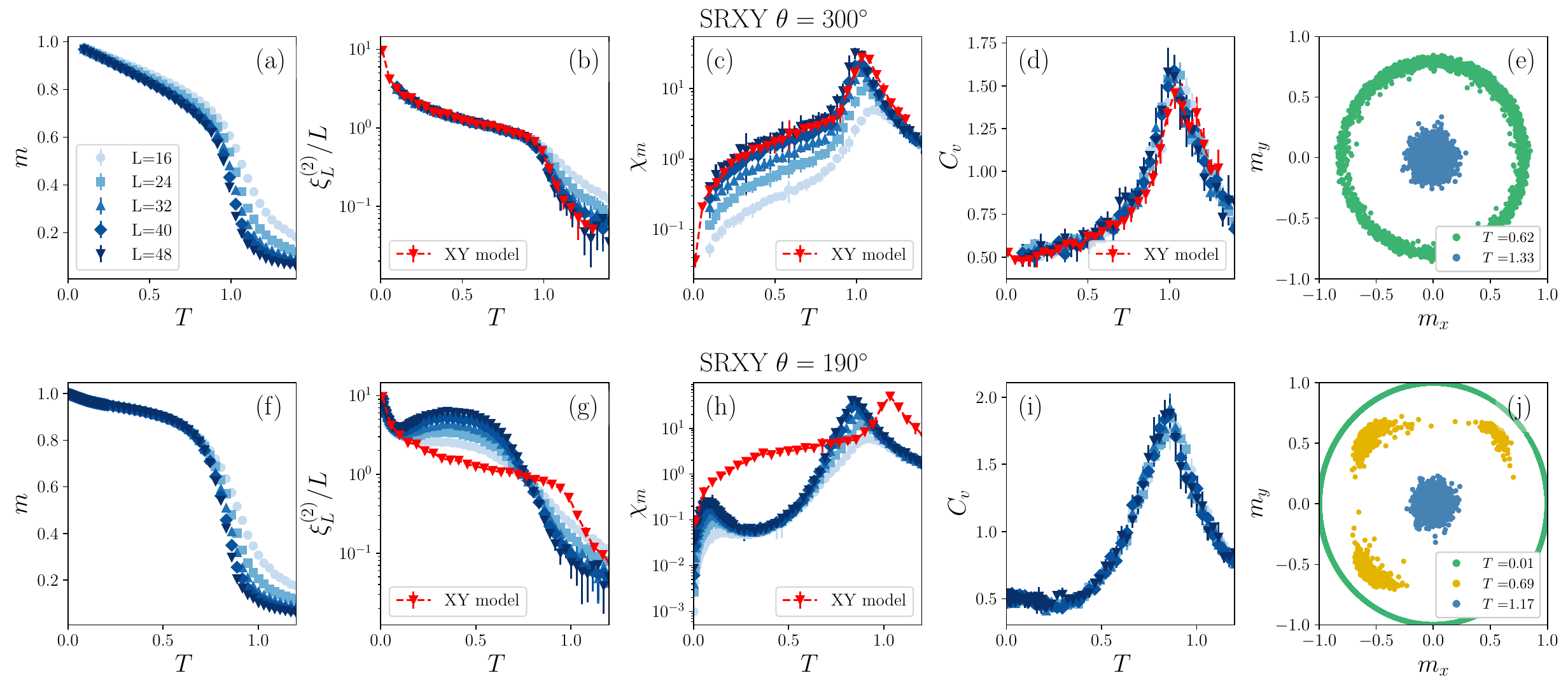}         \captionsetup{justification=RaggedRight} 
 \caption{\small
 Behavior of the quantities introduced in Sec.~\ref{sec:observables} for the SRXY model with VC angles (a)--(e) $\theta = 300^{\circ}$, and (f)--(j) $\theta = 190^{\circ}$. 
 For $\theta = 300^{\circ}$, all the observables in panels (a)--(e) display behaviors that are similar to those observed in the XY model presented in Fig.~\ref{fig:benchmarks}(k)--(o) (and here indicated by red triangles, for a comparison).
 For $\theta = 190^{\circ}$, evidence of an order-by-disorder transition is observed at low temperatures, specifically a transition from QLRO to LRO upon increasing the temperature. In panels  (g) and (h), the behavior of $\xi_L^{(2)}/L$ and $\chi_m$, respectively, are consistent at low temperatures with those of the XY model (i.e., with the presence of QLRO), but they deviate towards that expected in the presence of LRO as the temperature $T$ increases. The vectorial magnetization $\mathbf{m}$ in panel (e) is also consistent with the fact that at low temperatures the system is isotropic, while it recognizes the presence of four preferred directions only at higher temperatures.}
     \label{fig:M3-theta190}
\end{figure*}

In the next two sections, we will focus on the behavior of the SRXY model for two specific VC angles: $\theta = 300^\circ$, situated within the region where a transition from QLRO to DO occurs, and $\theta = 190^\circ$, where the aforementioned order-by-disorder phenomenon can be observed. The corresponding dependence of the relevant observables on the temperature $T$ are shown in Fig.~\ref{fig:M3-theta190}. 
%
%

\subsection{The role of redundant bonds for $\theta > 180^{\circ}$}\label{sec:srxy-redundant}

\begin{figure*}
    \centering
        \includegraphics[width = 0.15\linewidth]{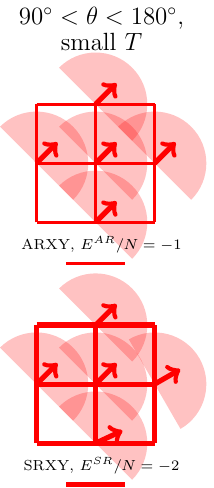}
        \put(-90,120){(a)}
        \qquad
         \includegraphics[width = 0.75\linewidth]{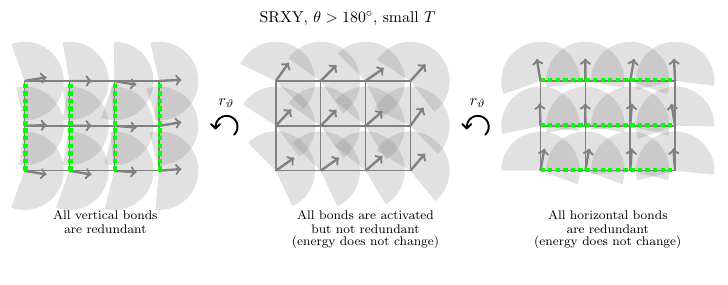}
        \put(-395,120){(b)}
        
         \captionsetup{justification=RaggedRight} 
    \caption{\small Illustration of the behavior of the SRXY model for $\theta \in [90^\circ, 180^\circ]$ and $\theta > 180^\circ$. (a) For $\theta \in [90^\circ, 180^\circ]$,
    in a low-temperature configuration with 
    all spins aligned up to small fluctuations, 
    there is no difference between the ARXY and SRXY models (apart from a factor 2 in the corresponding values of the energy, due to the normalization chosen in Eq.~\eqref{eq:arxy-energy}). In fact, all bonds are activated, an EUR exists, and the four directions that are preferred by the magnetization are the odd integer multiples of $45^\circ$. 
    Consequently, as in the case of the ARXY model (see Fig.~\ref{fig:arxy-pd}), there is a lobe with LRO at low temperatures in the phase diagrams in Fig.~\ref{fig:srxy-pd}(a)--(b). (b) For $\theta > 180^\circ$, the redundancy of bonds in the SRXY model makes it drastically different from the ARXY model. 
    Notably, a nearly aligned low-temperature configuration does not have a privileged alignment direction in the SRXY model. In fact, starting for example from the configuration $\{\bar{\phi}_i\approx0^{\circ}\}_i$, 
    all the vertical bonds are redundant; upon applying a global rotation $r_\vartheta$ of an angle $\vartheta \simeq 45^\circ$, all the bonds remain activated, but leaving zero redundant bonds, so that the energy of the system does not change (see Eq.~\eqref{eq:hamLasso}). With a further rotation, the horizontal bonds become redundant, but the energy is still unchanged. This mechanism effectively restores the $O(2)$ symmetry. 
}  \label{fig:illustration-SRXY}
\end{figure*}

The quantities plotted in Fig.~\ref{fig:srxy-pd} reveal that, unlike the NRXY and ARXY models, 
the SRXY model features
a single lobe of LRO for $90^{\circ}<\theta \lesssim 180^{\circ}$ and an extended QLRO phase for $\theta \gtrsim 180^\circ$. 
In order to highlight the occurrence of these different behaviors within the ordered phase, 
we focus on the behavior of the SRXY model for $\theta = 300^\circ$, which is reported in Fig.~\ref{fig:M3-theta190}(a)--(e). 
The close match of the observables in panels (a)--(d) with the corresponding profiles for the XY model, that we report in red symbols for $L=60$, allows one to conclude that the model exhibits only a QLRO and a DO phase. The behavior of the vectorial magnetization in panel (e) suggests a similar trend, forming a ring in the QLRO phase and an isotropic cloud centered around the origin in the DO phase. We refer to Figs.~\ref{fig:benchmarks}(k)--(o) for a comparison.
This restoration of the $O(2)$ symmetry in the SRXY model for $\theta > 180^\circ$ can be understood through the redundant bond mechanism introduced in Sec.~\ref{subsec:IIrec}, as explained below.

We start by analyzing a typical low-temperature configuration of the model for the VC angle $\theta$ either with $90^\circ < \theta < 180^\circ$ or $\theta > 180^\circ$, while applying the symmetry arguments already developed in the previous sections.
For $90^\circ<\theta<180^\circ$, 
Fig.~\ref{fig:illustration-SRXY}(a) illustrates that, assuming a configuration with aligned spins (up to small fluctuations) at low temperatures, there is actually no difference between the asymmetric and the symmetric reciprocal models in terms of the number of activated bonds, and thus the corresponding energy is the same, up to a factor of 2 arising from the choice of normalization in Eq.~\eqref{eq:arxy-energy} for the ARXY model.
When a rotation $r_{\vartheta}$ is applied to this spin configuration, the resulting change of the energy as a function of the angle $\vartheta$ is like the one shown in Fig.~\ref{fig:zeroT-mf}(c), exhibiting the $\mathbb{Z}_4$ symmetry and therefore indicating a transition from LRO to DO. 
Indeed, within this range of VC angles, the transition from LRO to DO is observed also in the behavior of $C_v$ reported in Fig.~\ref{fig:srxy-pd}(c), in terms of the occurrence of higher values of $C_v$ near the phase transition compared to the case with $\theta > 180^\circ$. In this respect, we remind (see also Figs.~\ref{fig:benchmarks}(d, n))
that the transition from LRO to DO is accompanied by a significantly more pronounced peak of $C_v$ (that scales logarithmically with system size) compared to the case of the transition between QLRO and DO (that is size-independent). 

The situation changes drastically for $\theta > 180^\circ$. In this case, applying the rotation $r_{\vartheta}$ to an aligned configuration does not change the energy of the SRXY model. The reason is the presence of the redundant bonds, as shown in Fig.~\ref{fig:illustration-SRXY}(b): for a hypothetical 
almost aligned configuration with $\{\bar{\phi}_i\approx0^{\circ}\}_i$ 
at low temperatures, all vertical bonds are redundant. However, the second ``virtual'' activation does not change the energy of the model (see Eq.~\eqref{eq:hamLasso}), contrary to what happens for the ARXY model. Rotating the system by an angle $\vartheta\simeq 45^{\circ}$ deactivates the redundancy of the vertical bonds, while the energy remains unchanged. This mechanism leads to a restoration of the $O(2)$ symmetry, which is reflected in the behavior of the susceptibility $\chi_m$ in Fig.~\ref{fig:srxy-pd}(b):  the two lobes with LRO that were present in the ARXY model for $\theta > 180^\circ$ are absent in the SRXY model, and they are replaced by a phase with QLRO.

\begin{figure*}
    \centering
        \raisebox{0.5cm}{\includegraphics[width = 0.21\linewidth]{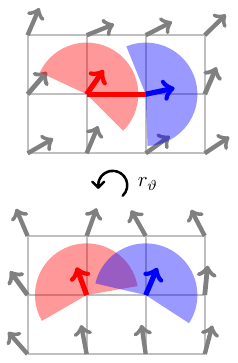}}
        \put(-120,152){(a)}
        \qquad
        \includegraphics[width = 0.65\linewidth]{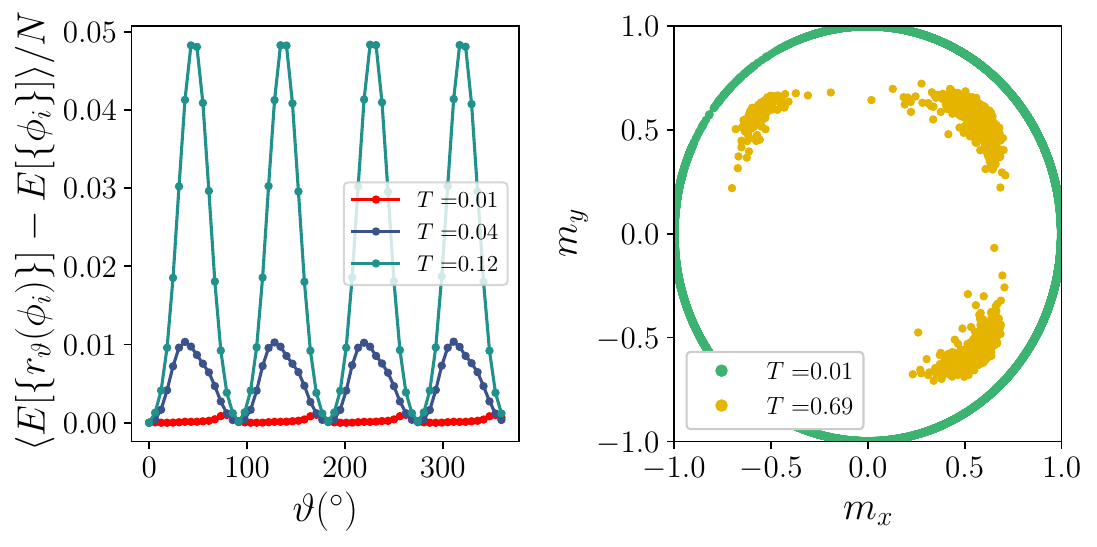}
        \put(-340,152){(b)}
        \put(-160,152){(c)}
         \captionsetup{justification=RaggedRight} 
    \caption{\small Illustration of the mechanism underlying the order-by-disorder transition observed in the
    SRXY model for $\theta = 190^\circ$. (a)
    As the temperature $T$ increases, the system exhibits more pronounced fluctuations even at the local level, making the presence of the redundant bonds less likely. In the upper panel, the bond between the red spin and the blue spin is activated only through the interaction of the red spin with the blue one. Thus, unlike what shown in Fig.~\ref{fig:illustration-SRXY}(b), the rotation $r_{\vartheta}$ of the spin configuration shown in the bottom panel can have the effect of deactivating a bond, thus increasing the energy. 
    (b) Energy difference due to a rotation $r_{\vartheta}$ of the spins as a function of $\vartheta$, for various temperatures $T$ close to the order-by-disorder transition. While at very low temperatures the energy difference vanishes, as expected for a system with $O(2)$ symmetry, finite values with $90^\circ$ periodicity emerge as $T$ increases. (c) Scatter plot of the vectorial magnetization $\mathbf{m}=(m_x,m_y)$. While at low temperatures the angular distribution of $\mathbf{m}$ is isotropic, upon increasing the temperature $T$,  $\mathbf{m}$ preferentially localizes along four directions: accordingly, disorder (in the form of fluctuations) reduces the symmetry from $O(2)$ to $\mathbb{Z}_4$. 
}  \label{fig:illustration-ObD}
\end{figure*}

\subsection{Order-by-disorder transition}\label{sec:obd}

As we have already noted in Sec.~\ref{sec:srxy-pd}, the behavior of the susceptibility $\chi_m$ in Fig.~\ref{fig:srxy-pd}(b) exhibits a reentrance at low temperatures for VC angles $\theta \gtrsim 180^\circ$. Correspondingly, upon increasing the temperature $T$, the system undergoes a transition from a phase with QLRO into one with LRO.
This behavior is further confirmed by the dependence of the observables $m$, $\xi_L^{(2)}/L$, $\chi_m$, $C_v$, and $\mathbf{m} = (m_x,m_y)$ shown in Figs.~\ref{fig:M3-theta190}(f)--(j) for this model with $\theta=190^{\circ}$. 
In fact, the susceptibility $\chi_m$ in Fig.~\ref{fig:M3-theta190}(h) initially increases, upon increasing the temperature $T$, in a size-dependent manner, matching the values observed for the XY model, reported with red symbols for $L=60$.
At higher values of $T$, $\chi_m$ then decreases and becomes size independent, as expected in a phase with LRO, before forming a peak at the transition to the DO phase. 
Similarly, in Fig.~\ref{fig:M3-theta190}(g) we observe that the curves for $\xi_L^{(2)}/L$ at various system sizes collapse at low temperatures, onto the same curve as for the 
XY model, which indicates the presence of a critical phase with QLRO. As the temperature increases, the system enters the phase with LRO, the curves corresponding to different values of $L$ begin to separate (as expected within a phase with LRO), but they intersect again at a single point corresponding to the temperature of transition to the DO phase. 
Another manifestation of the occurrence of the transition from QLRO to LRO is shown in Fig.~\ref{fig:M3-theta190}(j), where the vectorial magnetization $\mathbf{m} = (m_x,m_y)$ is finite and isotropic at very low temperatures. However, as the temperature increases, it becomes localized in four minima, before becoming isotropic again in the DO phase.

This transition from QLRO to LRO resembles an order-by-disorder transition, where, counterintuitively, order increases upon increasing the temperature~\cite{Melko2005,Honecker2011,Perkins2015}. 
The underlying mechanism driving this phenomenon can be rationalized in terms of the existence of redundant bonds,
and considering the rotation $r_{\vartheta}$ of the spin configuration as a tool to probe the internal symmetry of the system. 
As it has been already discussed in the previous section, an almost perfectly aligned configuration, such as the one in Fig.~\ref{fig:illustration-SRXY}(b), can present redundant bonds. However, this conclusion hinges on the assumption that the spins are aligned at low temperatures, which is actually  not the case for a system with $O(2)$ symmetry, because it cannot sustain an ordered state, which is destroyed by the spin-wave excitations at arbitrarily low temperatures. 
Nevertheless, the slow power-law decay of spatial correlations (see Eq.~\eqref{eq:correlation-scaling})  
implies that, at least locally, the fluctuations of the spins are small, and thus they can be considered almost aligned across large regions of space. 
By applying a rigid rotation $r_{\vartheta}$ to the spin configuration, we expect to observe no change in energy (indeed, this is what happens when $\theta$ is \emph{sufficiently} larger than $180^{\circ}$). However, as the temperature increases, local fluctuations become more pronounced, as illustrated in Fig.~\ref{fig:illustration-ObD}(a). 
As a consequence, as soon as 
$\theta$ is 
larger than $180^{\circ}$, some of the bonds that are not redundant might deactivate after a rotation $r_{\vartheta}$.
This is illustrated in Fig.~\ref{fig:illustration-ObD} for 
the VC angle $\theta=190^{\circ}$.  
In particular,
panel (a)
refers to the case of 
two neighboring spins that are connected by a non-redundant bond (activated only through the interaction of the red spin with the blue one): the rotation $r_{\vartheta}$ can 
lead to the deactivation of the bond, before it is activated again --- this time by the interaction of the blue spin with the red one. Panel (b) of 
Fig.~\ref{fig:illustration-ObD} shows the energy difference per spin after performing the rotation $r_{\vartheta}$ of the spin configuration with $\theta = 190^{\circ}$
at various temperatures close to the order-by-disorder transition. It can be seen that, as the temperature increases, regions emerge where the energy
difference becomes increasingly more
pronounced, with a dependence on $\vartheta$ showing a periodicity of $90^\circ$, thus again highlighting the $\mathbb{Z}_4$ symmetry. 
This is markedly different from the case shown in Fig.~\ref{fig:rotation} for the ARXY model, where an increase in temperature led to the smoothing of the square wave profile. 
Here, instead, at low temperatures no direction is favored, while as the temperature increases, a sort of EUR forms dynamically: 
fluctuations make the various directions not equivalent, and therefore favor the selection of one of 
the preferred directions. 
Specifically, the four lattice directions are disfavored in this case. Referring to the lower panel of Fig.~\ref{fig:illustration-ObD}(a), we observe that alignment around one of the lattice 
directions makes the system more susceptible to losing a bond following a small fluctuation. 
Consequently, the favored directions are the four odd integer multiples of $45^\circ$. 
This fact 
is also reflected in the vectorial magnetization shown in Fig.~\ref{fig:illustration-ObD}(c), which illustrates that at low temperatures there is no preferred direction, but one emerges as the temperature increases, demonstrating the transition from $O(2)$ to an effective $\mathbb{Z}_4$ symmetry.

\section{Conclusions}\label{sec:conclusions}

In this work, we explored the role of non-reciprocity in the XY model with a vision cone,
building on the analysis
initiated in Ref.~\cite{Loos2023}. By studying two reciprocal versions of the model (which we termed ARXY and SRXY models), we demonstrated that the non-reciprocity itself is \emph{not essential} for achieving LRO at low temperatures. It is rather the vision cone, which introduces a non-trivial coupling between the degrees of freedom of the system --- i.e., the two-component spins --- and the 
bond structure of the lattice, that causes the system to display a LRO phase characteristic of systems with $\mathbb{Z}_4$ symmetry. 

In particular, in Sec.~\ref{sec:arxy-results} we used numerical MC simulations to analyze
the LRO to DO transition in the ARXY model,
and introduced 
arguments based on the internal symmetry in order to rationalize
how the various quantities we study change
as a function of the temperature,
for various values of the vision cone. 

In Sec.~\ref{sec:results-srxy}
we studied a second reciprocal variant of the system, the SRXY model, in which bonds are activated regardless of whether they are within the vision cone of one or both of the
spins
connected by the bond. This redundant activation mechanism results in a radically different phase diagram, with a phase with QLRO for $\theta > 180^\circ$. In particular, for $\theta \gtrsim 180^\circ$, we observed that, starting from $T\gtrsim 0$, increasing the temperature 
drives the system from a QLRO phase to a LRO phase (see Fig.~\ref{fig:srxy-pd}(b)). 
The symmetry arguments mentioned above are able to 
elucidate
how this order-by-disorder transition, typically seen in quantum models~\cite{Melko2005,Honecker2011,Perkins2015},
can also occur in this classical context --- characterized by an energy that involves non-trivial couplings between the internal degrees of freedom and the underlying spatial structure.
It turns out that, in this case, the relevant mechanism involves a
fluctuation-induced reduction of symmetry (or degeneracy), from $O(2)$ to $\mathbb{Z}_4$.

A natural extension of our study would be towards off-lattice models, where the spin variables $\mathbf{s}_i$ represent the velocity vectors of active particles whose positions $\mathbf{r}_i$ evolve according to $\dot{\mathbf{r}}_i = \mathbf{s}_i$, and which tend to align 
their orientations with that of
particles in their spatial proximity~\cite{Vicsek1995}. In particular, it would be interesting to compare the phase diagram of active particle models with either reciprocal or non-reciprocal VC interactions.
Note that, in this case, both these models would be out of equilibrium, contrary to the on-lattice case, where the reciprocal variants of the model are at equilibrium. Accordingly, it would be interesting to analyze the EPR and the response to external perturbations in this context~\cite{Loffredo2023}. Similarly, future studies could address the critical dynamics of these models~\cite{halperin}, and the fate of collective phenomena such as flocking and the motility-induced phase separation (in the presence of additional positional interactions~\cite{marchetti_review}).
Finally, our findings call for a field-theoretical framework capable of capturing some of the features of the phase diagrams presented here. In particular, it would be valuable to further investigate the distinct roles that internal symmetry plays in the reciprocal and non-reciprocal versions of the model, as well as how these differences can modify or preserve certain aspects of the same universality class, as it was very recently addressed in Ref.~\cite{saha2024}.

\begin{acknowledgments}
We acknowledge the contribution of Daria Brtan to the initial stages of this project. AG acknowledges support from MIUR PRIN project “Coarse-grained description for non-equilibrium systems and transport phenomena (CO-NEST)” n.~201798CZL.
\end{acknowledgments}

\bibliography{references}

\providecommand{\noopsort}[1]{}\providecommand{\singleletter}[1]{#1}%
\begin{thebibliography}{59}%
\makeatletter
\providecommand \@ifxundefined [1]{%
 \@ifx{#1\undefined}
}%
\providecommand \@ifnum [1]{%
 \ifnum #1\expandafter \@firstoftwo
 \else \expandafter \@secondoftwo
 \fi
}%
\providecommand \@ifx [1]{%
 \ifx #1\expandafter \@firstoftwo
 \else \expandafter \@secondoftwo
 \fi
}%
\providecommand \natexlab [1]{#1}%
\providecommand \enquote  [1]{``#1''}%
\providecommand \bibnamefont  [1]{#1}%
\providecommand \bibfnamefont [1]{#1}%
\providecommand \citenamefont [1]{#1}%
\providecommand \href@noop [0]{\@secondoftwo}%
\providecommand \href [0]{\begingroup \@sanitize@url \@href}%
\providecommand \@href[1]{\@@startlink{#1}\@@href}%
\providecommand \@@href[1]{\endgroup#1\@@endlink}%
\providecommand \@sanitize@url [0]{\catcode `\\12\catcode `\$12\catcode `\&12\catcode `\#12\catcode `\^12\catcode `\_12\catcode `\%12\relax}%
\providecommand \@@startlink[1]{}%
\providecommand \@@endlink[0]{}%
\providecommand \url  [0]{\begingroup\@sanitize@url \@url }%
\providecommand \@url [1]{\endgroup\@href {#1}{\urlprefix }}%
\providecommand \urlprefix  [0]{URL }%
\providecommand \Eprint [0]{\href }%
\providecommand \doibase [0]{https://doi.org/}%
\providecommand \selectlanguage [0]{\@gobble}%
\providecommand \bibinfo  [0]{\@secondoftwo}%
\providecommand \bibfield  [0]{\@secondoftwo}%
\providecommand \translation [1]{[#1]}%
\providecommand \BibitemOpen [0]{}%
\providecommand \bibitemStop [0]{}%
\providecommand \bibitemNoStop [0]{.\EOS\space}%
\providecommand \EOS [0]{\spacefactor3000\relax}%
\providecommand \BibitemShut  [1]{\csname bibitem#1\endcsname}%
\let\auto@bib@innerbib\@empty
\bibitem [{\citenamefont {Onsager}(1944)}]{Onsager1944}%
  \BibitemOpen
  \bibfield  {author} {\bibinfo {author} {\bibfnamefont {L.}~\bibnamefont {Onsager}},\ }\bibfield  {title} {\bibinfo {title} {Crystal statistics. {I}. {A} two-dimensional model with an order-disorder transition},\ }\href {https://doi.org/10.1103/PhysRev.65.117} {\bibfield  {journal} {\bibinfo  {journal} {Phys. Rev.}\ }\textbf {\bibinfo {volume} {65}},\ \bibinfo {pages} {117} (\bibinfo {year} {1944})}\BibitemShut {NoStop}%
\bibitem [{\citenamefont {Mermin}\ and\ \citenamefont {Wagner}(1966)}]{Mermin1966}%
  \BibitemOpen
  \bibfield  {author} {\bibinfo {author} {\bibfnamefont {N.~D.}\ \bibnamefont {Mermin}}\ and\ \bibinfo {author} {\bibfnamefont {H.}~\bibnamefont {Wagner}},\ }\bibfield  {title} {\bibinfo {title} {Absence of ferromagnetism or antiferromagnetism in one- or two-dimensional isotropic {Heisenberg} models},\ }\href {https://doi.org/10.1103/PhysRevLett.17.1133} {\bibfield  {journal} {\bibinfo  {journal} {Phys. Rev. Lett.}\ }\textbf {\bibinfo {volume} {17}},\ \bibinfo {pages} {1133} (\bibinfo {year} {1966})}\BibitemShut {NoStop}%
\bibitem [{\citenamefont {Kosterlitz}\ and\ \citenamefont {Thouless}(1973)}]{Kosterlitz_1973}%
  \BibitemOpen
  \bibfield  {author} {\bibinfo {author} {\bibfnamefont {J.~M.}\ \bibnamefont {Kosterlitz}}\ and\ \bibinfo {author} {\bibfnamefont {D.~J.}\ \bibnamefont {Thouless}},\ }\bibfield  {title} {\bibinfo {title} {Ordering, metastability and phase transitions in two-dimensional systems},\ }\href {https://doi.org/10.1088/0022-3719/6/7/010} {\bibfield  {journal} {\bibinfo  {journal} {J. Phys. C}\ }\textbf {\bibinfo {volume} {6}},\ \bibinfo {pages} {1181} (\bibinfo {year} {1973})}\BibitemShut {NoStop}%
\bibitem [{\citenamefont {Kosterlitz}(1974)}]{kosterlitz1974}%
  \BibitemOpen
  \bibfield  {author} {\bibinfo {author} {\bibfnamefont {J.~M.}\ \bibnamefont {Kosterlitz}},\ }\bibfield  {title} {\bibinfo {title} {The critical properties of the two-dimensional {XY} model},\ }\href {https://doi.org/10.1088/0022-3719/7/6/005} {\bibfield  {journal} {\bibinfo  {journal} {J. Phys. C}\ }\textbf {\bibinfo {volume} {7}},\ \bibinfo {pages} {1046} (\bibinfo {year} {1974})}\BibitemShut {NoStop}%
\bibitem [{\citenamefont {Loos}\ \emph {et~al.}(2023)\citenamefont {Loos}, \citenamefont {Klapp},\ and\ \citenamefont {Martynec}}]{Loos2023}%
  \BibitemOpen
  \bibfield  {author} {\bibinfo {author} {\bibfnamefont {S.~A.~M.}\ \bibnamefont {Loos}}, \bibinfo {author} {\bibfnamefont {S.~H.~L.}\ \bibnamefont {Klapp}},\ and\ \bibinfo {author} {\bibfnamefont {T.}~\bibnamefont {Martynec}},\ }\bibfield  {title} {\bibinfo {title} {Long-range order and directional defect propagation in the nonreciprocal $\mathit{XY}$ model with vision cone interactions},\ }\href {https://doi.org/10.1103/PhysRevLett.130.198301} {\bibfield  {journal} {\bibinfo  {journal} {Phys. Rev. Lett.}\ }\textbf {\bibinfo {volume} {130}},\ \bibinfo {pages} {198301} (\bibinfo {year} {2023})}\BibitemShut {NoStop}%
\bibitem [{\citenamefont {Vicsek}\ \emph {et~al.}(1995)\citenamefont {Vicsek}, \citenamefont {Czir\'ok}, \citenamefont {Ben-Jacob}, \citenamefont {Cohen},\ and\ \citenamefont {Shochet}}]{Vicsek1995}%
  \BibitemOpen
  \bibfield  {author} {\bibinfo {author} {\bibfnamefont {T.}~\bibnamefont {Vicsek}}, \bibinfo {author} {\bibfnamefont {A.}~\bibnamefont {Czir\'ok}}, \bibinfo {author} {\bibfnamefont {E.}~\bibnamefont {Ben-Jacob}}, \bibinfo {author} {\bibfnamefont {I.}~\bibnamefont {Cohen}},\ and\ \bibinfo {author} {\bibfnamefont {O.}~\bibnamefont {Shochet}},\ }\bibfield  {title} {\bibinfo {title} {Novel type of phase transition in a system of self-driven particles},\ }\href {https://doi.org/10.1103/PhysRevLett.75.1226} {\bibfield  {journal} {\bibinfo  {journal} {Phys. Rev. Lett.}\ }\textbf {\bibinfo {volume} {75}},\ \bibinfo {pages} {1226} (\bibinfo {year} {1995})}\BibitemShut {NoStop}%
\bibitem [{\citenamefont {Toner}\ and\ \citenamefont {Tu}(1995)}]{Toner1995}%
  \BibitemOpen
  \bibfield  {author} {\bibinfo {author} {\bibfnamefont {J.}~\bibnamefont {Toner}}\ and\ \bibinfo {author} {\bibfnamefont {Y.}~\bibnamefont {Tu}},\ }\bibfield  {title} {\bibinfo {title} {Long-range order in a two-dimensional dynamical $\mathrm{XY}$ model: How birds fly together},\ }\href {https://doi.org/10.1103/PhysRevLett.75.4326} {\bibfield  {journal} {\bibinfo  {journal} {Phys. Rev. Lett.}\ }\textbf {\bibinfo {volume} {75}},\ \bibinfo {pages} {4326} (\bibinfo {year} {1995})}\BibitemShut {NoStop}%
\bibitem [{\citenamefont {Uchida}\ and\ \citenamefont {Golestanian}(2010)}]{Uchida2010}%
  \BibitemOpen
  \bibfield  {author} {\bibinfo {author} {\bibfnamefont {N.}~\bibnamefont {Uchida}}\ and\ \bibinfo {author} {\bibfnamefont {R.}~\bibnamefont {Golestanian}},\ }\bibfield  {title} {\bibinfo {title} {Synchronization and collective dynamics in a carpet of microfluidic rotors},\ }\href {https://doi.org/10.1103/PhysRevLett.104.178103} {\bibfield  {journal} {\bibinfo  {journal} {Phys. Rev. Lett.}\ }\textbf {\bibinfo {volume} {104}},\ \bibinfo {pages} {178103} (\bibinfo {year} {2010})}\BibitemShut {NoStop}%
\bibitem [{\citenamefont {Shankar}\ \emph {et~al.}(2022)\citenamefont {Shankar}, \citenamefont {Souslov}, \citenamefont {Bowick}, \citenamefont {Marchetti},\ and\ \citenamefont {Vitelli}}]{Shankar2022}%
  \BibitemOpen
  \bibfield  {author} {\bibinfo {author} {\bibfnamefont {S.}~\bibnamefont {Shankar}}, \bibinfo {author} {\bibfnamefont {A.}~\bibnamefont {Souslov}}, \bibinfo {author} {\bibfnamefont {M.~J.}\ \bibnamefont {Bowick}}, \bibinfo {author} {\bibfnamefont {M.~C.}\ \bibnamefont {Marchetti}},\ and\ \bibinfo {author} {\bibfnamefont {V.}~\bibnamefont {Vitelli}},\ }\bibfield  {title} {\bibinfo {title} {Topological active matter},\ }\href {https://doi.org/10.1038/s42254-022-00445-3} {\bibfield  {journal} {\bibinfo  {journal} {Nat. Rev. Phys.}\ }\textbf {\bibinfo {volume} {4}},\ \bibinfo {pages} {380} (\bibinfo {year} {2022})}\BibitemShut {NoStop}%
\bibitem [{\citenamefont {Saha}\ \emph {et~al.}(2019)\citenamefont {Saha}, \citenamefont {Ramaswamy},\ and\ \citenamefont {Golestanian}}]{Saha_2019}%
  \BibitemOpen
  \bibfield  {author} {\bibinfo {author} {\bibfnamefont {S.}~\bibnamefont {Saha}}, \bibinfo {author} {\bibfnamefont {S.}~\bibnamefont {Ramaswamy}},\ and\ \bibinfo {author} {\bibfnamefont {R.}~\bibnamefont {Golestanian}},\ }\bibfield  {title} {\bibinfo {title} {Pairing, waltzing and scattering of chemotactic active colloids},\ }\href {https://doi.org/10.1088/1367-2630/ab20fd} {\bibfield  {journal} {\bibinfo  {journal} {New J. Phys.}\ }\textbf {\bibinfo {volume} {21}},\ \bibinfo {pages} {063006} (\bibinfo {year} {2019})}\BibitemShut {NoStop}%
\bibitem [{\citenamefont {Nagy}\ \emph {et~al.}(2010)\citenamefont {Nagy}, \citenamefont {{\'A}kos}, \citenamefont {Biro},\ and\ \citenamefont {Vicsek}}]{Nagy2010}%
  \BibitemOpen
  \bibfield  {author} {\bibinfo {author} {\bibfnamefont {M.}~\bibnamefont {Nagy}}, \bibinfo {author} {\bibfnamefont {Z.}~\bibnamefont {{\'A}kos}}, \bibinfo {author} {\bibfnamefont {D.}~\bibnamefont {Biro}},\ and\ \bibinfo {author} {\bibfnamefont {T.}~\bibnamefont {Vicsek}},\ }\bibfield  {title} {\bibinfo {title} {Hierarchical group dynamics in pigeon flocks},\ }\href {https://doi.org/10.1038/nature08891} {\bibfield  {journal} {\bibinfo  {journal} {Nature}\ }\textbf {\bibinfo {volume} {464}},\ \bibinfo {pages} {890} (\bibinfo {year} {2010})}\BibitemShut {NoStop}%
\bibitem [{\citenamefont {Cavagna}\ \emph {et~al.}(2017)\citenamefont {Cavagna}, \citenamefont {Giardina}, \citenamefont {Jelic}, \citenamefont {Melillo}, \citenamefont {Parisi}, \citenamefont {Silvestri},\ and\ \citenamefont {Viale}}]{cavagna2017nonsymmetric}%
  \BibitemOpen
  \bibfield  {author} {\bibinfo {author} {\bibfnamefont {A.}~\bibnamefont {Cavagna}}, \bibinfo {author} {\bibfnamefont {I.}~\bibnamefont {Giardina}}, \bibinfo {author} {\bibfnamefont {A.}~\bibnamefont {Jelic}}, \bibinfo {author} {\bibfnamefont {S.}~\bibnamefont {Melillo}}, \bibinfo {author} {\bibfnamefont {L.}~\bibnamefont {Parisi}}, \bibinfo {author} {\bibfnamefont {E.}~\bibnamefont {Silvestri}},\ and\ \bibinfo {author} {\bibfnamefont {M.}~\bibnamefont {Viale}},\ }\bibfield  {title} {\bibinfo {title} {Nonsymmetric interactions trigger collective swings in globally ordered systems},\ }\href {https://doi.org/10.1103/PhysRevLett.118.138003} {\bibfield  {journal} {\bibinfo  {journal} {Phys. Rev. Lett.}\ }\textbf {\bibinfo {volume} {118}},\ \bibinfo {pages} {138003} (\bibinfo {year} {2017})}\BibitemShut {NoStop}%
\bibitem [{\citenamefont {Yllanes}\ \emph {et~al.}(2017)\citenamefont {Yllanes}, \citenamefont {Leoni},\ and\ \citenamefont {Marchetti}}]{Yllanes_2017}%
  \BibitemOpen
  \bibfield  {author} {\bibinfo {author} {\bibfnamefont {D.}~\bibnamefont {Yllanes}}, \bibinfo {author} {\bibfnamefont {M.}~\bibnamefont {Leoni}},\ and\ \bibinfo {author} {\bibfnamefont {M.~C.}\ \bibnamefont {Marchetti}},\ }\bibfield  {title} {\bibinfo {title} {How many dissenters does it take to disorder a flock?},\ }\href {https://doi.org/10.1088/1367-2630/aa8ed7} {\bibfield  {journal} {\bibinfo  {journal} {New J. Phys.}\ }\textbf {\bibinfo {volume} {19}},\ \bibinfo {pages} {103026} (\bibinfo {year} {2017})}\BibitemShut {NoStop}%
\bibitem [{\citenamefont {Tan}\ \emph {et~al.}(2022)\citenamefont {Tan}, \citenamefont {Mietke}, \citenamefont {Li}, \citenamefont {Chen}, \citenamefont {Higinbotham}, \citenamefont {Foster}, \citenamefont {Gokhale}, \citenamefont {Dunkel},\ and\ \citenamefont {Fakhri}}]{tan2022odd}%
  \BibitemOpen
  \bibfield  {author} {\bibinfo {author} {\bibfnamefont {T.~H.}\ \bibnamefont {Tan}}, \bibinfo {author} {\bibfnamefont {A.}~\bibnamefont {Mietke}}, \bibinfo {author} {\bibfnamefont {J.}~\bibnamefont {Li}}, \bibinfo {author} {\bibfnamefont {Y.}~\bibnamefont {Chen}}, \bibinfo {author} {\bibfnamefont {H.}~\bibnamefont {Higinbotham}}, \bibinfo {author} {\bibfnamefont {P.~J.}\ \bibnamefont {Foster}}, \bibinfo {author} {\bibfnamefont {S.}~\bibnamefont {Gokhale}}, \bibinfo {author} {\bibfnamefont {J.}~\bibnamefont {Dunkel}},\ and\ \bibinfo {author} {\bibfnamefont {N.}~\bibnamefont {Fakhri}},\ }\bibfield  {title} {\bibinfo {title} {Odd dynamics of living chiral crystals},\ }\href {https://doi.org/https://doi.org/10.1038/s41586-022-04889-6} {\bibfield  {journal} {\bibinfo  {journal} {Nature}\ }\textbf {\bibinfo {volume} {607}},\ \bibinfo {pages} {287} (\bibinfo {year} {2022})}\BibitemShut {NoStop}%
\bibitem [{\citenamefont {Dadhichi}\ \emph {et~al.}(2020)\citenamefont {Dadhichi}, \citenamefont {Kethapelli}, \citenamefont {Chajwa}, \citenamefont {Ramaswamy},\ and\ \citenamefont {Maitra}}]{Dadhichi2020}%
  \BibitemOpen
  \bibfield  {author} {\bibinfo {author} {\bibfnamefont {L.~P.}\ \bibnamefont {Dadhichi}}, \bibinfo {author} {\bibfnamefont {J.}~\bibnamefont {Kethapelli}}, \bibinfo {author} {\bibfnamefont {R.}~\bibnamefont {Chajwa}}, \bibinfo {author} {\bibfnamefont {S.}~\bibnamefont {Ramaswamy}},\ and\ \bibinfo {author} {\bibfnamefont {A.}~\bibnamefont {Maitra}},\ }\bibfield  {title} {\bibinfo {title} {Nonmutual torques and the unimportance of motility for long-range order in two-dimensional flocks},\ }\href {https://doi.org/10.1103/PhysRevE.101.052601} {\bibfield  {journal} {\bibinfo  {journal} {Phys. Rev. E}\ }\textbf {\bibinfo {volume} {101}},\ \bibinfo {pages} {052601} (\bibinfo {year} {2020})}\BibitemShut {NoStop}%
\bibitem [{\citenamefont {Besse}\ \emph {et~al.}(2022)\citenamefont {Besse}, \citenamefont {Chat\'e},\ and\ \citenamefont {Solon}}]{Besse_2022}%
  \BibitemOpen
  \bibfield  {author} {\bibinfo {author} {\bibfnamefont {M.}~\bibnamefont {Besse}}, \bibinfo {author} {\bibfnamefont {H.}~\bibnamefont {Chat\'e}},\ and\ \bibinfo {author} {\bibfnamefont {A.}~\bibnamefont {Solon}},\ }\bibfield  {title} {\bibinfo {title} {Metastability of constant-density flocks},\ }\href {https://doi.org/10.1103/PhysRevLett.129.268003} {\bibfield  {journal} {\bibinfo  {journal} {Phys. Rev. Lett.}\ }\textbf {\bibinfo {volume} {129}},\ \bibinfo {pages} {268003} (\bibinfo {year} {2022})}\BibitemShut {NoStop}%
\bibitem [{\citenamefont {Montbri\'o}\ and\ \citenamefont {Paz\'o}(2018)}]{Montbri2018}%
  \BibitemOpen
  \bibfield  {author} {\bibinfo {author} {\bibfnamefont {E.}~\bibnamefont {Montbri\'o}}\ and\ \bibinfo {author} {\bibfnamefont {D.}~\bibnamefont {Paz\'o}},\ }\bibfield  {title} {\bibinfo {title} {Kuramoto model for excitation-inhibition-based oscillations},\ }\href {https://doi.org/10.1103/PhysRevLett.120.244101} {\bibfield  {journal} {\bibinfo  {journal} {Phys. Rev. Lett.}\ }\textbf {\bibinfo {volume} {120}},\ \bibinfo {pages} {244101} (\bibinfo {year} {2018})}\BibitemShut {NoStop}%
\bibitem [{\citenamefont {Sompolinsky}\ and\ \citenamefont {Kanter}(1986)}]{Sompolinsky1986}%
  \BibitemOpen
  \bibfield  {author} {\bibinfo {author} {\bibfnamefont {H.}~\bibnamefont {Sompolinsky}}\ and\ \bibinfo {author} {\bibfnamefont {I.}~\bibnamefont {Kanter}},\ }\bibfield  {title} {\bibinfo {title} {Temporal association in asymmetric neural networks},\ }\href {https://doi.org/10.1103/PhysRevLett.57.2861} {\bibfield  {journal} {\bibinfo  {journal} {Phys. Rev. Lett.}\ }\textbf {\bibinfo {volume} {57}},\ \bibinfo {pages} {2861} (\bibinfo {year} {1986})}\BibitemShut {NoStop}%
\bibitem [{\citenamefont {Scheibner}\ \emph {et~al.}(2020)\citenamefont {Scheibner}, \citenamefont {Souslov}, \citenamefont {Banerjee}, \citenamefont {Sur{\'o}wka}, \citenamefont {Irvine},\ and\ \citenamefont {Vitelli}}]{Scheibner2020}%
  \BibitemOpen
  \bibfield  {author} {\bibinfo {author} {\bibfnamefont {C.}~\bibnamefont {Scheibner}}, \bibinfo {author} {\bibfnamefont {A.}~\bibnamefont {Souslov}}, \bibinfo {author} {\bibfnamefont {D.}~\bibnamefont {Banerjee}}, \bibinfo {author} {\bibfnamefont {P.}~\bibnamefont {Sur{\'o}wka}}, \bibinfo {author} {\bibfnamefont {W.~T.~M.}\ \bibnamefont {Irvine}},\ and\ \bibinfo {author} {\bibfnamefont {V.}~\bibnamefont {Vitelli}},\ }\bibfield  {title} {\bibinfo {title} {Odd elasticity},\ }\href {https://doi.org/10.1038/s41567-020-0795-y} {\bibfield  {journal} {\bibinfo  {journal} {Nat. Phys.}\ }\textbf {\bibinfo {volume} {16}},\ \bibinfo {pages} {475} (\bibinfo {year} {2020})}\BibitemShut {NoStop}%
\bibitem [{\citenamefont {Brandenbourger}\ \emph {et~al.}(2019)\citenamefont {Brandenbourger}, \citenamefont {Locsin}, \citenamefont {Lerner},\ and\ \citenamefont {Coulais}}]{Brandenbourger2019}%
  \BibitemOpen
  \bibfield  {author} {\bibinfo {author} {\bibfnamefont {M.}~\bibnamefont {Brandenbourger}}, \bibinfo {author} {\bibfnamefont {X.}~\bibnamefont {Locsin}}, \bibinfo {author} {\bibfnamefont {E.}~\bibnamefont {Lerner}},\ and\ \bibinfo {author} {\bibfnamefont {C.}~\bibnamefont {Coulais}},\ }\bibfield  {title} {\bibinfo {title} {Non-reciprocal robotic metamaterials},\ }\href {https://doi.org/10.1038/s41467-019-12599-3} {\bibfield  {journal} {\bibinfo  {journal} {Nat. Commun.}\ }\textbf {\bibinfo {volume} {10}},\ \bibinfo {pages} {4608} (\bibinfo {year} {2019})}\BibitemShut {NoStop}%
\bibitem [{\citenamefont {Li}\ \emph {et~al.}(2011)\citenamefont {Li}, \citenamefont {Wang}, \citenamefont {Han}, \citenamefont {Tian}, \citenamefont {Xi},\ and\ \citenamefont {Wang}}]{Li_2011}%
  \BibitemOpen
  \bibfield  {author} {\bibinfo {author} {\bibfnamefont {Y.-J.}\ \bibnamefont {Li}}, \bibinfo {author} {\bibfnamefont {S.}~\bibnamefont {Wang}}, \bibinfo {author} {\bibfnamefont {Z.-L.}\ \bibnamefont {Han}}, \bibinfo {author} {\bibfnamefont {B.-M.}\ \bibnamefont {Tian}}, \bibinfo {author} {\bibfnamefont {Z.-D.}\ \bibnamefont {Xi}},\ and\ \bibinfo {author} {\bibfnamefont {B.-H.}\ \bibnamefont {Wang}},\ }\bibfield  {title} {\bibinfo {title} {Optimal view angle in the three-dimensional self-propelled particle model},\ }\href {https://doi.org/10.1209/0295-5075/93/68003} {\bibfield  {journal} {\bibinfo  {journal} {EPL}\ }\textbf {\bibinfo {volume} {93}},\ \bibinfo {pages} {68003} (\bibinfo {year} {2011})}\BibitemShut {NoStop}%
\bibitem [{\citenamefont {Nguyen}\ \emph {et~al.}(2015)\citenamefont {Nguyen}, \citenamefont {Lee},\ and\ \citenamefont {Ngo}}]{Nguyen2015}%
  \BibitemOpen
  \bibfield  {author} {\bibinfo {author} {\bibfnamefont {P.~T.}\ \bibnamefont {Nguyen}}, \bibinfo {author} {\bibfnamefont {S.-H.}\ \bibnamefont {Lee}},\ and\ \bibinfo {author} {\bibfnamefont {V.~T.}\ \bibnamefont {Ngo}},\ }\bibfield  {title} {\bibinfo {title} {Effect of vision angle on the phase transition in flocking behavior of animal groups},\ }\href {https://doi.org/10.1103/PhysRevE.92.032716} {\bibfield  {journal} {\bibinfo  {journal} {Phys. Rev. E}\ }\textbf {\bibinfo {volume} {92}},\ \bibinfo {pages} {032716} (\bibinfo {year} {2015})}\BibitemShut {NoStop}%
\bibitem [{\citenamefont {Barberis}\ and\ \citenamefont {Peruani}(2016)}]{Barberis2016}%
  \BibitemOpen
  \bibfield  {author} {\bibinfo {author} {\bibfnamefont {L.}~\bibnamefont {Barberis}}\ and\ \bibinfo {author} {\bibfnamefont {F.}~\bibnamefont {Peruani}},\ }\bibfield  {title} {\bibinfo {title} {Large-scale patterns in a minimal cognitive flocking model: Incidental leaders, nematic patterns, and aggregates},\ }\href {https://doi.org/10.1103/PhysRevLett.117.248001} {\bibfield  {journal} {\bibinfo  {journal} {Phys. Rev. Lett.}\ }\textbf {\bibinfo {volume} {117}},\ \bibinfo {pages} {248001} (\bibinfo {year} {2016})}\BibitemShut {NoStop}%
\bibitem [{\citenamefont {Durve}\ \emph {et~al.}(2018)\citenamefont {Durve}, \citenamefont {Saha},\ and\ \citenamefont {Sayeed}}]{Durve2018}%
  \BibitemOpen
  \bibfield  {author} {\bibinfo {author} {\bibfnamefont {M.}~\bibnamefont {Durve}}, \bibinfo {author} {\bibfnamefont {A.}~\bibnamefont {Saha}},\ and\ \bibinfo {author} {\bibfnamefont {A.}~\bibnamefont {Sayeed}},\ }\bibfield  {title} {\bibinfo {title} {Active particle condensation by non-reciprocal and time-delayed interactions},\ }\href {http://dx.doi.org/10.1140/epje/i2018-11653-4} {\bibfield  {journal} {\bibinfo  {journal} {Eur. Phys. J. E}\ }\textbf {\bibinfo {volume} {41}},\ \bibinfo {pages} {49} (\bibinfo {year} {2018})}\BibitemShut {NoStop}%
\bibitem [{\citenamefont {Couzin}\ \emph {et~al.}(2002)\citenamefont {Couzin}, \citenamefont {Krause}, \citenamefont {James}, \citenamefont {Ruxton},\ and\ \citenamefont {Franks}}]{Couzin2002}%
  \BibitemOpen
  \bibfield  {author} {\bibinfo {author} {\bibfnamefont {I.~D.}\ \bibnamefont {Couzin}}, \bibinfo {author} {\bibfnamefont {J.}~\bibnamefont {Krause}}, \bibinfo {author} {\bibfnamefont {R.}~\bibnamefont {James}}, \bibinfo {author} {\bibfnamefont {G.~D.}\ \bibnamefont {Ruxton}},\ and\ \bibinfo {author} {\bibfnamefont {N.~R.}\ \bibnamefont {Franks}},\ }\bibfield  {title} {\bibinfo {title} {Collective memory and spatial sorting in animal groups},\ }\href {https://doi.org/10.1006/jtbi.2002.3065} {\bibfield  {journal} {\bibinfo  {journal} {J. Theor. Biol.}\ }\textbf {\bibinfo {volume} {218}},\ \bibinfo {pages} {1–11} (\bibinfo {year} {2002})}\BibitemShut {NoStop}%
\bibitem [{\citenamefont {Hildenbrandt}\ \emph {et~al.}(2010)\citenamefont {Hildenbrandt}, \citenamefont {Carere},\ and\ \citenamefont {Hemelrijk}}]{Hildenbrandt2010}%
  \BibitemOpen
  \bibfield  {author} {\bibinfo {author} {\bibfnamefont {H.}~\bibnamefont {Hildenbrandt}}, \bibinfo {author} {\bibfnamefont {C.}~\bibnamefont {Carere}},\ and\ \bibinfo {author} {\bibfnamefont {C.}~\bibnamefont {Hemelrijk}},\ }\bibfield  {title} {\bibinfo {title} {Self-organized aerial displays of thousands of starlings: a model},\ }\href {https://doi.org/10.1093/beheco/arq149} {\bibfield  {journal} {\bibinfo  {journal} {Behav. Ecol.}\ }\textbf {\bibinfo {volume} {21}},\ \bibinfo {pages} {1349–1359} (\bibinfo {year} {2010})}\BibitemShut {NoStop}%
\bibitem [{\citenamefont {Solon}\ \emph {et~al.}(2022)\citenamefont {Solon}, \citenamefont {Chat\'e}, \citenamefont {Toner},\ and\ \citenamefont {Tailleur}}]{Solon2022polarflocks}%
  \BibitemOpen
  \bibfield  {author} {\bibinfo {author} {\bibfnamefont {A.}~\bibnamefont {Solon}}, \bibinfo {author} {\bibfnamefont {H.}~\bibnamefont {Chat\'e}}, \bibinfo {author} {\bibfnamefont {J.}~\bibnamefont {Toner}},\ and\ \bibinfo {author} {\bibfnamefont {J.}~\bibnamefont {Tailleur}},\ }\bibfield  {title} {\bibinfo {title} {Susceptibility of polar flocks to spatial anisotropy},\ }\href {https://doi.org/10.1103/PhysRevLett.128.208004} {\bibfield  {journal} {\bibinfo  {journal} {Phys. Rev. Lett.}\ }\textbf {\bibinfo {volume} {128}},\ \bibinfo {pages} {208004} (\bibinfo {year} {2022})}\BibitemShut {NoStop}%
\bibitem [{\citenamefont {Chatterjee}\ \emph {et~al.}(2022)\citenamefont {Chatterjee}, \citenamefont {Mangeat},\ and\ \citenamefont {Rieger}}]{Chatterjee2022polarflocks}%
  \BibitemOpen
  \bibfield  {author} {\bibinfo {author} {\bibfnamefont {S.}~\bibnamefont {Chatterjee}}, \bibinfo {author} {\bibfnamefont {M.}~\bibnamefont {Mangeat}},\ and\ \bibinfo {author} {\bibfnamefont {H.}~\bibnamefont {Rieger}},\ }\bibfield  {title} {\bibinfo {title} {Polar flocks with discretized directions: The active clock model approaching the {Vicsek} model},\ }\href {https://doi.org/10.1209/0295-5075/ac6e4b} {\bibfield  {journal} {\bibinfo  {journal} {EPL}\ }\textbf {\bibinfo {volume} {138}},\ \bibinfo {pages} {41001} (\bibinfo {year} {2022})}\BibitemShut {NoStop}%
\bibitem [{\citenamefont {Solon}(2022)}]{solon-journal-club}%
  \BibitemOpen
  \bibfield  {author} {\bibinfo {author} {\bibfnamefont {A.}~\bibnamefont {Solon}},\ }\bibfield  {title} {\bibinfo {title} {Flocking without moving},\ }\href {http://dx.doi.org/10.36471/JCCM_December_2022_01} {\bibfield  {journal} {\bibinfo  {journal} {Journal Club for Condensed Matter Physics}\ } (\bibinfo {year} {2022})}\BibitemShut {NoStop}%
\bibitem [{\citenamefont {Rouzaire}\ \emph {et~al.}(2024)\citenamefont {Rouzaire}, \citenamefont {Pearce}, \citenamefont {Pagonabarraga},\ and\ \citenamefont {Levis}}]{rouzaire2024nonreciprocal}%
  \BibitemOpen
  \bibfield  {author} {\bibinfo {author} {\bibfnamefont {Y.}~\bibnamefont {Rouzaire}}, \bibinfo {author} {\bibfnamefont {D.~J.}\ \bibnamefont {Pearce}}, \bibinfo {author} {\bibfnamefont {I.}~\bibnamefont {Pagonabarraga}},\ and\ \bibinfo {author} {\bibfnamefont {D.}~\bibnamefont {Levis}},\ }\href {https://arxiv.org/abs/2401.12637} {\bibinfo {title} {Non-reciprocal interactions reshape topological defect annihilation}} (\bibinfo {year} {2024}),\ \Eprint {https://arxiv.org/abs/2401.12637} {arXiv:2401.12637 [cond-mat.stat-mech]} \BibitemShut {NoStop}%
\bibitem [{\citenamefont {Melko}\ \emph {et~al.}(2005)\citenamefont {Melko}, \citenamefont {Paramekanti}, \citenamefont {Burkov}, \citenamefont {Vishwanath}, \citenamefont {Sheng},\ and\ \citenamefont {Balents}}]{Melko2005}%
  \BibitemOpen
  \bibfield  {author} {\bibinfo {author} {\bibfnamefont {R.~G.}\ \bibnamefont {Melko}}, \bibinfo {author} {\bibfnamefont {A.}~\bibnamefont {Paramekanti}}, \bibinfo {author} {\bibfnamefont {A.~A.}\ \bibnamefont {Burkov}}, \bibinfo {author} {\bibfnamefont {A.}~\bibnamefont {Vishwanath}}, \bibinfo {author} {\bibfnamefont {D.~N.}\ \bibnamefont {Sheng}},\ and\ \bibinfo {author} {\bibfnamefont {L.}~\bibnamefont {Balents}},\ }\bibfield  {title} {\bibinfo {title} {Supersolid order from disorder: Hard-core bosons on the triangular lattice},\ }\href {https://doi.org/10.1103/PhysRevLett.95.127207} {\bibfield  {journal} {\bibinfo  {journal} {Phys. Rev. Lett.}\ }\textbf {\bibinfo {volume} {95}},\ \bibinfo {pages} {127207} (\bibinfo {year} {2005})}\BibitemShut {NoStop}%
\bibitem [{\citenamefont {Honecker}\ \emph {et~al.}(2011)\citenamefont {Honecker}, \citenamefont {Cabra}, \citenamefont {Everts}, \citenamefont {Pujol},\ and\ \citenamefont {Stauffer}}]{Honecker2011}%
  \BibitemOpen
  \bibfield  {author} {\bibinfo {author} {\bibfnamefont {A.}~\bibnamefont {Honecker}}, \bibinfo {author} {\bibfnamefont {D.~C.}\ \bibnamefont {Cabra}}, \bibinfo {author} {\bibfnamefont {H.-U.}\ \bibnamefont {Everts}}, \bibinfo {author} {\bibfnamefont {P.}~\bibnamefont {Pujol}},\ and\ \bibinfo {author} {\bibfnamefont {F.}~\bibnamefont {Stauffer}},\ }\bibfield  {title} {\bibinfo {title} {Order by disorder and phase transitions in a highly frustrated spin model on the triangular lattice},\ }\href {https://doi.org/10.1103/PhysRevB.84.224410} {\bibfield  {journal} {\bibinfo  {journal} {Phys. Rev. B}\ }\textbf {\bibinfo {volume} {84}},\ \bibinfo {pages} {224410} (\bibinfo {year} {2011})}\BibitemShut {NoStop}%
\bibitem [{\citenamefont {Rousochatzakis}\ \emph {et~al.}(2015)\citenamefont {Rousochatzakis}, \citenamefont {Reuther}, \citenamefont {Thomale}, \citenamefont {Rachel},\ and\ \citenamefont {Perkins}}]{Perkins2015}%
  \BibitemOpen
  \bibfield  {author} {\bibinfo {author} {\bibfnamefont {I.}~\bibnamefont {Rousochatzakis}}, \bibinfo {author} {\bibfnamefont {J.}~\bibnamefont {Reuther}}, \bibinfo {author} {\bibfnamefont {R.}~\bibnamefont {Thomale}}, \bibinfo {author} {\bibfnamefont {S.}~\bibnamefont {Rachel}},\ and\ \bibinfo {author} {\bibfnamefont {N.~B.}\ \bibnamefont {Perkins}},\ }\bibfield  {title} {\bibinfo {title} {Phase diagram and quantum order by disorder in the {Kitaev} ${K}_{1}\ensuremath{-}{K}_{2}$ honeycomb magnet},\ }\href {https://doi.org/10.1103/PhysRevX.5.041035} {\bibfield  {journal} {\bibinfo  {journal} {Phys. Rev. X}\ }\textbf {\bibinfo {volume} {5}},\ \bibinfo {pages} {041035} (\bibinfo {year} {2015})}\BibitemShut {NoStop}%
\bibitem [{\citenamefont {Marro}\ and\ \citenamefont {Dickman}(1999)}]{Marro_Dickman_1999}%
  \BibitemOpen
  \bibfield  {author} {\bibinfo {author} {\bibfnamefont {J.}~\bibnamefont {Marro}}\ and\ \bibinfo {author} {\bibfnamefont {R.}~\bibnamefont {Dickman}},\ }\href@noop {} {\emph {\bibinfo {title} {Nonequilibrium Phase Transitions in Lattice Models}}},\ Collection Alea-Saclay: Monographs and Texts in Statistical Physics\ (\bibinfo  {publisher} {Cambridge University Press},\ \bibinfo {year} {1999})\BibitemShut {NoStop}%
\bibitem [{\citenamefont {Avni}\ \emph {et~al.}(2023)\citenamefont {Avni}, \citenamefont {Fruchart}, \citenamefont {Martin}, \citenamefont {Seara},\ and\ \citenamefont {Vitelli}}]{VitellinonreciprocalIsing}%
  \BibitemOpen
  \bibfield  {author} {\bibinfo {author} {\bibfnamefont {Y.}~\bibnamefont {Avni}}, \bibinfo {author} {\bibfnamefont {M.}~\bibnamefont {Fruchart}}, \bibinfo {author} {\bibfnamefont {D.}~\bibnamefont {Martin}}, \bibinfo {author} {\bibfnamefont {D.}~\bibnamefont {Seara}},\ and\ \bibinfo {author} {\bibfnamefont {V.}~\bibnamefont {Vitelli}},\ }\href@noop {} {\bibinfo {title} {{The non-reciprocal Ising model}}} (\bibinfo {year} {2023}),\ \Eprint {https://arxiv.org/abs/2311.05471} {arXiv:2311.05471 [cond-mat.stat-mech]} \BibitemShut {NoStop}%
\bibitem [{\citenamefont {Seara}\ \emph {et~al.}(2023)\citenamefont {Seara}, \citenamefont {Piya},\ and\ \citenamefont {Tabatabai}}]{Seara_2023}%
  \BibitemOpen
  \bibfield  {author} {\bibinfo {author} {\bibfnamefont {D.~S.}\ \bibnamefont {Seara}}, \bibinfo {author} {\bibfnamefont {A.}~\bibnamefont {Piya}},\ and\ \bibinfo {author} {\bibfnamefont {A.~P.}\ \bibnamefont {Tabatabai}},\ }\bibfield  {title} {\bibinfo {title} {Non-reciprocal interactions spatially propagate fluctuations in a {2D Ising} model},\ }\href {https://doi.org/10.1088/1742-5468/accce7} {\bibfield  {journal} {\bibinfo  {journal} {J. Stat. Mech.}\ }\textbf {\bibinfo {volume} {2023}},\ \bibinfo {pages} {043209} (\bibinfo {year} {2023})}\BibitemShut {NoStop}%
\bibitem [{\citenamefont {Tobochnik}\ and\ \citenamefont {Chester}(1979)}]{Tobochnik1979}%
  \BibitemOpen
  \bibfield  {author} {\bibinfo {author} {\bibfnamefont {J.}~\bibnamefont {Tobochnik}}\ and\ \bibinfo {author} {\bibfnamefont {G.~V.}\ \bibnamefont {Chester}},\ }\bibfield  {title} {\bibinfo {title} {Monte {Carlo} study of the planar spin model},\ }\href {https://doi.org/10.1103/PhysRevB.20.3761} {\bibfield  {journal} {\bibinfo  {journal} {Phys. Rev. B}\ }\textbf {\bibinfo {volume} {20}},\ \bibinfo {pages} {3761} (\bibinfo {year} {1979})}\BibitemShut {NoStop}%
\bibitem [{\citenamefont {Archambault}\ \emph {et~al.}(1997)\citenamefont {Archambault}, \citenamefont {Bramwell},\ and\ \citenamefont {Holdsworth}}]{Archambault1997}%
  \BibitemOpen
  \bibfield  {author} {\bibinfo {author} {\bibfnamefont {P.}~\bibnamefont {Archambault}}, \bibinfo {author} {\bibfnamefont {S.~T.}\ \bibnamefont {Bramwell}},\ and\ \bibinfo {author} {\bibfnamefont {P.~C.~W.}\ \bibnamefont {Holdsworth}},\ }\bibfield  {title} {\bibinfo {title} {Magnetic fluctuations in a finite two-dimensional model},\ }\href {https://doi.org/10.1088/0305-4470/30/24/005} {\bibfield  {journal} {\bibinfo  {journal} {J. Phys. A: Math. Gen.}\ }\textbf {\bibinfo {volume} {30}},\ \bibinfo {pages} {8363} (\bibinfo {year} {1997})}\BibitemShut {NoStop}%
\bibitem [{\citenamefont {Cooper}\ \emph {et~al.}(1982)\citenamefont {Cooper}, \citenamefont {Freedman},\ and\ \citenamefont {Preston}}]{cooper1982}%
  \BibitemOpen
  \bibfield  {author} {\bibinfo {author} {\bibfnamefont {F.}~\bibnamefont {Cooper}}, \bibinfo {author} {\bibfnamefont {B.}~\bibnamefont {Freedman}},\ and\ \bibinfo {author} {\bibfnamefont {D.}~\bibnamefont {Preston}},\ }\bibfield  {title} {\bibinfo {title} {Solving \(\phi_{1,2}^4\) field theory with {Monte Carlo}},\ }\href {https://doi.org/https://doi.org/10.1016/0550-3213(82)90240-1} {\bibfield  {journal} {\bibinfo  {journal} {Nucl. Phys. B}\ }\textbf {\bibinfo {volume} {210}},\ \bibinfo {pages} {210} (\bibinfo {year} {1982})}\BibitemShut {NoStop}%
\bibitem [{\citenamefont {Amit}\ and\ \citenamefont {Martin-Mayor}(2005)}]{amit2005}%
  \BibitemOpen
  \bibfield  {author} {\bibinfo {author} {\bibfnamefont {D.~J.}\ \bibnamefont {Amit}}\ and\ \bibinfo {author} {\bibfnamefont {V.}~\bibnamefont {Martin-Mayor}},\ }\href {https://doi.org/10.1142/5715} {\emph {\bibinfo {title} {Field Theory, the Renormalization Group, and Critical Phenomena}}},\ \bibinfo {edition} {3rd}\ ed.\ (\bibinfo  {publisher} {World Scientific},\ \bibinfo {year} {2005})\BibitemShut {NoStop}%
\bibitem [{\citenamefont {Martynec}\ \emph {et~al.}(2020)\citenamefont {Martynec}, \citenamefont {Klapp},\ and\ \citenamefont {Loos}}]{Martynec2020}%
  \BibitemOpen
  \bibfield  {author} {\bibinfo {author} {\bibfnamefont {T.}~\bibnamefont {Martynec}}, \bibinfo {author} {\bibfnamefont {S.~H.~L.}\ \bibnamefont {Klapp}},\ and\ \bibinfo {author} {\bibfnamefont {S.~A.~M.}\ \bibnamefont {Loos}},\ }\bibfield  {title} {\bibinfo {title} {Entropy production at criticality in a nonequilibrium {Potts} model},\ }\href {https://doi.org/10.1088/1367-2630/abb5f0} {\bibfield  {journal} {\bibinfo  {journal} {New J. Phys.}\ }\textbf {\bibinfo {volume} {22}},\ \bibinfo {pages} {093069} (\bibinfo {year} {2020})}\BibitemShut {NoStop}%
\bibitem [{\citenamefont {Noa}\ \emph {et~al.}(2019)\citenamefont {Noa}, \citenamefont {Harunari}, \citenamefont {de~Oliveira},\ and\ \citenamefont {Fiore}}]{noa2019}%
  \BibitemOpen
  \bibfield  {author} {\bibinfo {author} {\bibfnamefont {C.~E.~F.}\ \bibnamefont {Noa}}, \bibinfo {author} {\bibfnamefont {P.~E.}\ \bibnamefont {Harunari}}, \bibinfo {author} {\bibfnamefont {M.~J.}\ \bibnamefont {de~Oliveira}},\ and\ \bibinfo {author} {\bibfnamefont {C.~E.}\ \bibnamefont {Fiore}},\ }\bibfield  {title} {\bibinfo {title} {Entropy production as a tool for characterizing nonequilibrium phase transitions},\ }\href {https://doi.org/10.1103/PhysRevE.100.012104} {\bibfield  {journal} {\bibinfo  {journal} {Phys. Rev. E}\ }\textbf {\bibinfo {volume} {100}},\ \bibinfo {pages} {012104} (\bibinfo {year} {2019})}\BibitemShut {NoStop}%
\bibitem [{\citenamefont {Tom\'e}\ and\ \citenamefont {de~Oliveira}(2012)}]{Tania2012}%
  \BibitemOpen
  \bibfield  {author} {\bibinfo {author} {\bibfnamefont {T.}~\bibnamefont {Tom\'e}}\ and\ \bibinfo {author} {\bibfnamefont {M.~J.}\ \bibnamefont {de~Oliveira}},\ }\bibfield  {title} {\bibinfo {title} {Entropy production in nonequilibrium systems at stationary states},\ }\href {https://doi.org/10.1103/PhysRevLett.108.020601} {\bibfield  {journal} {\bibinfo  {journal} {Phys. Rev. Lett.}\ }\textbf {\bibinfo {volume} {108}},\ \bibinfo {pages} {020601} (\bibinfo {year} {2012})}\BibitemShut {NoStop}%
\bibitem [{\citenamefont {Schnakenberg}(1976)}]{Schnakenberg1976}%
  \BibitemOpen
  \bibfield  {author} {\bibinfo {author} {\bibfnamefont {J.}~\bibnamefont {Schnakenberg}},\ }\bibfield  {title} {\bibinfo {title} {Network theory of microscopic and macroscopic behavior of master equation systems},\ }\href {https://doi.org/10.1103/RevModPhys.48.571} {\bibfield  {journal} {\bibinfo  {journal} {Rev. Mod. Phys.}\ }\textbf {\bibinfo {volume} {48}},\ \bibinfo {pages} {571} (\bibinfo {year} {1976})}\BibitemShut {NoStop}%
\bibitem [{\citenamefont {Li}\ \emph {et~al.}(2020)\citenamefont {Li}, \citenamefont {Yang}, \citenamefont {Xie}, \citenamefont {Tu}, \citenamefont {Liao},\ and\ \citenamefont {Xiang}}]{li2020}%
  \BibitemOpen
  \bibfield  {author} {\bibinfo {author} {\bibfnamefont {Z.-Q.}\ \bibnamefont {Li}}, \bibinfo {author} {\bibfnamefont {L.-P.}\ \bibnamefont {Yang}}, \bibinfo {author} {\bibfnamefont {Z.~Y.}\ \bibnamefont {Xie}}, \bibinfo {author} {\bibfnamefont {H.-H.}\ \bibnamefont {Tu}}, \bibinfo {author} {\bibfnamefont {H.-J.}\ \bibnamefont {Liao}},\ and\ \bibinfo {author} {\bibfnamefont {T.}~\bibnamefont {Xiang}},\ }\bibfield  {title} {\bibinfo {title} {Critical properties of the two-dimensional $q$-state clock model},\ }\href {https://doi.org/10.1103/PhysRevE.101.060105} {\bibfield  {journal} {\bibinfo  {journal} {Phys. Rev. E}\ }\textbf {\bibinfo {volume} {101}},\ \bibinfo {pages} {060105} (\bibinfo {year} {2020})}\BibitemShut {NoStop}%
\bibitem [{\citenamefont {Baxter}(2007)}]{baxter2007exactly}%
  \BibitemOpen
  \bibfield  {author} {\bibinfo {author} {\bibfnamefont {R.}~\bibnamefont {Baxter}},\ }\href@noop {} {\emph {\bibinfo {title} {Exactly Solved Models in Statistical Mechanics}}},\ Dover books on physics\ (\bibinfo  {publisher} {Dover Publications},\ \bibinfo {year} {2007})\BibitemShut {NoStop}%
\bibitem [{\citenamefont {Surungan}\ \emph {et~al.}(2019)\citenamefont {Surungan}, \citenamefont {Masuda}, \citenamefont {Komura},\ and\ \citenamefont {Okabe}}]{Surungan_2019}%
  \BibitemOpen
  \bibfield  {author} {\bibinfo {author} {\bibfnamefont {T.}~\bibnamefont {Surungan}}, \bibinfo {author} {\bibfnamefont {S.}~\bibnamefont {Masuda}}, \bibinfo {author} {\bibfnamefont {Y.}~\bibnamefont {Komura}},\ and\ \bibinfo {author} {\bibfnamefont {Y.}~\bibnamefont {Okabe}},\ }\bibfield  {title} {\bibinfo {title} {{Berezinskii–Kosterlitz–Thouless} transition on regular and {Villain} types of q-state clock models},\ }\href {https://doi.org/10.1088/1751-8121/ab226d} {\bibfield  {journal} {\bibinfo  {journal} {J. Phys. A: Math. Theor.}\ }\textbf {\bibinfo {volume} {52}},\ \bibinfo {pages} {275002} (\bibinfo {year} {2019})}\BibitemShut {NoStop}%
\bibitem [{\citenamefont {Tuan}\ \emph {et~al.}(2022)\citenamefont {Tuan}, \citenamefont {Long}, \citenamefont {Nui}, \citenamefont {Minh}, \citenamefont {Trung~Kien},\ and\ \citenamefont {Viet}}]{tuan2022}%
  \BibitemOpen
  \bibfield  {author} {\bibinfo {author} {\bibfnamefont {L.~M.}\ \bibnamefont {Tuan}}, \bibinfo {author} {\bibfnamefont {T.~T.}\ \bibnamefont {Long}}, \bibinfo {author} {\bibfnamefont {D.~X.}\ \bibnamefont {Nui}}, \bibinfo {author} {\bibfnamefont {P.~T.}\ \bibnamefont {Minh}}, \bibinfo {author} {\bibfnamefont {N.~D.}\ \bibnamefont {Trung~Kien}},\ and\ \bibinfo {author} {\bibfnamefont {D.~X.}\ \bibnamefont {Viet}},\ }\bibfield  {title} {\bibinfo {title} {Binder ratio in the two-dimensional $q$-state clock model},\ }\href {https://doi.org/10.1103/PhysRevE.106.034138} {\bibfield  {journal} {\bibinfo  {journal} {Phys. Rev. E}\ }\textbf {\bibinfo {volume} {106}},\ \bibinfo {pages} {034138} (\bibinfo {year} {2022})}\BibitemShut {NoStop}%
\bibitem [{\citenamefont {Drouin-Touchette}(2022)}]{Drouin-Touchette2022}%
  \BibitemOpen
  \bibfield  {author} {\bibinfo {author} {\bibfnamefont {V.}~\bibnamefont {Drouin-Touchette}},\ }\href {https://arxiv.org/abs/2207.13748} {\bibinfo {title} {The {Kosterlitz-Thouless} phase transition: an introduction for the intrepid student}} (\bibinfo {year} {2022}),\ \Eprint {https://arxiv.org/abs/2207.13748} {arXiv:2207.13748 [cond-mat.stat-mech]} \BibitemShut {NoStop}%
\bibitem [{\citenamefont {Hasenbusch}(2005)}]{Hasenbusch_2005}%
  \BibitemOpen
  \bibfield  {author} {\bibinfo {author} {\bibfnamefont {M.}~\bibnamefont {Hasenbusch}},\ }\bibfield  {title} {\bibinfo {title} {The two-dimensional {XY} model at the transition temperature: a high-precision {Monte Carlo} study},\ }\href {https://doi.org/10.1088/0305-4470/38/26/003} {\bibfield  {journal} {\bibinfo  {journal} {J. Phys. A: Math. Gen.}\ }\textbf {\bibinfo {volume} {38}},\ \bibinfo {pages} {5869} (\bibinfo {year} {2005})}\BibitemShut {NoStop}%
\bibitem [{\citenamefont {Loffredo}\ \emph {et~al.}(2023)\citenamefont {Loffredo}, \citenamefont {Venturelli},\ and\ \citenamefont {Giardina}}]{Loffredo2023}%
  \BibitemOpen
  \bibfield  {author} {\bibinfo {author} {\bibfnamefont {E.}~\bibnamefont {Loffredo}}, \bibinfo {author} {\bibfnamefont {D.}~\bibnamefont {Venturelli}},\ and\ \bibinfo {author} {\bibfnamefont {I.}~\bibnamefont {Giardina}},\ }\bibfield  {title} {\bibinfo {title} {Collective response to local perturbations: how to evade threats without losing coherence},\ }\href {https://doi.org/10.1088/1478-3975/acc5cc} {\bibfield  {journal} {\bibinfo  {journal} {Phys. Biol.}\ }\textbf {\bibinfo {volume} {20}},\ \bibinfo {pages} {035003} (\bibinfo {year} {2023})}\BibitemShut {NoStop}%
\bibitem [{\citenamefont {Hohenberg}\ and\ \citenamefont {Halperin}(1977)}]{halperin}%
  \BibitemOpen
  \bibfield  {author} {\bibinfo {author} {\bibfnamefont {P.~C.}\ \bibnamefont {Hohenberg}}\ and\ \bibinfo {author} {\bibfnamefont {B.~I.}\ \bibnamefont {Halperin}},\ }\bibfield  {title} {\bibinfo {title} {Theory of dynamic critical phenomena},\ }\href {https://doi.org/10.1103/RevModPhys.49.435} {\bibfield  {journal} {\bibinfo  {journal} {Rev. Mod. Phys.}\ }\textbf {\bibinfo {volume} {49}},\ \bibinfo {pages} {435} (\bibinfo {year} {1977})}\BibitemShut {NoStop}%
\bibitem [{\citenamefont {Marchetti}\ \emph {et~al.}(2013)\citenamefont {Marchetti}, \citenamefont {Joanny}, \citenamefont {Ramaswamy}, \citenamefont {Liverpool}, \citenamefont {Prost}, \citenamefont {Rao},\ and\ \citenamefont {Simha}}]{marchetti_review}%
  \BibitemOpen
  \bibfield  {author} {\bibinfo {author} {\bibfnamefont {M.~C.}\ \bibnamefont {Marchetti}}, \bibinfo {author} {\bibfnamefont {J.~F.}\ \bibnamefont {Joanny}}, \bibinfo {author} {\bibfnamefont {S.}~\bibnamefont {Ramaswamy}}, \bibinfo {author} {\bibfnamefont {T.~B.}\ \bibnamefont {Liverpool}}, \bibinfo {author} {\bibfnamefont {J.}~\bibnamefont {Prost}}, \bibinfo {author} {\bibfnamefont {M.}~\bibnamefont {Rao}},\ and\ \bibinfo {author} {\bibfnamefont {R.~A.}\ \bibnamefont {Simha}},\ }\bibfield  {title} {\bibinfo {title} {Hydrodynamics of soft active matter},\ }\href {https://doi.org/10.1103/RevModPhys.85.1143} {\bibfield  {journal} {\bibinfo  {journal} {Rev. Mod. Phys.}\ }\textbf {\bibinfo {volume} {85}},\ \bibinfo {pages} {1143} (\bibinfo {year} {2013})}\BibitemShut {NoStop}%
\bibitem [{\citenamefont {Saha}\ and\ \citenamefont {Mohanty}(2024)}]{saha2024}%
  \BibitemOpen
  \bibfield  {author} {\bibinfo {author} {\bibfnamefont {S.~K.}\ \bibnamefont {Saha}}\ and\ \bibinfo {author} {\bibfnamefont {P.~K.}\ \bibnamefont {Mohanty}},\ }\href {https://arxiv.org/abs/2412.19664} {\bibinfo {title} {Non-reciprocal interactions preserve the universality class of {Potts} model}} (\bibinfo {year} {2024}),\ \Eprint {https://arxiv.org/abs/2412.19664} {arXiv:2412.19664 [cond-mat.stat-mech]} \BibitemShut {NoStop}%
\bibitem [{\citenamefont {Newman}\ and\ \citenamefont {Barkema}(1999)}]{Newman-Barkema}%
  \BibitemOpen
  \bibfield  {author} {\bibinfo {author} {\bibfnamefont {M.~E.~J.}\ \bibnamefont {Newman}}\ and\ \bibinfo {author} {\bibfnamefont {G.~T.}\ \bibnamefont {Barkema}},\ }\href {https://doi.org/10.1093/oso/9780198517962.001.0001} {\emph {\bibinfo {title} {Monte Carlo Methods in Statistical Physics}}}\ (\bibinfo  {publisher} {Oxford University Press},\ \bibinfo {year} {1999})\BibitemShut {NoStop}%
\bibitem [{\citenamefont {Metropolis}\ \emph {et~al.}(1953)\citenamefont {Metropolis}, \citenamefont {Rosenbluth}, \citenamefont {Rosenbluth}, \citenamefont {Teller},\ and\ \citenamefont {Teller}}]{Metropolis1953}%
  \BibitemOpen
  \bibfield  {author} {\bibinfo {author} {\bibfnamefont {N.}~\bibnamefont {Metropolis}}, \bibinfo {author} {\bibfnamefont {A.~W.}\ \bibnamefont {Rosenbluth}}, \bibinfo {author} {\bibfnamefont {M.~N.}\ \bibnamefont {Rosenbluth}}, \bibinfo {author} {\bibfnamefont {A.~H.}\ \bibnamefont {Teller}},\ and\ \bibinfo {author} {\bibfnamefont {E.}~\bibnamefont {Teller}},\ }\bibfield  {title} {\bibinfo {title} {Equation of state calculations by fast computing machines},\ }\href {https://doi.org/10.1063/1.1699114} {\bibfield  {journal} {\bibinfo  {journal} {J. Chem. Phys.}\ }\textbf {\bibinfo {volume} {21}},\ \bibinfo {pages} {1087} (\bibinfo {year} {1953})}\BibitemShut {NoStop}%
\bibitem [{\citenamefont {Berche}\ \emph {et~al.}(2003)\citenamefont {Berche}, \citenamefont {Fari{\~{n}}as-S{\'a}nchez}, \citenamefont {Holovatch},\ and\ \citenamefont {Paredes}}]{Berche2003}%
  \BibitemOpen
  \bibfield  {author} {\bibinfo {author} {\bibfnamefont {B.}~\bibnamefont {Berche}}, \bibinfo {author} {\bibfnamefont {A.~I.}\ \bibnamefont {Fari{\~{n}}as-S{\'a}nchez}}, \bibinfo {author} {\bibfnamefont {Y.}~\bibnamefont {Holovatch}},\ and\ \bibinfo {author} {\bibfnamefont {R.}~\bibnamefont {Paredes}},\ }\bibfield  {title} {\bibinfo {title} {Influence of quenched dilution on the quasi-long-range ordered phase of the 2d {XY} model},\ }\href {https://doi.org/10.1140/epjb/e2003-00310-5} {\bibfield  {journal} {\bibinfo  {journal} {Eur. Phys. J. B}\ }\textbf {\bibinfo {volume} {36}},\ \bibinfo {pages} {91} (\bibinfo {year} {2003})}\BibitemShut {NoStop}%
\bibitem [{\citenamefont {Okabe}\ and\ \citenamefont {Surungan}(2005)}]{Okabe2005}%
  \BibitemOpen
  \bibfield  {author} {\bibinfo {author} {\bibfnamefont {Y.}~\bibnamefont {Okabe}}\ and\ \bibinfo {author} {\bibfnamefont {T.}~\bibnamefont {Surungan}},\ }\bibfield  {title} {\bibinfo {title} {Phase transition of two-dimensional diluted {XY} and clock models},\ }\href {https://doi.org/10.1143/PTPS.157.132} {\bibfield  {journal} {\bibinfo  {journal} {Progr. Theor. Phys. Supp.}\ }\textbf {\bibinfo {volume} {157}},\ \bibinfo {pages} {132} (\bibinfo {year} {2005})}\BibitemShut {NoStop}%
\bibitem [{\citenamefont {Pathria}\ and\ \citenamefont {Beale}(2022)}]{PATHRIA2022487}%
  \BibitemOpen
  \bibfield  {author} {\bibinfo {author} {\bibfnamefont {R.}~\bibnamefont {Pathria}}\ and\ \bibinfo {author} {\bibfnamefont {P.~D.}\ \bibnamefont {Beale}},\ }\bibfield  {title} {\bibinfo {title} {13 - {P}hase transitions: exact (or almost exact) results for various models},\ }in\ \href {https://doi.org/https://doi.org/10.1016/B978-0-08-102692-2.00022-3} {\emph {\bibinfo {booktitle} {Statistical Mechanics}}},\ \bibinfo {editor} {edited by\ \bibinfo {editor} {\bibfnamefont {R.}~\bibnamefont {Pathria}}\ and\ \bibinfo {editor} {\bibfnamefont {P.~D.}\ \bibnamefont {Beale}}}\ (\bibinfo  {publisher} {Academic Press},\ \bibinfo {year} {2022})\ \bibinfo {edition} {4th}\ ed.,\ pp.\ \bibinfo {pages} {487--554}\BibitemShut {NoStop}%
\end{thebibliography}%

\appendix

\renewcommand{\thefigure}{A\arabic{figure}} 
\setcounter{figure}{0}

\section{Details of the Monte Carlo simulation}\label{sec:simulations}

We report here the details of the 
MC
simulations. Once the model, system size, and temperature are set, the MC dynamics is initiated in two possible ways (which turn out to lead to identical results): 
(i) In a disordered configuration (with vanishing magnetization), but with a magnetic field selecting one of the four directions in which the system might order. As the simulation proceeds, the strength of the magnetic field is made to vanish. (ii)
In an ordered configuration, with all spins aligned along one of the directions of possible order.
%
%
This is to prevent the system from getting trapped in a metastable state with two or more domain walls (at low temperatures, when 
we expect a LRO phase with spins aligned except for fluctuations). 
%
%
After allowing for a thermalization time, estimated by observing when the quantities of interest at temperatures close to the phase transition have reached
a steady value, the sampling begins. 
%
%
Between each sampling, we perform a number of sweeps proportional to $L^2$, due to the scaling of the autocorrelation time $\tau$, which scales as $\tau \sim \xi^2$ for local update algorithms. With the sampled variables, such as the magnetization and the energy, we calculate the observables described in Sec.~\ref{sec:observables}, and estimate the corresponding statistical error by using a bootstrap method that accounts for the residual correlation between data points via a blocking technique~\cite{Newman-Barkema}.

\section{Scaling of the second-moment correlation length}
\label{sec:corr-length-scaling}

\begin{figure}
    \centering
        \includegraphics[width = 0.98\linewidth]{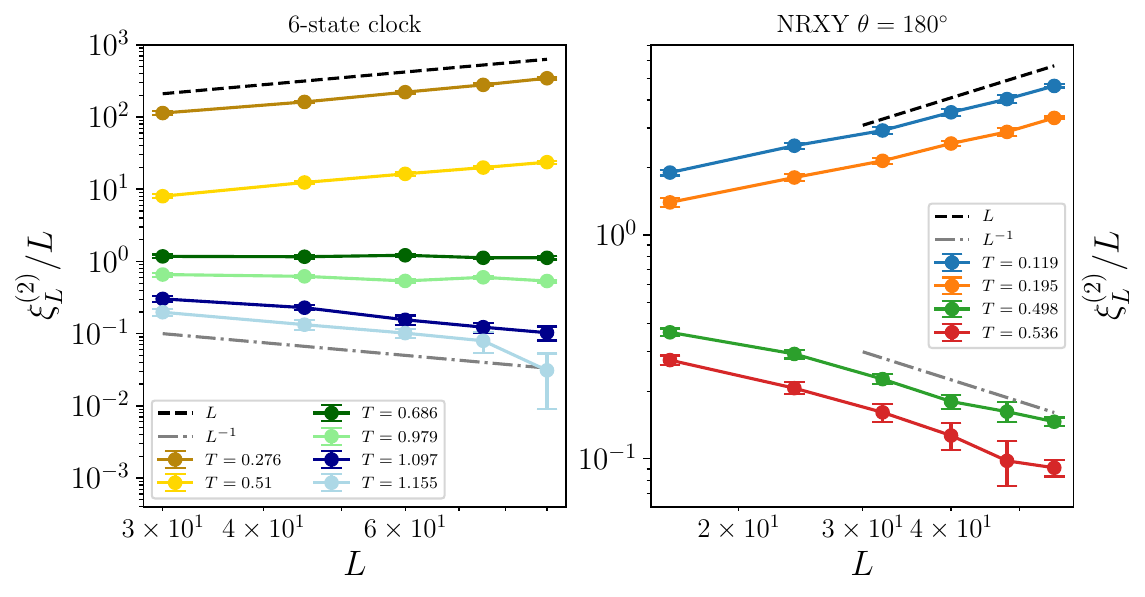}
        \put(-238,110){(a)}
        \put(-119,110){(b)}
        \captionsetup{justification=RaggedRight} 
        \caption{\small \rev{Dependence of the second-moment correlation length $\xi_L^{(2)}$ on the system size $L$ for selected values of the temperature. (a) In the 6-state clock model we observe, depending on the temperature $T$, three different scalings of $\xi_L^{(2)}/L$ as a function of $L$, namely as $L$ (dashed line, to guide the eyes), constant, and $1/L$ (dash-dotted line), corresponding to phases with LRO, QLRO and DO, respectively --- see Eq.~\eqref{eq:corrlengthScaling}. 
    In Fig.~\ref{fig:benchmarks}(g) the same quantity is plotted as a function of the temperature, for different system sizes. (b)~In the NRXY model with $\theta = 180^{\circ}$ we observe that, depending on the temperature, $\xi_L^{(2)}/L$ scales as $L$ (dashed line) and $1/L$ (dash-dotted line) upon increasing $L$, in the LRO and DO phase, respectively. In Fig.~\ref{fig:nrec-m}(e) the same quantity is plotted as a function of the temperature, for different system sizes.}}
        \label{fig:scalingCorrelationLength}

\end{figure}

\rev{Here we present the scaling behavior of the second-moment correlation length $\xi_L^{(2)}$ defined in Eq.~\eqref{eq:corrLength} as a function of the system size $L$, for the two significant cases of the 6-state clock model and and the NRXY model with $\theta = 180^{\circ}$. As discussed in Sec.~\ref{sec:observables}, $\xi_L^{(2)}$ is expected to scale as 
Eq.~\eqref{eq:corrlengthScaling}
within the three possible phases encountered in this work, i.e., LRO, QLRO, and DO.}

\rev{
In particular, the 6-state clock model exhibits all these three phases within different temperature ranges, as illustrated in Fig.~\ref{fig:benchmarks}(f)--(j). In Fig.~\ref{fig:scalingCorrelationLength}(a), we report $\xi_L^{(2)}/L$ as a function of the system size $L$ for various temperatures. Specifically, upon increasing $L$,
$\xi_L^{(2)}/L$ grows as $L$  (dashed line) 
at low temperatures (LRO phase), remains constant in the intermediate temperature regime (QLRO phase), while it decreases as $1/L$ (dash-dotted line) at high temperatures (DO phase).
}

\rev{
In Fig.~\ref{fig:scalingCorrelationLength}(b), instead, we present the corresponding plot for the NRXY model at $\theta = 180^{\circ}$.
Here, we highlight the two observed scaling behaviors: $\xi_L^{(2)}/L\sim L$ (dashed line) in the LRO phase, and $\xi_L^{(2)}/L\sim 1/L$ (dash-dotted line) in the DO phase.
}

\begin{figure}
    \centering
        \includegraphics[width = 0.98\linewidth]{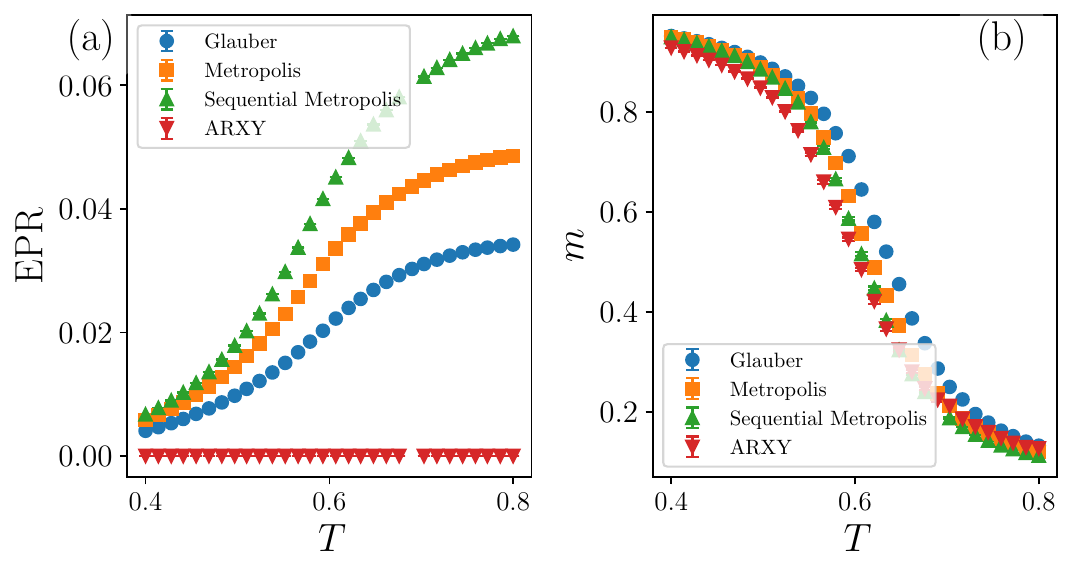}
         \captionsetup{justification=RaggedRight} 
    \caption{\small
    Temperature dependence of (a) the entropy production rate 
    and (b) the magnetization $m$ for various update protocols in the MC dynamics (see the legend) of the NRXY model with $\theta = 225^{\circ}$. The dependence of these quantities on the protocols demonstrates that the latter lead to distinct steady states.
    In particular, it turns out that the EPR in panel (a) is the smallest for the Glauber dynamics (which has the lowest acceptance ratio probability), and the largest when implemented with sequential sweeps, while it takes intermediate values for the Metropolis dynamics. Clearly, the EPR vanishes for the equilibrium ARXY model. The magnetization $m$ in panel (b) (as well as other observables) varies quantitatively, depending on the update protocol. However,
    these differences do not affect qualitatively the resulting phase diagram of the model.
    } \label{fig:EPR-protocols}

\end{figure}

\section{NRXY model with different update protocols}\label{sec:protocols}

In a non-equilibrium model such as the NRXY
model, the choice of the MC update protocol is generically expected to affect the stationary state reached by the dynamics and, consequently, the quantities measured in such a state.
In principle, there are infinitely many ways to implement MC dynamics --- for instance, one
could choose the Glauber transition rate or the Metropolis one~\cite{Metropolis1953}, and updates can be performed either sequentially across the lattice or by randomly selecting sites for updates. 
In general,
these choices can lead to different entropy production rates,
see Eq.~\eqref{eq:epr}. 
To illustrate this point, in Fig.~\ref{fig:EPR-protocols}(a) we present simulations of the NRXY model at various temperatures for $\theta = 225^{\circ}$,
and for a few selected choices of update protocols. 
Here we observe that the EPR increases when employing Glauber-like dynamics, which is characterized by lower acceptance rates, especially at high temperatures, compared to Metropolis dynamics. The EPR increases even further when updates are performed sequentially --- this behavior can be attributed to the faster propagation of domain wall defects (see
the video included in the Supplementary Material),
%
%
which enhances the speed of irreversible dynamics.

However, beyond the quantitative differences, Fig.~\ref{fig:EPR-protocols}(a) shows that the qualitative trend of the EPR is quite robust upon changing the update protocols. 
(In passing, we recall that for the SRXY and ARXY models investigated in this work, the EPR vanishes identically due to their equilibrium nature.)  
Similarly, most other static (i.e., equal-time) observables that we measured in this work for the NRXY model turn out to exhibit only minor quantitative (but not qualitative) differences
depending on the update rule.
%
%
For example, in Fig.~\ref{fig:EPR-protocols}(b) we report the magnetization $m$ obtained using various  update protocols for simulating the NRXY model, showing their overall qualitative consistency. In particular, these values of $m$ are also very similar to
those of the ARXY model (discussed in Sec.~\ref{sec:arxy-results}) for the same value of $\theta$, which (being an equilibrium model) reaches the same equilibrium state regardless of the chosen update protocol.

\begin{figure}
    \centering
        \includegraphics[width = 0.94\linewidth]{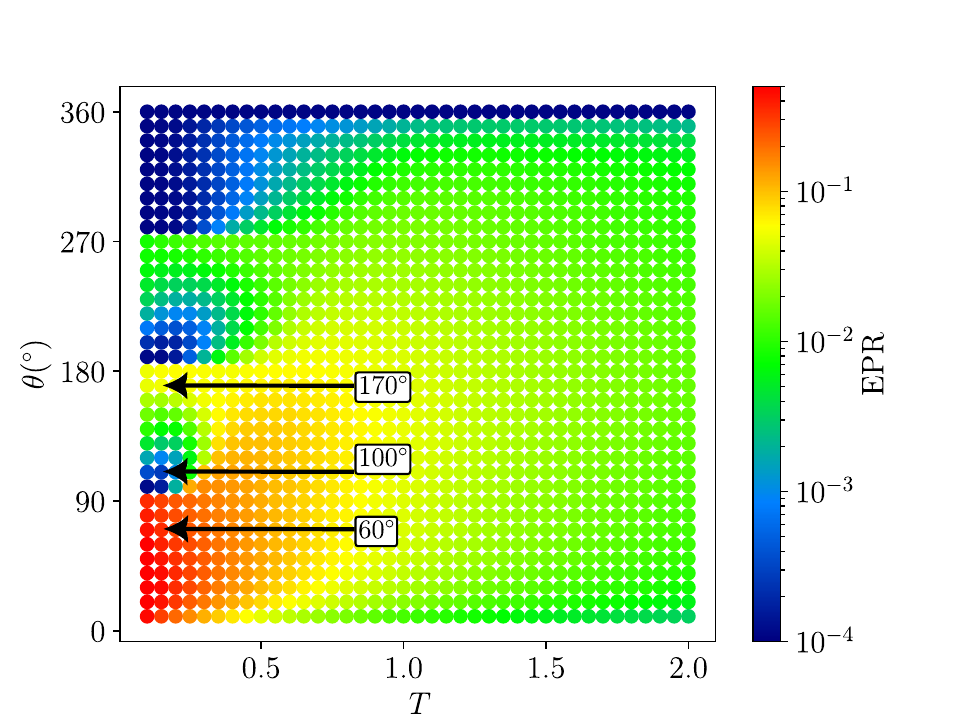}
                \put(-232,147){(a)}
        \!\!\!\!\!\!\!\!\!
        \includegraphics[width = 0.94\linewidth]{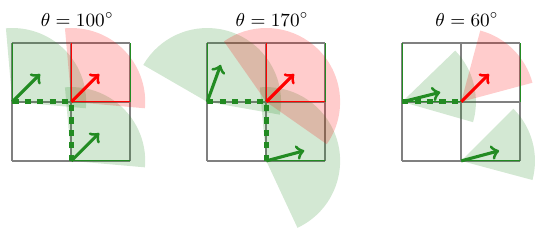}
    \put(-227,20){(b)}
    \put(-144,20){(c)}
    \put(-61,20){(d)}
         \captionsetup{justification=RaggedRight} 
    \caption{\small
    (a) Phase diagram of the EPR for the non-reciprocal XY model, obtained using Glauber updates with random site selection for a system of size $L=100$. The EPR vanishes at $\theta = 360^{\circ}$ (corresponding to the standard equilibrium XY model) and within the lobes with LRO, particularly in their lower parts, where the EUR is almost $360^{\circ}$.
    (b) Typical low-temperature configuration for $\theta = 100^{\circ}$. The green spins, which make, for the update of the red spin, the difference between the NRXY model and the ARXY model, are almost aligned. As a result, they approximately act as a constant magnetic field, 
    which allows one to write the  variation
    of the selfish energy 
    during an update as the 
    variation
    of a global energy, see Eq.~\eqref{eq:delta-global}. This renders the model effectively an equilibrium one.
    (c) Typical low-temperature configuration for $\theta = 170^{\circ}$. Since the smaller EUR allows for a greater range of motion for the spins even within the LRO phase, the constant magnetic field approximation is no longer applicable, and the system begins to exhibit its non-reciprocal nature. (d) Typical configuration for $\theta = 60^{\circ}$. Since, for this value of the vision cone $\theta$, there is no LRO even at low temperatures, the approximation in terms of the presence of a magnetic field is no longer viable and, as a consequence, the EPR takes its largest values within this region of the phase diagram. 
 } \label{fig:nrec-epr}
 \end{figure}

\section{Analysis of the EPR in the NRXY model}\label{sec:epr}

As we discussed in Sec.~\ref{subsec:nrec}, the stationary state reached by the 
NRXY model is a non-equilibrium one, characterized by a non-vanishing EPR. As shown in App.~\ref{sec:protocols}, the actual value of the EPR depends on the chosen update protocol, while  its overall behavior does not. Here, we investigate the dependence of the EPR (see Eq.~\eqref{eq:epr}) on the value of the VC amplitude $\theta$.

\begin{figure*}
    \centering
        \includegraphics[width=0.99\linewidth]{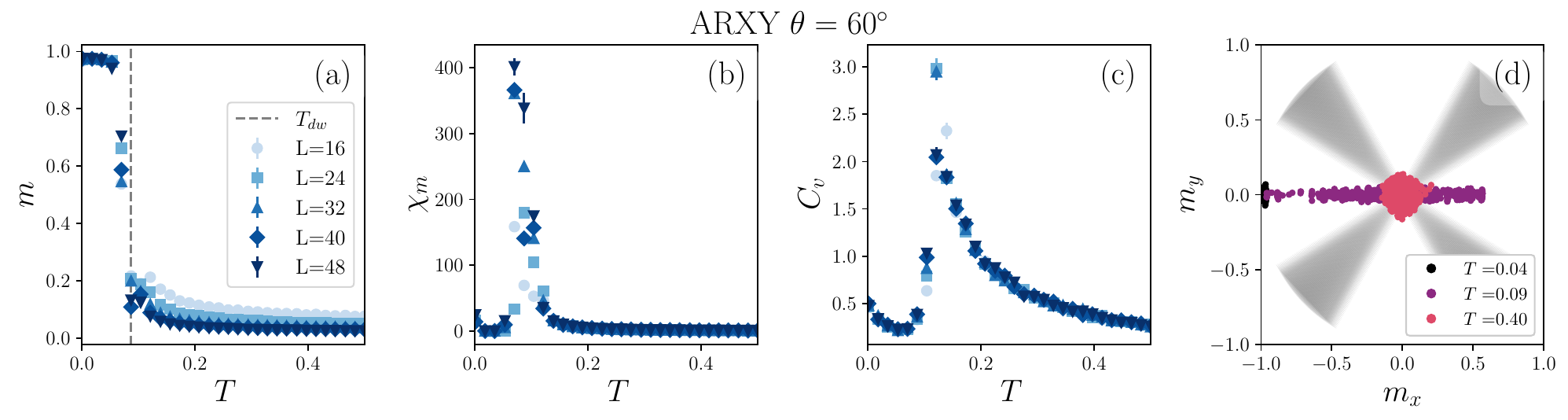}
         \captionsetup{justification=RaggedRight} 
    \caption{\small MC results for the 
    ARXY
    model with VC angle $\theta=60^{\circ}$. The magnetization in panel (a) becomes nonzero at a finite temperature, despite the system being effectively one-dimensional. We estimate the temperature (vertical dashed line) at which this occurs in
    App.~\ref{sec:M2-60}.
    Correspondingly, panels (b) and (c) show that the magnetic susceptibility and specific heat exhibit non-analytic behavior at this transition. In panel (d), the vectorial magnetization illustrates alignment along the horizontal direction at low temperatures, transitioning to alignment with no preferred orientation along the horizontal directions, and eventually becoming disordered as the temperature increases. For a detailed discussion, we refer to App.~\ref{sec:M2-60}.
    }
    \label{fig:M2-theta60}
\end{figure*}

Figure \ref{fig:nrec-epr}(a) shows the EPR as a function of the temperature $T$ and of $\theta$, determined from the MC simulation of the model using Glauber dynamics with random updates and with $L=100$.  
Interestingly enough, we observe that, generically, the EPR is rather small and it either vanishes or approaches zero within the lobes in the phase diagram (see Fig.~\ref{fig:nrec-m}(a))
corresponding to LRO, particularly in their lower parts. (Note that, as expected, the EPR vanishes identically for $\theta = 360^{\circ}$ because, correspondingly, the NRXY model becomes the equilibrium XY model.) 
In order to explain this fact, we observe that the spin configurations within these lobes are characterized by almost perfectly aligned spins (indeed, the corresponding large EUR implies limited freedom of rotation at low temperatures). 
As shown in Fig.~\ref{fig:nrec-epr}(b) for the choice $\theta = 100^{\circ}$, the bonds activated by the green spins --- which, in the NRXY model, are irrelevant for updating the red spin (hence the dashed green lines) --- mark the difference between the selfish ($E_{i}^{\text{NR}}$, see Eq.~\eqref{eq:selfish}) and asymmetric reciprocal ($E^{\text{AR}}$, see Eq.~\eqref{eq:arxy-energy}) energy functionals. 
%
%
Accordingly, these green spins act like a magnetic field $\mathbf{h}$ aligned with the overall magnetization of the model ($\mathbf{h} = h \hat{\mathbf m}$, with $\hat{\mathbf m} = \mathbf{m}/|\mathbf{m}|$), whose fluctuations diminish as the EUR extends (i.e., approaching the lower parts of the lobes). 
As a result, the difference $\Delta_i E^\text{NR}_i$ in the selfish energy $E^\text{NR}_i$ due to an update in which spin $i$ is 
%
%
updated from being $\mathbf{s}_i$ to $\mathbf{s}'_i$
can be written as the variation of a global functional: 
\begin{equation}\label{eq:delta-global}
    \Delta_i E^\text{NR}_i = \Delta_i E^\text{AR}+ \mathbf{h} \cdot (\mathbf{s}'_i - \mathbf{s}_i) = \Delta_i (E^\text{AR} -  \mathcal{E}^\text{mag}),
\end{equation}
%
%
where $\mathcal{E}^\text{mag}=-\mathbf{h} \cdot \mathbf{m}$. 
Correspondingly, the model behaves effectively as an equilibrium one, with vanishing EPR. 
This behavior is fundamentally different from other non-reciprocal two-species lattice models~\cite{VitellinonreciprocalIsing}, where even a minimal non-reciprocity at low temperatures leads to phenomena that are markedly different from those observed in the corresponding reciprocal versions.

Upon increasing the value of $\theta$ within each lobe, i.e., upon moving from its bottom part towards its upper part, the freedom of movement of the green spins in Fig.~\ref{fig:nrec-epr}(b) 
increases due to a decrease of the EUR, 
as shown in Fig.~\ref{fig:nrec-epr}(c). Correspondingly, the effective magnetic field $h$ fluctuates more, and its very notion is no longer useful.
Consequently, the difference in the selfish energy can no longer be described by a global functional, as in Eq.~\eqref{eq:delta-global}. At this point, the model begins to feel its non-reciprocal nature, and the EPR is no longer zero. This effect is even more pronounced for $\theta = 60^{\circ}$, as shown in Fig.~\ref{fig:nrec-epr}(d). In this case, no LRO phase exists even at low temperature (as we discuss in App.~\ref{sec:M2-60}), making the description in terms of an almost constant magnetic field no longer appropriate; this is indeed the region with the largest EPR (see Fig.~\ref{fig:nrec-epr}(a)).

\section{ARXY model for $\theta = 60^{\circ}$}\label{sec:M2-60}

Here we comment on the numerical results for the ARXY model with $\theta=60^\circ$, which are summarized in Fig.~\ref{fig:M2-theta60}. 
%
%
We first note that,
as shown by the behavior
of the magnetization $m$ for the NRXY reported in Fig.~\ref{fig:nrec-m}(a) and for the ARXY in Fig.~\ref{fig:arxy-pd}(a), for $\theta \in (0^{\circ}, 90^\circ)$ there is no lobe with LRO at low temperatures. 
This was rationalized in Ref.~\cite{Loos2023} in terms of an XY model with randomly diluted bonds, in which each bond can be activated with probability $p$~\cite{Berche2003, Okabe2005}.
In this model, the QLRO phase at low temperature can only form if the number of activated bonds exceeds the percolation threshold $p_c$ (equal to 1/2 for the model on a square lattice). 

In the ARXY model for $\theta < 90^{\circ}$, the fraction of activated bonds is equal to 1/2 only for configurations in which the spins are aligned along the lattice direction,
where the system effectively behaves as a collection of $L$ almost independent  one-dimensional chains of length $L$: in particular, in the limit $\theta\to0$, this is approximated by one-dimensional Ising chains.
As it is well-known, one-dimensional models do not exhibit LRO at finite temperature and, similarly, no LRO phase emerges here. On the other hand, they have a critical point at $T=0$ characterized by exponential singularities of the correlation length~\cite{PATHRIA2022487}.

However, the LRO, which is absent in the thermodynamic limit, might appear as a finite-size effect for large but finite systems. This can be understood in terms of the presence of domain walls which destroy the ordered state if their free energy $F_\text{dw}$ is negative. Indeed, considering a single line of the ARXY model, a domain-wall defect has an energy $E_\text{dw} = J/2 = 1/2$ and an entropy $S_\text{dw} = \log{L}$. Accordingly, the corresponding free energy is given by $F_\text{dw} = 1/2 - T \log{L}$ and it is negative for $T_\text{dw}(L) > 1/(2\log{L})$: as shown in Fig.~\ref{fig:M2-theta60}(a), the system, in fact, acquires a finite magnetization $m$ for $T<T_\text{dw}(L=56)$. Correspondingly, the susceptibility in Fig.~\ref{fig:M2-theta60}(b) and the specific heat in Fig.~\ref{fig:M2-theta60}(c) exhibit non-analytical behavior at the corresponding temperature.
Finally, Fig.~\ref{fig:M2-theta60}(d) shows the vectorial magnetization $\mathbf{m}$, revealing the aforementioned organization into horizontal stripes at low temperatures (the equivalent ordering in vertical stripes can be obtained by changing the initial conditions in the MC dynamics). In this regime, all rows align in the same direction. As the temperature increases, different rows adopt distinct magnetization orientations until the system reaches a fully disordered configuration.

\end{document}